\documentclass[11pt]{article}

\usepackage{times}
\usepackage[version=3]{mhchem}
\usepackage{authblk}
\usepackage{hyperref} %hidelinks
\usepackage[capitalize]{cleveref}
\usepackage{booktabs}
\usepackage[referable]{threeparttablex}
\usepackage{longtable}
\usepackage{amssymb}

\usepackage{setspace}

\usepackage{multirow}
\usepackage{adjustbox}
\usepackage{tabularx}

\usepackage{geometry}
\usepackage{pdflscape}

\usepackage[title]{appendix}

\usepackage{lineno}

\usepackage{natbib}
\bibpunct[]{(}{)}{;}{a}{,}{,}

\graphicspath{{Fig/}}

\title{\textbf{The composition of Mars}}
\author[1]{Takashi Yoshizaki\thanks{Corresponding author. E-mail: \href{takashiy@tohoku.ac.jp}{takashiy@tohoku.ac.jp}}}
\author[1,2,3]{William F. McDonough}
\affil[1]{Department of Earth Science, Graduate School of Science, Tohoku University, Sendai, Miyagi 980-8578, Japan}
\affil[2]{Department of Geology, University of Maryland, College Park, MD 20742, USA}
\affil[3]{Research Center of Neutrino Sciences, Tohoku University, Sendai, Miyagi 980-8578, Japan}

\begin{document}
	
	\maketitle
	
%	\doublespacing
	
	\section*{Abstract}
	
	Comparing compositional models of the terrestrial planets provides insights into physicochemical processes that produced planet-scale similarities and differences. The widely accepted compositional model for Mars assumes Mn and more refractory elements are in CI chondrite proportions in the planet, including Fe, Mg, and Si, which along with O make up $>$90\% of the mass of Mars. However, recent improvements in our understandings on the composition of the solar photosphere and meteorites challenge the use of CI chondrite as an analog of Mars. Here we present an alternative model composition for Mars that avoids such an assumption and is based on data from Martian meteorites and spacecraft observations. Our modeling method was previously applied to predict the Earth's composition. The model establishes the absolute abundances of refractory lithophile elements in the bulk silicate Mars (BSM) at 2.26 times higher than that in CI carbonaceous chondrites. Relative to this chondritic composition, Mars has a systematic depletion in moderately volatile lithophile elements as a function of their condensation temperatures. Given this finding, we constrain the abundances of siderophile and chalcophile elements in the bulk Mars and its core. The Martian volatility trend is consistent with $ \leq $7 wt\% S in its core, which is significantly lower than that assumed in most core models (i.e., $ > $10 wt\% S). Furthermore, the occurrence of ringwoodite at the Martian core-mantle boundary might have contributed to the partitioning of O and H into the Martian core.
	\clearpage
	
%	\tableofcontents
	
	\clearpage
	
%		\linenumbers
		
	\section{Introduction}
	\label{sec:intro}
	
	Mars is the second best-explored planet in our Solar System, given multiple space missions and cosmochemical studies on Martian meteorites \citep{mcsween2014mars}. Therefore comparison of physical and chemical properties of Mars with those of the Earth can provide important insights into the origin and evolution of the rocky planets, especially conditions for habitable planet formation. Radioisotope dating of Martian meteorites demonstrates that its accretion and evolution occurred earlier than that of the Earth \citep{dauphas2011hf,kruijer2017early,bouvier2018evidence}. The rapid formation of Mars is consistent with a pebble accretion and/or runaway and oligarchic growth model, depending upon model assumptions \citep{dauphas2011hf,johansen2015growth,levison2015growing}. Thus, a comparison of the composition of Mars and the Earth \citep{mcdonough1995composition,mcdonough2014compositional} will provide insights into processes of planetary formation and evolution. \bigskip
	
	Compositional modeling of terrestrial planets requires determining the abundances and distribution of elements, given limited chemical data from their silicate shell, knowledge of the behavior of elements in different P-T-composition-\textit{f}\ce{O2} conditions (\cref{tab:elem_class}), and constraints from their geodetic properties. A compositional model for the bulk planet and its core and mantle can be used to understand the many and markedly different processes involved in its accretion and differentiation. Models for the chemical composition of Mars \citep[e.g.,][]{,morgan1979chemical,longhi1992bulk,wanke1994chemistry,lodders1997oxygen,sanloup1999simple,halliday2001accretion,burbine2004determining,taylor2013bulk} have been reviewed recently by \citet{taylor2013bulk}. Limited cosmochemical constraints and seismic data from Mars make it difficult to evaluate critically these competing models. Importantly, most existing models assume Mars' major element composition equates to that in CI carbonaceous chondrites. Chondrites are undifferentiated assemblage of metal and silicate \citep{scott2014chondrites}. Chondritic meteorites, especially the CI carbonaceous chondrites, are chemically similar to the solar photosphere \citep[e.g.,][]{,palme2014solar}, which is taken to reflect the Sun's abundances of non-gaseous elements. At $ > $99\% of mass of the Solar System, understanding the Sun's composition and that of chondrites, the building blocks of the terrestrial planets, is key to understanding the sources and processes involved in making the planets. \bigskip
	
	The W{\"a}nke and Dreibus family of models \citep{wanke1981constitution,wanke1987chemistry,dreibus1984accretion,dreibus1987volatiles,wanke1988chemical,wanke1994chemistry} is the most widely accepted compositional model of Mars. It is based on the chemical composition of Martian meteorites and assumes that Mn and more refractory elements (\cref{tab:elem_class}) are in CI-like proportions in Mars, and considers CI chondrites as one of the components of Mars' building materials. Abundances of other less refractory elements are determined from chemical correlations with refractory or major elements. Many studies use a similar approach  \citep{longhi1992bulk,halliday2001accretion,taylor2013bulk}. \citet{taylor2013bulk} revisited and updated the W{\"a}nke and Dreibus model using the more abundant, recent chemical data for Martian meteorites and spacecraft observations, and found no significant difference with the W{\"a}nke and Dreibus model. This model is the standard for most geochemical and geophysical modeling \citep[e.g.,][]{,sohl1997interior,khan2018geophysical} and experimental work \citep[e.g.,][]{,bertka1997mineralogy,bertka1998density}. \bigskip

  As an alternative approach, \citet{morgan1979chemical} determined Martian refractory and moderately volatile element abundances based on K/Th (i.e., volatile/refractory) ratio of 620, which was measured for the Martian surface by a gamma-ray survey (GRS) in the Mars 5 orbiter mission in 1970s \citep{surkov1977gamma}. This K/Th value is significantly lower than recently observed values \citep[$ \sim $5300;][]{,taylor2006bulk,taylor2006variations}. The \citet{morgan1979chemical} model also assumes CI-like Mg/Si ratio for Mars as did by \citet{wanke1994chemistry}. \bigskip
	
	The chemical composition of Mars has also been estimated using a mixing ratio of chondritic meteorites that can reproduce the O-isotopic composition of Martian meteorites \citep{lodders1997oxygen,sanloup1999simple,burbine2004determining}. This approach estimates high volatile abundances for the terrestrial planets \citep[e.g., K/Th $ \sim $ 16400 for Mars;][]{,lodders1997oxygen}, which are not consistent with the measured planetary surface compositions \citep{surkov1986venus,surkov1987uranium,mcdonough1995composition,taylor2006bulk,taylor2006variations,peplowski2011radioactive,prettyman2015concentrations}. \bigskip
	
	Here we present an alternative compositional model for Mars. Given the recent improvements in our understandings on the composition of the solar photosphere and meteoritic samples (\cref{sec:recent}), we avoid the CI chondrite assumption. We base our model on data from Martian meteorites and spacecraft observations and use a method that was previously applied to predict the Earth's composition. We determine a unique composition for bulk silicate Mars (BSM) and a best fit, non-unique model for its core composition. By establishing the systematic depletion in volatile elements in the BSM, we show that the core has $\leq$7 wt\% S along with O and H as light elements. We discuss similarities and differences between the Earth and Mars and possible causes of these differences elsewhere. \bigskip

	\section{Recent developments in understanding the Solar System}
	\label{sec:recent}
	
	Over the last decade technological advances and insights have revealed markedly new perspectives about the Sun's composition and restrictions in the radial distribution of certain chondritic materials. Multiple challenges have been advanced regarding the solar photosphere's composition, weakening the use of CI chondrites as a proxy for the bulk solar composition. Spectroscopic observations of the solar photosphere and assumptions about local thermodynamic equilibrium in modeling the photosphere's composition are used to constrain its elemental abundances \citep[e.g.,][]{,asplund2009chemical}. The {\it solar metallicity problem} notes the significant difference in estimates of the sun's metallicity (\(Z_\odot\), abundance of elements in the Sun heavier than He) from spectroscopic observations versus helioseismology \citep[e.g.,][]{,basu2008helioseismology,Haxton13,bergemann2014solar}. The former method suggests 30 to 40\% lower metal content in the Sun and the finding is at 5$ \sigma $ to 15$ \sigma $ outside the limits set by helioseismology for the Sun's surface He abundance, the sound speed in the convective zone, and the depth of the convective zone boundary. Data from experiments on the opacity of metals in high temperature plasma \citep{bailey2015higher,nagayama2019systematic}, the composition of solar wind particles \citep{schmelz2012composition}, and measurements of the solar neutrino flux \citep{Haxton13,Borexino18} are in mutual agreement with findings from helioseismology, regarding the Sun's metallicity. One solution to the problem is to have a significant increase in the Sun's abundance of Mg, Si, S and Fe, which leads to a hotter core temperature \citep{basu2008helioseismology,asplund2009chemical,bergemann2014solar}. \bigskip
	
	It is also important to recognize the accretion settings of different chondritic parent bodies. Isotopic distinctions (e.g., O, Ni, Cr, Ti, Mo, W) are now clearly established for the non-carbonaceous and carbonaceous (NC and CC, respectively) meteorite groups, including the chondrites \citep{warren2011stable,dauphas2016mass,kruijer2017age}. These differences likely originated because of a limited radial transport in the accretion disk, which may have been controlled by an early-formed young Jupiter \citep{walsh2011low,kruijer2017age,raymond2017origin,desch2017effect}. As members of the inner Solar System NC group, Mars and Earth are isotopically most similar to ordinary and enstatite chondrites, respectively, whereas CI chondrites are isotopically a part of the CC group meteorites, which are taken to sample the outer Solar System building blocks. Trace element chemistry of NC and CC meteorites supports this isotopic divide of two groups \citep{dauphas2015thulium,barrat2016origin}. The $ < $10\% variation in relative abundances of refractory elements among NC and CC chondrites \citep[e.g.,][]{,masuda1957simple,coryell1963procedure,larimer1988refractory,wasson1988compositions,bouvier2008lu} is used to constrain compositional models of the Earth by accepting these chondritic refractory element ratios \citep{mcdonough1995composition,palme2014cosmochemical}. Thus, we recommend not using a CI-chondrite compositional model for the inner terrestrial planets, including Mars, especially for non-refractory elements. Here we develop a compositional model for Mars that is based on Martian rocks and is free of the CI chondrite assumption for non-refractory elements. As a reference frame for comparing meteorites and planetary compositions, including for the production of chondrite-normalized patterns (e.g., rare earth element (REE) diagrams), the CI composition	provides a useful standard. In this paper, we use the CI reference frame without it having any implications for the origin of materials that formed Mars and other terrestrial planets.
	\bigskip
		
	%Thus composition of Mars needs to be constrained based directly on data from Martian rocks. Here we develop a new compositional model of Mars constrained by data from Martian meteorites and satellite observations. We applied a cosmochemical approach which is similar to that used for the compositional modeling of the Earth \citep{mcdonough1995composition}. The model is chemically and physically self-consistent, specifically matching planetary geodetic observations (mass, density and moment of inertia (MOI)). Based on the chemical modeling, we provide new constraints on the interior structure of Mars. This paper describes a detailed method used in the compositional modeling of Mars, which is compared with that of the Earth \citep{mcdonough1995composition} in a companion paper \citep{yoshizaki2019mars_short}.  \bigskip
	
	\section{Data}
	\label{sec:data}
	
	A compilation of chemical and isotopic data of Martian meteorites was used in this study, with most data coming from the Martian Meteorite Compendium \citep{righter2017martian} and the online MetBase database (\href{https://metbase.org/}{https://metbase.org/}). Shergottites, especially poikilitic (previously identified as lherzolitic) samples, were used to estimate the composition of the bulk silicate Mars (BSM). These samples have experienced crystal accumulation, but their composition maintain critical imprints of their primordial origins. We complement this suite of rocks with data from olivine and basaltic shergottites in order to elucidate melt-residue correlations. In most cases we excluded nakhlites, chassignites and other ungrouped Martian meteorites (e.g., Allan Hills (ALH) 84001) from our analyses because these samples are variably thermally modified cumulates, or show more complex lithologies rather than being simple melt-derived rocks \citep{mcsween2008martian,day2018martian}. Paired Martian meteorites (e.g., Northwest Africa (NWA) 2975 and NWA 2986) are treated as one sample. Several analyses which are unrepresentative due to sample heterogeneity or terrestrial contamination are also excluded from our dataset. Average values of elemental abundances in each Martian meteorite are calculated and used in the compositional modeling. We also use data from the Martian surface as measured by spacecraft missions using GRS \citep{taylor2006bulk,taylor2006variations,boynton2008elemental}. For some elemental abundances, we adopted estimates by recent studies \citep[e.g.,][; see \cref{sec:BSM_comp}]{,yang2015siderophile,wang2017chalcophile,tait2018chondritic}. Errors are reported as 1 standard deviations, unless otherwise noted. For a CI chondritic composition (\cref{tab:CI_comp}), we adopted the value proposed by \citet{palme2014solar} for most elements, with some modifications for halogens \citep{clay2017halogens}, Mo, Tl, Bi \citep{wang2015mass}, highly siderophile elements \citep{day2016highly}, and U \citep{wipperfurth2018earth}.	\bigskip
	
	We note that our compositional modeling of Mars is based on limited chemical data from Martian rocks that might not be representative samples of the Martian crust or mantle \citep[e.g.,][ and references therein]{taylor2013bulk}. As is likely the case for the Earth, there is likely no residual domain of the primitive BSM remaining. In addition, multiple evidence indicate compositional heterogeneity in the Martian mantle and crust. Incompatible element compositions \citep{mclennan2003large}, radiogenic isotope systematics \citep[e.g., initial \ce{^{87}Sr}/\ce{^{86}Sr} and \ce{^{143}Nd}/\ce{^{144}Nd};][]{norman1999composition,blichert1999lu,symes2008age} and redox state \citep{wadhwa2001redox,hui2011petrogenesis}	of shergottites indicate the presence of at least two distinct sources in the Martian mantle, whose origin remains unconstrained \citep[e.g.,][]{symes2008age,brandon2012evolution,mcsween2015petrology}. Furthermore,  radiogenic isotope systematics of nakhlites and chassignites indicate a third reservoir which was moderately depleted in incompatible elements \citep{jones2003constraints,foley2005early,treiman2005nakhlite,mccubbin2013petrogenetic,day2018martian,udry20181}. Lower K, Th and Fe abundances and the lower K/Th ratio of shergottites than that of the Martian crust measured by Mars Odyssey GRS, are also consistent with diverse sources in the Martian mantle \citep{mclennan2003large,taylor2006bulk}. These observations may reflect a highly heterogeneous mantle and possibly unrepresentative sampling of the Martian crust by Martian meteorites. The compositional model of Mars proposed in the current work is the best estimate based on data available today, but it is potentially influenced by these problems.
			
	\section{Composition of the bulk silicate Mars}
	\label{sec:BSM_comp}
	
	\subsection{Refractory lithophile elements}
	\label{sec:RLE}
	
	Refractory lithophile elements (RLE; \cref{tab:elem_class}) remain in the silicate shell during core-mantle differentiation and their relative abundances show limited variation (generally $<$10\%) among chondritic meteorites \citep[e.g.,][]{,masuda1957simple,coryell1963procedure,larimer1988refractory,wasson1988compositions,bouvier2008lu}. Thus, if you establish the absolute concentration of one RLE, you can calculate the abundances of all of the others based on chondritic ratios (e.g., Sm/Nd, Lu/Hf, Ca/Al). \bigskip
	%Although recently challenged by non-chondritic \ce{^{142}Nd}/\ce{^{144}Nd} and Nb/Ta values of the Earth \citep{boyet2005142nd,campbell2012evidence}, the present compositional models for Earth has chondritic relative abundances of RLE \citep{mcdonough1995composition,palme2014cosmochemical}. \bigskip
	
\begin{figure}[p]
	\centering
	\includegraphics[width=1\linewidth]{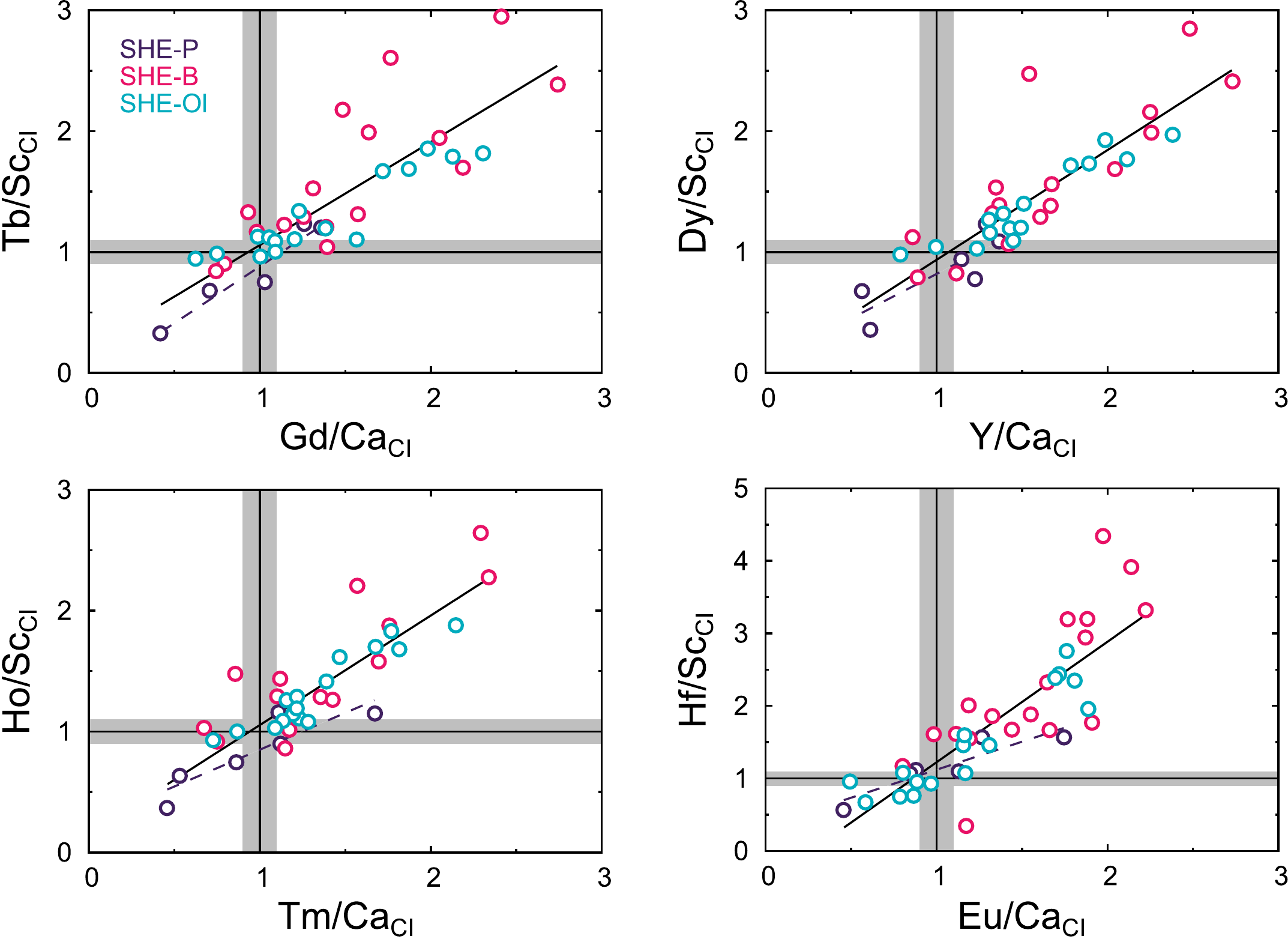}
	\caption{Ratios versus ratios of refractory lithophile elements in shergottites. The values are normalized to CI chondrite abundance (\cref{tab:CI_comp}). Horizontal and vertical gray bands show CI ratio $\pm$ 10\%. Trend lines for all shergottites and poikilitic shergottites, shown in solid black and broken purple lines, respectively, cross CI chondritic compositions, showing that these chemical trends reflect melt-residue differentiation in the Martian silicate mantle. SHE-P, SHE-B and SHE-Ol are poikilitic, basaltic and olivine-phyric shergottites, respectively.}
	\label{fig:RLE_ratio}
\end{figure}

	As a first step in determining the composition of Mars, we independently tested if Mars has chondritic ratios of the RLE. We used ratio-ratio plots to demonstrate that the trends cross at the intersection of the primitive joint chondritic compositions when using two pairs of refractory lithophile elements with different incompatibilities (\cref{fig:RLE_ratio}). Poikilitic shergottites are interpreted to be olivine-pyroxene cumulates derived from basaltic magma, whereas other shergottites are considered to have formed as lava flows \citep[e.g.,][]{mcsween2014mars}. As has been observed for terrestrial komatiites \citep[e.g.,][]{sun1982chemical,arndt1986crustally}, we find that cumulates and their co-existing melts (basaltic shergottites) fall along differentiation trends that project back to residual peridotite compositions (\cref{fig:RLE_ratio}). Thus, these chemical trends see through previous melt-residue differentiation events back to their primordial compositions, which are chondritic RLE ratios, within uncertainties. Given this finding, all RLE are taken to be in chondritic relative proportions in the BSM and, by implication, in bulk Mars too. From this starting point, we can directly determine the BSM abundances of most elements, whose concentrations are correlated with ratios or abundances of refractory lithophile elements in Martian rocks, due to chemical trends resulting from melt-residue differentiation.  \bigskip
	
\begin{figure}[p]
	\centering
	\includegraphics[width=1\linewidth]{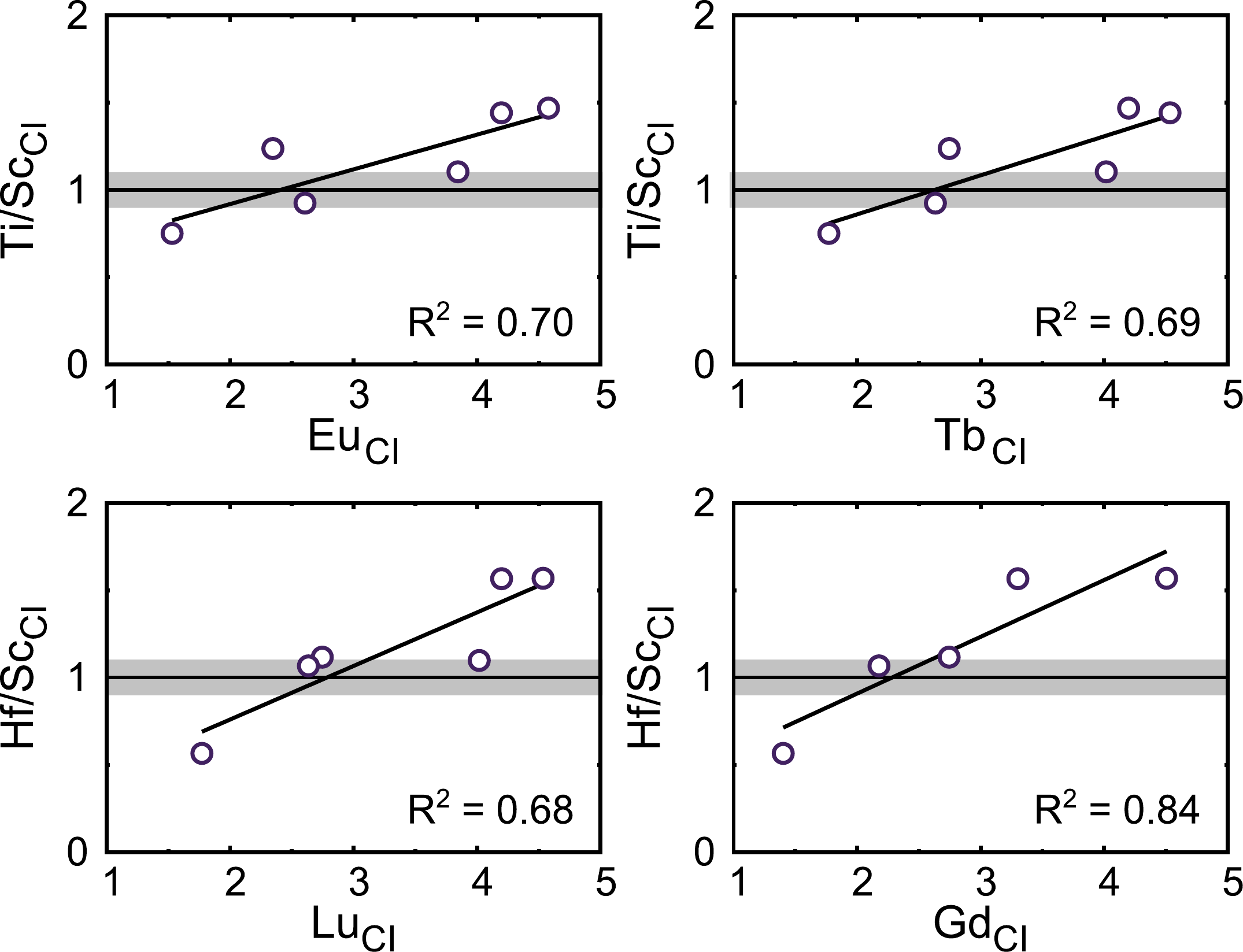}
	\caption{Ratios versus abundances of refractory lithophile elements in poikilitic shergottites. The values are normalized to CI chondrite ratios/abundance (\cref{tab:CI_comp}). Horizontal gray bands show CI ratio $\pm$ 10\%. The correlations among multiple element combinations indicate that refractory lithophile element abundance in the BSM is 2.26 times higher than in the CI chondrites (cf. 2.75 $\times$ CI in the BSE \citep{mcdonough1995composition}).}
	\label{fig:RLE}
\end{figure}

	The absolute abundance of the RLE are established using variation diagrams involving a single RLE versus an RLE ratio (\cref{fig:RLE}). The poikilitic shergottite are the best recorders of melt depletion trends, providing an accurate estimate of the primitive BSM composition. Using these melt depletion trends for multiple element combinations, we estimate the absolute abundance of the RLE in the BSM at $ \sim$2.26 times that in CI chondrites with $ \sim $10\% uncertainties \citep[cf. 2.75 $ \times $ CI in the bulk silicate Earth (BSE);][]{mcdonough1995composition}. The uncertainty for the absolute abundances of RLE in the BSM was determined from an analysis of the uncertainties associated with the intercepts of Hf/Sc or Ti/Sc and the heaviest nine REE. The Hf/Sc and Ti/Sc ratios were used because they represented the greatest differences in D-values with Hf and Ti being the most incompatible and Sc being the least incompatible. Doing so provides the greatest leverage in the regression analyses and their uncertainties. The estimated RLE abundance in the BSM is $ \sim $20 \% higher than that of the \citet{wanke1994chemistry} model.
	\bigskip

	\subsection{Major elements (Mg, Si, Fe)}
	\label{sec:major}
	
\begin{figure}[p]
	\centering
	\includegraphics[width=1\linewidth]{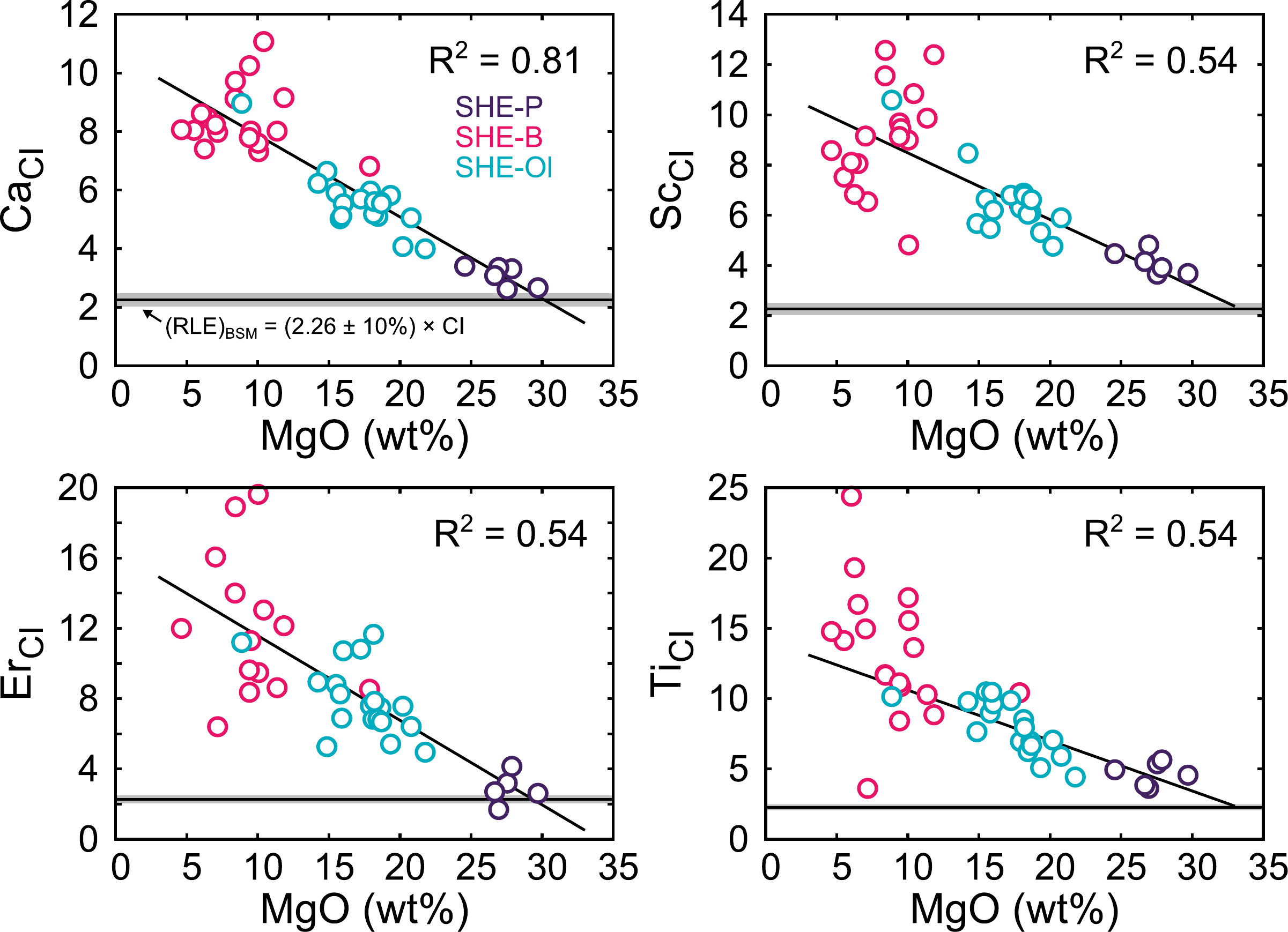}
	\caption{Magnesium oxide contents versus CI-normalized abundances of refractory lithophile elements in shergottites. Horizontal gray bands show the CI-normalized abundance of refractory lithophile elements in the BSM with 10\% uncertainties. Intercepts of the correlation lines and the BSM composition suggest 31.0 $\pm$ 2.0 wt\% MgO in the BSM. SHE-P, SHE-B and SHE-Ol are poikilitic, basaltic and olivine-phyric shergottites, respectively.}
	\label{fig:Mg_RLE}
\end{figure}

\begin{figure}[p]
	\centering
	\includegraphics[width=1\linewidth]{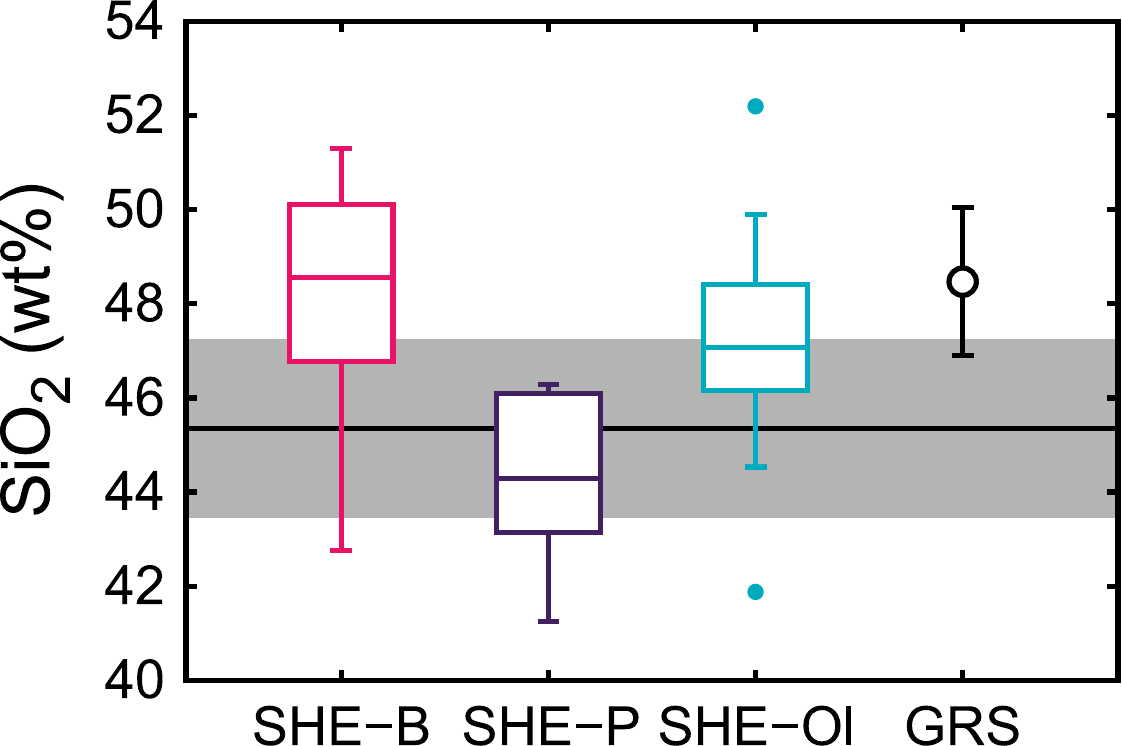}
	\caption{Silicon oxide data from shergottites and GRS survey of Martian surface. The box plots show median, minimum, maximum, first and last quartile of \ce{SiO2} contents in shergottites. Outliers which are more than 1.5 times the interquartile range from the end of boxes are shown in dots. The Martian surface composition is determined by gamma-ray spectroscopy (GRS) \citep{boynton2007concentration}, whose error is in 1 standard deviation. Horizontal line and gray bands show BSM abundance of \ce{SiO2} of 45.5 $ \pm $ 1.8 wt\%.  SHE-P, SHE-B and SHE-Ol are poikilitic, basaltic and olivine-phyric shergottites, respectively.}
	\label{fig:SiO2}
\end{figure}

	Magnesia correlates negatively with RLE abundances in shergottites (\cref{fig:Mg_RLE}) and trends for multiple RLE establish the MgO abundance at 31.0 $ \pm $ 2.0 wt\% for the BSM. There is limited variation in silica contents in shergottites (\cref{fig:SiO2}), reflecting silicon's bulk distribution coefficient of $ \sim $1 during silicate melt production. The \ce{SiO2} content of the Martian surface as measured by GRS \citep{boynton2007concentration} overlaps with the range found in basaltic shergottites, confirming the bulk crust of Mars is basaltic \citep{mcsween2009elemental}. By averaging the \ce{SiO2} abundances in shergottites, we estimate 45.5 $ \pm $ 1.8 wt\% \ce{SiO2} for the BSM. Assuming no Si or Mg in the Martian core, our BSM and bulk Mars model compositions have a Mg/Si value of 0.88 $ \pm $ 0.07, which agrees with an estimate based on Si isotope systematics \citep[0.86 $ \pm $ 0.05;][]{,dauphas2015planetary}. Uncertainty in the Mg/Si value for the BSM does, however, overlap with average values for ordinary ($\sim$0.82) and carbonaceous chondrites ($\sim$0.92). \bigskip
	
	As compared to the Earth's mantle, the Martian mantle is more oxidized and Martian meteorites, including the least evolved samples, have distinctly lower Mg\# (atomic ratio of Mg/(Mg + Fe)) relative to basalts from the Earth \citep[e.g.,][]{,wadhwa2001redox,wadhwa2008redox,herd2002oxygen}. Estimates of the Mg\# of the Martian mantle range between 0.7 and 0.8, and the W\"{a}nke and Dreibus model have Mg\# = 0.75 (\cref{tab:BSM_comparison}).	The melt-rock partition coefficient for FeO ($ D_{\mathrm{FeO}}$) reported by \citet{taylor2013bulk} is 0.95 $ \pm $ 0.06. From this value, the author concluded that the Martian mantle FeO content is 18.1 $ \pm $ 2.2 wt\% (2$ \sigma_{\mathrm{m}} $), which leads to Mg\# comparable to the estimate of \citet{wanke1994chemistry}.
	 \bigskip
	
	Multiple studies suggest a higher Mg\# for the BSM than the W\"{a}nke and Dreibus model. Melting experiments by \citet{agee2004experimental} showed that the FeO abundance and high \ce{CaO}/\ce{Al2O3} ratio of shergottites cannot be reproduced by mantle differentiation from the BSM with Mg\# = 0.75, but it can be derived from a more Fe-poor mantle with an Mg\# $ \sim $ 0.80. Olivines in primitive depleted shergottites reach Mg\# of 0.86 \citep{usui2008petrogenesis}, which requires extensive ($ > $50\%) melting of the Martian primitive mantle if the BSM has Mg\# = 0.75, which is unlikely to be achieved when shergottites formed \citep{white2006experimental}. On the other hand, with an Fe-poor Martian primitive mantle (Mg\# $ \sim $ 0.80), much lower degrees of mantle melting can produce the high-Mg olivine \citep{white2006experimental,collinet2015melting,mccoy2016experimentally_conf}. \citet{collinet2015melting} concluded that the Martian mantle has heterogeneous Mg\# ranging between 0.75--0.82. The Fe-poor BSM model is multiply supported by geochemical and geophysical modeling \citep{borg2003petrogenetic,draper2005crystallization,minitti2006new}. Given these observations, we prefer to adopt the FeO-poor (Mg\# = 0.79 $ \pm $ 0.02) primitive Martian mantle composition, which gives an FeO content of 14.7 $ \pm $ 1.0 wt\% in the BSM. Collectively, these findings for the RLE (Al, Ca and Ti) oxides, magnesia, silica and ferrous iron sets the BSM composition at a total of 97.8 wt\% (\cref{tab:BSM_major}). \bigskip

	\subsection{Non-refractory lithophile elements}
	\label{sec:non-RLE}
	
	\subsubsection{Manganese, chromium and vanadium}
	\label{sec:MnCrV}
	
	The MnO concentration in the BSM \citep{wanke1994chemistry,taylor2013bulk} is established foremost from the nearly constant FeO/MnO ratio in Martian meteorites (shergottite average 39.4 $ \pm $ 6.7) and the Martian mantle's FeO content, producing a BSM having 0.37 $ \pm $ 0.07 wt\% MnO. An alternative approach uses experimental studies on the partitioning of Mn at 1--2 GPa during melting \citep{baratoux2011thermal,filiberto2011fe2+}. The measured olivine/melt partition coefficient for MnO \citep{takahashi1983melting,herzberg1996melting,walter1998melting,wasylenki2003near,le2011mineralogical} is 0.93 $ \pm $ 0.04 \citep{taylor2013bulk}. Using this value and a shergottite average MnO content (0.48 $ \pm $ 0.06 wt\%) predicts 0.44 $ \pm  $ 0.06 wt\% MnO in the BSM. The MnO content of the BSM obtained using these methods are in agreement within uncertainties.  \bigskip
	
	Chromium correlates with Al ($ R^2 $ = 0.72) in poikilitic and olivine-phyric shergottites \cref{fig:Cr}). Where the Cr-Al trend crosses the Al content of the BSM yields 0.88 $ \pm $ 0.15 wt\% \ce{Cr2O3} in the BSM. The CI-normalized abundances in the BSM for Cr and major elements (Mg, Si and Fe), with similar condensation temperatures \citep{lodders2003solar}, is used to conclude that the Martian inventory of Cr is hosted solely in the mantle.  \bigskip
	
	In the solar nebula, vanadium behaves as a refractory lithophile element \citep{lodders2003solar}. During the formation of the Earth's core, V behaved equally lithophile and siderophile \citep{ringwood1990system,wade2005core,wood2006accretion,wood2008core,corgne2008metal,siebert2013terrestrial}, consequently half of the Earth's inventory of V is considered to be in the core \citep{mcdonough2014compositional}. Estimates of the V content of the BSM using correlation diagrams yields $\sim$130 ppm V, which is an equivalent concentration to that of the other RLE ($ \sim $2.26 $ \times $ CI) (note, here and throughout the paper ppm and ppb will refer to parts per million and billion by weight, respectively). Thus, we conclude that V behaved exclusively as a lithophile element during Mars' core formation and is wholly concentrated in the BSM. \bigskip

	\subsubsection{Sodium and potassium}
	\label{sec:NaK}
	
	A log-log abundance plot of Na and Al in poikilitic and olivine-phyric shergottites show a well-defined slope $ \sim $1 correlation ($ R^2 = $ 0.83; \cref{fig:Na}), indicating nearly equal incompatibility of these elements during a partial melting of the Martian mantle. The limited variation in Na/Al ratio in shergottites (0.44 $ \pm $ 0.10) is consistent with a \ce{Na2O} content of 0.59 $ \pm $ 0.13 wt\% in the BSM. \bigskip
	
	The radioactive elements, K, Th and U, have similar partitioning behavior during mantle melting. Consequently, the Martian surface K/Th value of 5300 $ \pm $ 220, measured by GRS aboard Mars Odyssey \citep{taylor2006bulk,taylor2006variations}, is taken as the bulk K/Th value. Both Th and U are RLE and this K/Th value gives 0.043 $ \pm $ 0.005 wt\% \ce{K2O} (360 ppm K; \cref{tab:BSM_all}) in the BSM. Our BSM model has Na/K of 12 $ \pm $ 3, comparable to the Earth's ratio (11) and overlapping within errors of the lower chondritic value \citep[9.1 $ \pm $ 1.3;][]{,wasson1988compositions}. Importantly, this ratio of elements increases as a function of the relative condensation temperature, such that Na/K increases from CI to CM to CO/CV chondrites as refractory to volatile element ratios increase \citep{wasson1988compositions}. This trend is consistent with higher Na/K in the Earth and Mars. 
	%The Mn/Na ratio of the BSM (0.66 $ \pm $ 0.20) is slightly higher than that of chondritic meteorites \citep[0.41 $ \pm $ 0.06;][]{,oneill2008collisional}, but in a range of the estimated bulk Earth composition \citep[0.5--0.7;][]{,siebert2018chondritic} (\cref{fig:MnNa_MnMg}). \bigskip
	
	\subsubsection{Rubidium}
	
\begin{figure}[p]
	\centering
	\includegraphics[width=1\linewidth]{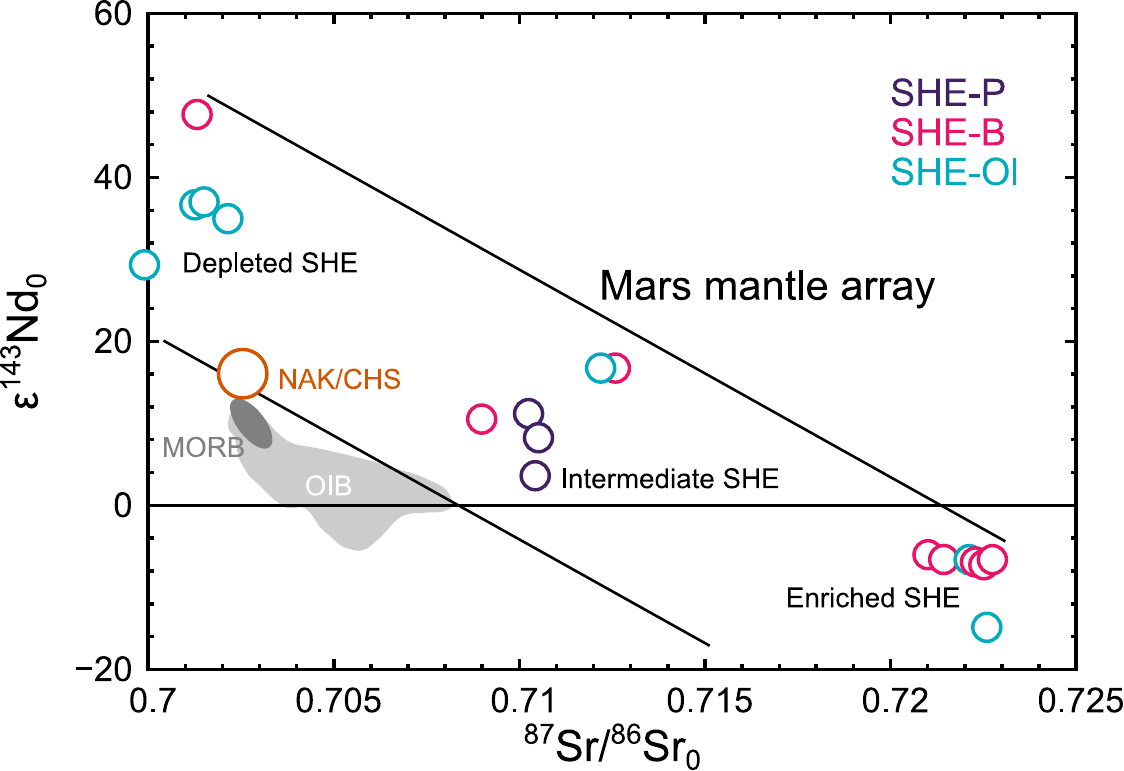}
	\caption{Initial \ce{^{87}Sr}/\ce{^{86}Sr} versus $ \mathrm{\varepsilon}$\ce{^{143}Nd} of Martian meteorites. Modified after \citet{day2018martian}. Compositions of individual shergottite samples and a compositional range for nakhlites and chassignites are shown. Isotopic compositions of terrestrial basalts (mid-ocean ridge basalt (MORB) and ocean island basalts (OIB) are also shown. Assuming that Martian meteorites represent isotopic composition of the Martian mantle and Rb/Sr fractionation at 4.56 Ga, the BSM abundance of Rb is estimated to be 1.6 $ \pm $ 0.8 ppm Rb. $ \varepsilon $\ce{^{143}Nd} = [(\ce{^{143}Nd}/\ce{^{144}Nd_{0}})$ \mathrm{_{sample}} $/(\ce{^{143}Nd}/\ce{^{144}Nd_{0}})$ \mathrm{_{chondritic}} $ $ - $ 1] $ \times ~ 10^4$. Data are from \citet{day2018martian} and references therein.}
	\label{fig:Nd-Sr}
\end{figure}

	Martian meteorites show a negative trend in a plot of initial $ \varepsilon $\ce{^{143}Nd} versus \ce{^{87}Sr}/\ce{^{86}Sr} values, establishing the Martian mantle array (\cref{fig:Nd-Sr}). The Martian mantle array is shifted to higher \ce{^{87}Sr}/\ce{^{86}Sr} values compared to the Earth's mantle array, consistent with Mars' proportionally higher content of volatile elements. The Martian mantle array yields an initial \ce{^{87}Sr}/\ce{^{86}Sr} value of 0.709--0.721 in the BSM assuming an initial $ \varepsilon $\ce{^{143}Nd} = 0 for the planet. If Rb/Sr fractionation took place at the earliest stages of accretion, then Mars' initial \ce{^{87}Sr}/\ce{^{86}Sr} value  is consistent with a Rb/Sr of 0.08 $ \pm $ 0.04, and thus 1.5 $ \pm $ 0.8 ppm Rb in the BSM. Analogous to Na/K, K/Rb ratio for Mars is 300, which is higher than the CI value (250) and not as volatile depleted as the Earth's value (400). \bigskip
	
	Alternatively, a log-log concentration plot of Rb versus La (slope = 1.0 $ \pm $ 0.1; $ R^2 $ = 0.84; \cref{fig:Rb}) reveals their similar partition coefficients during mantle melting. Using La (an RLE) abundance in the BSM, we predict 0.9 $ \pm $ 0.4 ppm Rb in the BSM, comparable to that based on the Nd- and Sr-isotopic systematics. An average of these estimated values yields 1.2 $ \pm $ 0.4 ppm Rb in the BSM. \bigskip
	
	\subsubsection{Cesium}
	
	In a log-log concentration plot, Cs and La show a trend with a slope of 1.0 $ \pm $ 0.1 ($ R^2 $ = 0.78; \cref{fig:Cs}). The Cs/La ratio in shergottites (0.14 $ \pm $ 0.10) leads to 0.08 $ \pm $ 0.06 ppm Cs in the BSM. This value is consistent with an average Rb/Cs ratio of shergottites (16.0 $ \pm $ 3.0) which indicates 0.07 $ \pm $ 0.03 ppm Cs in the BSM. \bigskip
	
	\subsubsection{Lithium}
	
	Lithium correlates with Nb and the LREE in Martian meteorites (\cref{fig:Li}). Intercepts of the Li-RLE trends and the BSM abundances of the RLE ($ \sim $2.26 $ \times $ CI) yields Li concentration in the BSM of 1.8 $ \pm $ 0.4 ppm.
	
	\subsubsection{Boron}
	
	Boron and Ca are correlated in poikilitic and olivine-phyric shergottites (\cref{fig:B}), except for five high-B ($ > $10 ppm B) olivine-phyric shergottite samples discussed by \citet{day2018martian}. 	
	We agree that these five high-B samples are possibly affected by terrestrial weathering \citep{curtis1980revision,yang2015siderophile,day2018martian} and exclude them from consideration. The average B/Ca (0.41 $ \pm $ 0.25) in poikilitic and olivine-phyric shergottites leads to  0.84 $ \pm $ 0.53 ppm B in the BSM. \bigskip	
	
	\subsubsection{Gallium}
	
	Gallium and Al are positively correlated in shergottites ($ R^2 $ = 0.84; \cref{fig:Ga}). An average Ga/Al ratio for shergottites (4.6 $ \pm $ 0.9) yields 8.7 $ \pm $ 1.9 ppm Ga in the BSM. Similarly, the Earth's average Ga/Al ratio for basalts and mantle rocks is 4.3 \citep{mcdonough1990comment}, the same as that for Mars. The abundance of Ga in the BSM and BSE points to its lithophile behavior during core--mantle differentiation and these cores having negligible quantities of Ga. 
	
	%	Slope of log-log plot ($ 0.75 \pm 0.05 $) indicate Ga is more compatible than Al
	
	\subsubsection{Halogens}
	
	%	Log-log concentration plot indicate that Cl is more compatible than Sr
	
	Efforts to estimate the abundances of the halogens (Cl, F, Br, I) are fraught with challenges, because halogens are fluid-mobile and thus readily lost during magma degassing and by alteration processes \citep[e.g.,][ and references therein]{,filiberto2019volatiles}. Many have attempted to filter Martian meteorite data affected by such secondary processes \citep{dreibus1985mars,dreibus1987volatiles,treiman2003chemical,filiberto2009martian,filiberto2016constraints,filiberto2019volatiles}. \bigskip
	
	\citet{filiberto2009martian} and \citet{filiberto2016constraints} constrained Cl/La value for Martian basalt to be 51 $ \pm $ 17, which is higher than the terrestrial Cl/La ratio (21 $ \pm $ 6). This Cl/La ratio and the BSM abundance of La (0.55 $ \pm $ 0.05 ppm) yields 28 $ \pm $ 10 ppm Cl in the BSM, which is our preferred value. We also observed a well-defined correlation between Cl and  Sr in Martian meteorites ($ R^2 $ = 0.79; \cref{fig:Cl}), which corresponds to 32 $ \pm $ 21 ppm of Cl in the BSM. Although errors in the Cl abundance estimated by these two methods are large, these values are consistent in general. \citet{taylor2010k} used GRS data to establish the Cl/K ratio of 1.3 $ \pm $ 0.2, which is a factor of six higher than the CI value (\cref{tab:CI_comp}). Chlorine abundance in the Martian surface reflects secondary deposition and thus high Cl/K ratios are not  an indicator of the primitive Cl content of the BSM \citep{keller2006equatorial, filiberto2019volatiles}. \bigskip
	
	There are limited F data for Martian meteorites. The few shergottites samples with coupled F and B data show a correlation ($ R^2 $ = 0.80; \cref{fig:F}), indicating 16 $ \pm $ 12 ppm F in the BSM.  \citet{filiberto2016constraints} reported shergottites and terrestrial basalts with a mean Cl/F of 0.8 $ \pm $ 0.4. Given terrestrial and Martian basalts are modified by magma degassing, which decreases the initial value, whereas alteration increases the ratio, \citet{filiberto2016constraints} suggested Cl/F $ \sim $ 1 in the Martian mantle. Using our estimate of the BSM abundance for Cl, we obtained the BSM abundance for F of $ \sim $30 ppm, our preferred value.  \bigskip
	
	An average Cl/Br ratio in Martian meteorites \citep[224 $ \pm $ 140;][]{,filiberto2016constraints} would suggest 0.12 $ \pm $ 0.09 ppm Br in the BSM.  In contrast, the chondritic Cl/Br ratio is 600, which is comparable to the BSE value \citep[e.g.,][ and references therein]{,mcdonough1995composition,palme2014cosmochemical,clay2017halogens}. If we assume $ \sim $0.12 ppm Br in the BSM, it would result in a marked spike in the volatility curve singularly for Br. This raises our suspicion and leads us to question the resiliency of the Martian meteoritic record for Br. Thus we conclude Br content of $ \sim $0.05 ppm in the BSM based on the chondritic Cl/Br ratio. \bigskip
	
	There is limited data for I in Martian meteorites, the most volatile of the lithophile elements (\cref{tab:elem_class}). We do not observe any clear correlation of the halides with other elements. Chondritic meteorites show limited variation of I/Cl ratios \citep{clay2017halogens} comparable to that observed for the BSE composition \citep[e.g.,][ and references therein]{,mcdonough1995composition,palme2014cosmochemical,clay2017halogens}. Assuming the BSM has CI-like I/Cl ratio \citep[(5.0 $ \pm $ 1.7) $ \times $ 10$ ^{-4} $;][]{clay2017halogens}, we estimate 14 $ \pm $ 7 ppb I in the BSM. \bigskip

	\subsection{Siderophile and chalcophile elements in the BSM}
	\label{sec:RSE}
	\subsubsection{Refractory and major siderophile elements in the BSM}
	
	These elements include W and Mo along with Fe, Ni, and Co (\cref{tab:elem_class}). Nickel in poikilitic and olivine-phyric shergottites correlate with Mg  (\cref{fig:Ni}) and their average Ni/Mg ratio is 19.4 $ \pm $ 5.0, consistent with 0.046 $ \pm $ 0.012 wt\% NiO in the BSM. Likewise these rocks have an average Co/Ni ratio of 0.27 $ \pm $ 0.10, consistent with 96 $ \pm $ 44 ppm Co in the BSM. \bigskip
	
	Molybdenum abundance in shergottites is highly variable between 0.05--0.7, and is not well correlated with lithophile elements. The origin of the wide variation in Mo contents in shergottites is not constrained. \citet{yang2015siderophile} attributes this variation to hydrothermal processes, whereas \citet{noll1996role} argues that Mo is not fluid-mobile. Based on a broad Mo-Ce co-variation, \citet{righter2011moderately} and \citet{yang2015siderophile} estimated the BSM abundance of Mo is 0.08--0.6 ppm. We estimate the BSM to have between 0.1--0.8 ppm Mo. \bigskip
	
	In a log-log plot for all shergottites, W and Th shows a linear trend with a slope 1.1 $ \pm $ 0.2 (\cref{fig:W}), indicating their similar incompatibilities. Using an average W/Th ratio in shergottites (1.0 $ \pm $ 0.5), W concentration in the BSM is estimated to be 0.07 $ \pm $ 0.04 ppm. Our BSM model has Hf/W of (3.47 $ \pm $ 1.90), which is consistent with an estimate by \citet{dauphas2011hf}. \bigskip
	
	\subsubsection{Highly siderophile elements}
	\label{sec:HSE}
	
	Highly siderophile elements (HSE) include Re, Os, Ir, Pt, Ru, Rh, Pd and Au. \citet{brandon2012evolution} and \citet{tait2018chondritic} estimated the BSM abundances of the HSE based on co-variation between the HSE and MgO in shergottites. Although Re, Pd and Pt are poorly correlated with Mg, \citet{tait2018chondritic} assumed MgO $ \sim $ 35 wt\% in the BSM and obtained flat, chondrite-like HSE patterns in the BSM, with absolute abundances at (0.010 $ \pm $ 0.003) $ \times $ CI. The MgO composition of the BSM used by \citet{tait2018chondritic} is slightly higher than our estimate (MgO $ \sim $ 31.0 wt\%). However, this difference is negligible for the predicted abundances of the HSE in the BSM and so we adopt the \citet{tait2018chondritic} estimate. \bigskip
	
\begin{landscape}
	\begin{figure}[p]
		\centering
		\includegraphics[width=1\linewidth]{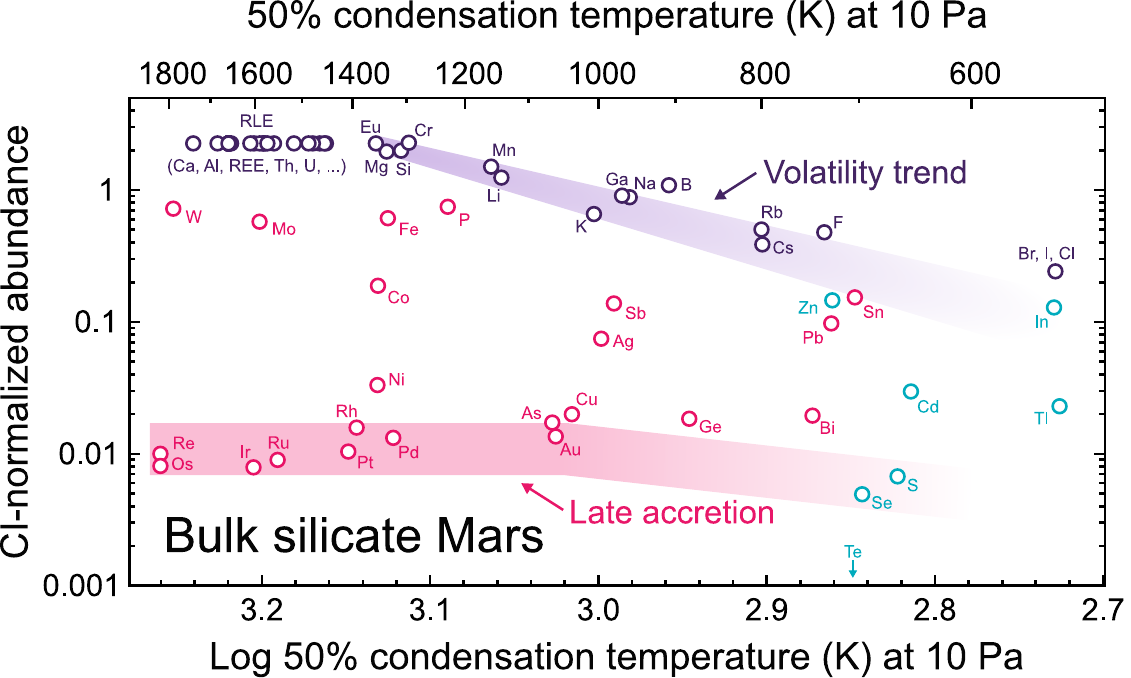}
		\caption{Abundance of lithophile (purple), siderophile (pink) and chalcophile (blue) elements in the bulk silicate Mars (\cref{tab:BSM_all}) are normalized to CI chondrites (\cref{tab:CI_comp}) and plotted against log of the 50\% condensation temperature (K) at 10 Pa (see \cref{tab:elem_class} for updates to condensation temperatures of elements).}
		\label{fig:volatility_trend}
	\end{figure}
\end{landscape}

	Similarly, Au abundances show a broad correlation with MgO in shergottites (\cref{fig:Au}). This correlation indicates $ \sim $2 ppb Au in the BSM, which corresponds to the BSM abundance of $ \sim $ 0.01 $ \times $ CI (\cref{fig:volatility_trend}). Thus it is likely that the bulk of the BSM budget of Au originated in the late accretion of materials with a chondritic HSE composition. \bigskip
	
	There are limited data for Rh in Martian meteorites with a wide variation, and Rh is not correlated with other elements in Martian meteorites. If we take the mean Rh concentration in the Martian meteorites ($ \sim $2 ppb) as the BSM abundance of Rh, this value corresponds to $ \sim $ 0.02 $ \times $ CI in the BSM, which is close to that of other HSE \citep[\cref{fig:volatility_trend};][]{,tait2018chondritic}. \bigskip
	
	\subsubsection{Moderately volatile siderophile elements}
	\label{sec:MVE_siderophile}
	
	Moderately-volatile, siderophile elements include P, Cu, Ge, As, Ag, Sb, Sn, Pb and Bi. A well-defined positive correlation for P versus Y ($ R^2 $ = 0.88), with a slope of 1.1 $ \pm $ 0.1, is seen for shergottites (\cref{fig:P}). Their average P/Y value in shergottites (0.021 $ \pm $ 0.005) is used to estimate the  \ce{P2O5} at 0.17 $ \pm $ 0.05 wt\% in the BSM. \bigskip
	
	\citet{wang2017chalcophile} estimated Cu in the BSM abundance at 2.0 $ \pm $ 0.4 ppm, based on Cu and MgO correlations in shergottites. Similarly, we find positive Cu versus Ti correlations in shergottites (\cref{fig:Cu}), except for some anomalously high Cu/Ti samples (Cu/Ti $ > $ 0.005), and their average Cu/Ti value (0.0026 $ \pm $ 0.0005) yields 2.6 $ \pm $ 0.6 ppm Cu is in the BSM, which is comparable to the estimate by \citet{wang2017chalcophile}. \bigskip
	
	%The BSM abundances of the rest of moderately volatile siderophile elements are uncertain, because they do not show any clear correlation with lithophile elements. 
	
	Germanium does not correlate with lithophile elements in Martian meteorites, as mantle melting produces little variation in Ge content in the melt and residue \citep{ringwood1966chemical}. Germanium contents in shergottites ranges from 0.8 $ \pm $ 0.4 ppm in poikilitic shergottites to 1.5 $ \pm $ 0.3 ppm in olivine-phyric shergottites and 0.7--2.0 ppm basaltic shergottites. From a weak negative correlation between Ge and MgO in poikilitic and olivine-phyric shergottites (\cref{fig:Ge}), we estimate Ge abundance in the BSM is 0.6 $ \pm $ 0.4 ppm.	\bigskip
	
	In terrestrial rocks, As correlates with Ce \citep{noll1996role}. In contrast, As in shergottites does not correlate with tested lithophile elements \citep{yang2015siderophile}. Arsenic is a fluid-mobile element \citep{noll1996role}; thus, the lack of any As-Ce correlation in shergottites might be due to hydrothermal processes \citep{yang2015siderophile}. By attributing elevated As abundance in some Martian meteorites to the hydrothermal modification and ignoring such data, \citet{yang2015siderophile} derived 30 $ \pm $ 25  ppb As in the BSM. \bigskip
	
	We do not observe any correlation of Sb and Ag with lithophile elements. \citet{yang2015siderophile} estimated the BSM abundance of Sb to be 0.01--0.03 ppm using a roughly constrained Sb/Pr ratio of Martian meteorites and the BSM abundance of Pr of 0.17 ppm from \citet{lodders1997oxygen}. We accept their estimate given that 0.21 ppm Pr in our model does not make significant changes in the estimation of the BSM abundance of Sb. A log-log plot of Sb versus Ag concentrations in shergottites (slope $ \sim $ 1; \cref{fig:Sb}) indicates their similar compatibility during differentiation. The average Sb/Ag ratio of 1.4 $ \pm $ 1.0, with Ag being 0.02 ppm in the BSM with $\sim$90\% uncertainty.  \bigskip
	
	\citet{yang2015siderophile} estimated the BSM abundance of Sn, Cd and In based on their broad correlation with lithophile elements in Martian meteorites and a model of BSM abundances for lithophile elements by \citet{lodders1997oxygen}. The observed correlation resulted in more than a factor of two uncertainties in their estimates. Although abundances of lithophile elements in our model are different from that of \citet{lodders1997oxygen}, this difference does not produce any significant decrease in the estimated BSM abundances of Cd, In and Sn in \citet{yang2015siderophile}. Thus, we accept the BSM abundances of these elements given by \citet{yang2015siderophile}. \bigskip
	
	The BSM abundance of Pb is constrained from U-Pb isotopic systematics of Martian meteorites. Martian meteorites have a wide ranges of $ \mu $ (= \ce{^{238}U}/\ce{^{204}Pb}) values and Pb isotopic heterogeneity \citep{nakamura1982origin,misawa1997u,borg2005constraints,bouvier2005age,bouvier2008case,bouvier2009martian,chen1986formation,gaffney2007uranium,moriwaki2017coupled,bellucci2018pb}. The recent review of the Martian $ \mu $ value ($\sim$3.6), the average \ce{^{208}Pb}/\ce{^{204}Pb}, \ce{^{207}Pb}/\ce{^{204}Pb} and \ce{^{206}Pb}/\ce{^{204}Pb} in Martian meteorites of 33, 12.6 and 13, respectively, and 18 ppb U (\cref{tab:BSM_all}) yields 0.26 $ \pm $ 0.05 ppm Pb in the BSM. \bigskip
	
	Bismuth is correlated with Th in shergottites (except for high-Bi samples EETA 79001, Tissint and NWA 5990; \cref{fig:Bi}) \citep{yang2015siderophile}, which yields 2 $ \pm $ 1 ppb Bi in the BSM. \bigskip
	
	\subsubsection{Moderately volatile chalcophile elements}
	\label{sec:MVE_chalcophile}
	
	Moderately-volatile, chalcophile elements include Zn, Te, Se, S and Cd. As mentioned in \cref{sec:MVE_siderophile}, we adopt the BSM abundances of Cd estimated by \citet{yang2015siderophile}. \bigskip
	
	Zinc and Lu positively correlate in shergottites ($ R^2 $ = 0.63) as do Zn and Ti ($ R^2 $ = 0.56), which yield  45 $ \pm $ 15 ppm and 40 $ \pm $ 15 ppm Zn, respectively, in the BSM (\cref{fig:Zn}). These two estimates agree within errors. 	\bigskip
	
	\citet{taylor2013bulk} estimated Zn in the BSM at 18.9 $ \pm $ 1.5 ppm, whereas as \citet{yang2015siderophile} proposed 50--70 ppm. \citet{taylor2013bulk} estimate was based on a Zn--Sc correlation, which is poorly correlated in our dataset ($ R^2 $ = 0.27). The relatively flat Zn--Sc trend observed by \citet{taylor2013bulk} leads to a poor control on an accuracy estimate. \citet{yang2015siderophile} assumed an olivine-rich Martian mantle and bulk mineral-melt partition coefficients for Zn of $\sim$1 during mantle melting \citep{le2011mineralogical,davis2013experimentally}. The positive Zn-Lu correlation in shergottites (\cref{fig:Zn}) reflects the slightly incompatible nature of Zn in the Martian mantle.  Thus, \citet{yang2015siderophile} overestimates the Zn abundance in the BSM. \bigskip
	
	There is a wide range of estimates for the abundances of S-Se-Te in the BSM \citep[][ and references therein]{,franz2019sulfur,wang2017chalcophile}. The limited variation in S/Se/Te ratios indicates little degassing loss of S, Se and Te, which led \citet{wang2017chalcophile} to derive the BSM abundance of these elements at 360 $ \pm $ 120 ppm, 100 $ \pm $ 27 ppb and 0.50 $ \pm $ 0.25 ppb, respectively. Here we adopt their estimates. \bigskip

	\subsubsection{Volatile chalcophile elements}
	\label{sec:volatile_sidero_chalco}
	
	Volatile siderophile and chalcophile elements include In, Tl and Hg. 	As described in \cref{sec:MVE_siderophile}, we adopt the BSM abundances of In estimated by \citet{yang2015siderophile} in our BSM model. Thallium is positively correlated with Sm ($ R^2 $ = 0.73; \cref{fig:Tl}) in shergottites with a slope of 1.29 $ \pm $ 0.38 in a log-log slope, indicating 4 $ \pm $ 2 ppb Tl in the BSM. 	\bigskip
	
	%There is limited numbers of Hg data for Martian meteorites. \citet{ehmann1967abundance} and \citet{weinke1978chemical} reported 0.7 ppm and 0.23 ppm Hg in Nakhla meteorite, respectively. \citet{treiman1997trace} found high Hg concentration in a nakhlite Lafayette and attributed it to the terrestrial contamination. In addition, since Hg is a highly volatile element, it be can easily lost during source magma of Martian meteorites. Therefore it is very difficult to constrain the BSM abundance of Hg based on data from Martian meteorites. Note that even the solar abundance of Hg is still poorly constrained since Hg in the solar photosphere cannot be detected \citep{lauretta1999cosmochemical,grevesse2015elemental} and Hg content in CI chondrites varies by up to three orders of magnitude between different CI samples and even between different measurements of the same samples \citep{meier2016mercury}. By assuming that Hg is completely chalcophile, \citet{palme2014cosmochemical} estimated the BSM abundance of Hg of $ \sim $6 ppb based on Hg/Se $ \sim $ 0.075 in the continental crust of East China \citep{gao1998chemical}. With a Hg contents of 0.23 ppm in Nakhlite \citep{weinke1978chemical}, which is a value less likely affected by terrestrial contamination, Nakhlite have Hg/Se $ \sim $ 0.05. Assuming that the BSM has Hg/Se ratio between that of Nakhlite and the BSM (0.05--0.08), the BSM abundance of Hg is estimated to be $ \sim $7 ppb, but the estimated value is associated with significant uncertainties. \bigskip
	Limited data exist for Hg ($\leq$0.7 ppm) in Martian meteorites  \citep{ehmann1967abundance,weinke1978chemical,treiman1997trace}. It is readily contaminated by terrestrial sources, it is a highly volatile element, and it is lost during magma transport \citep[e.g.,][]{treiman1997trace}. Therefore it is very difficult to constrain the BSM abundance of Hg. Estimates of the solar abundance of Hg (few hundred ppb) are poorly constrained and Hg is undetectable in the solar photosphere \citep{lauretta1999cosmochemical,grevesse2015elemental,meier2016mercury}. The BSM abundance of Hg is estimated to be $ \sim $7 ppb, with  uncertainty ranging from $\leq$1 to a few tens of ppb. \bigskip

	%	\begin{itemize}
	%		\item Abundances of halogen in the BSM is difficult to estimate because they can be easily volatilize and lost from lava flow surfaces (under oxidizing conditions) \citep{taylor2013bulk}
	%		\item Additional important problems in halogens is that they are fluid-mobile so that their concentration in the Martian meteorites might be affected by the secondary processes
	%		\item As proposed for the elevated abundances of Cl and Br in the Nakhla meteorite, composition of these elements in Martian meteorites can be modified due to terrestrial contamination \citep{dreibus2003comparison}
	%		%		\item Although Nakhla is the only fall meteorite among the nakhlites, this sample show clearly elevated Br content compared to other nakhlites collected from hot and cold deserts \citep{dreibus2003comparison}
	%	\end{itemize}
	
	\subsection{Atmophile elements}
	\label{sec:atmophile}
	
	The Mars' volatility trend (\cref{fig:volatility_trend}) cannot provide strong constraints on the planetary abundances of atmophile elements (H, C, N, O, noble gases; \cref{tab:elem_class}), given their distinctive behaviors compared with less volatile elements that are retained in rocks. However, approximate estimates of these elements in the BSM were predicted from the Martian volatility trend, following the practice used in \citet{mcdonough2014compositional}. An extrapolation of the Martian volatility trend to the lower temperature indicates their abundances to be 0.001--0.01$ \times $CI. \bigskip
	
	Based on the water contents of Martian meteorites and their constituting hydrous phases (e.g., apatite, amphibole, glass), the BSM is estimated to have $ \sim $140 ppm \ce{H2O} \citep[e.g.,][]{,mccubbin2016geologic,mccubbin2016heterogeneous,filiberto2019volatiles}. This \ce{H2O} content corresponds to $ \sim $16 ppm H in the BSM, which is lower than that is expected based on the volatility trend. The low H abundance in the BSM might reflect H incorporation into the core (see \cref{sec:comp_core_model}) and/or loss of H from the mantle by degassing. Estimates for the water content of the Martian mantle range between 14--250 ppm, which likely reflects a  heterogeneous water distribution in the mantle \citep[e.g.,][]{,mccubbin2010hydrous,mccubbin2016geologic,mccubbin2016heterogeneous}. The BSM might have contained $ \sim $1000 ppm \ce{H2O} if volatile-rich chondritic materials were added during a late accretion (i.e., 0.6--0.7\% by mass of Mars), as indicated by highly siderophile element abundances and Os isotope systematics in shergottites \citep{tait2018chondritic}. \bigskip
	
	%Using \ce{CO2}/\ce{H2O} values in olivine-hosted melt inclusions from the olivine-phyric shergottite Yamato 980458 \citep{usui2012origin}, which records a primitive magma composition \citep{usui2008petrogenesis}, and the BSM abundance of $ \sim $140 ppm \ce{H2O} \citep{mccubbin2016geologic,mccubbin2016heterogeneous,filiberto2019volatiles}, the BSM is estimated to have $ \sim $500 ppm \ce{CO2}, which corresponds $\sim$150 ppm C. However, \citet{filiberto2019volatiles} notes that this method overestimates the C abundance since H might have been preferentially lost from the inclusions.	\bigskip	

	\citet{filiberto2019volatiles} estimated the BSM abundances of C and N, based on mean C/H and N/H ratios in CI and CM carbonaceous chondrites \citep[2.1 and 0.11, respectively;][]{,alexander2012provenances} and a BSM estimate for its \ce{H2O} content. This method assumes the BSM inventory of these atmophile elements is dominantly by a late addition component \citep[e.g.,][]{,tait2018chondritic} with CI- and CM-chondritic chemical composition \citep{filiberto2019volatiles}. These assumptions lead to a BSM with $ \sim $140 ppm \ce{H2O}, $ \sim $32 ppm C and $ \sim $1.6 ppm N. These estimates have large uncertainties, given the degree of mantle degassing is essentially unconstrained. \bigskip

	Martian meteorites contain several noble gas components of distinct origins \citep{bogard2001martian,ott2019noble}. It is likely that significant fractions of Martian noble gas budgets are located in the Martian interior \citep{dauphas2014chemical}, but their abundances are poorly constrained. Chassigny meteorite shows high abundance of noble gases with low Ar/Xe and Kr/Xe ratios, which is considered to reflect mixing of a composition of the Martian interior component \citep{ott1988noble,bogard2001martian,mathew2001early}. Further studies are needed to understand noble gas budgets and composition in the Martian interior.

	\section{Composition of the Martian core}
	\label{sec:comp_core}
	
	\subsection{Compositional model of the Martian core}
	\label{sec:comp_core_model}
	
\begin{landscape}
	\begin{figure}[p]
		\centering
		\includegraphics[width=1\linewidth]{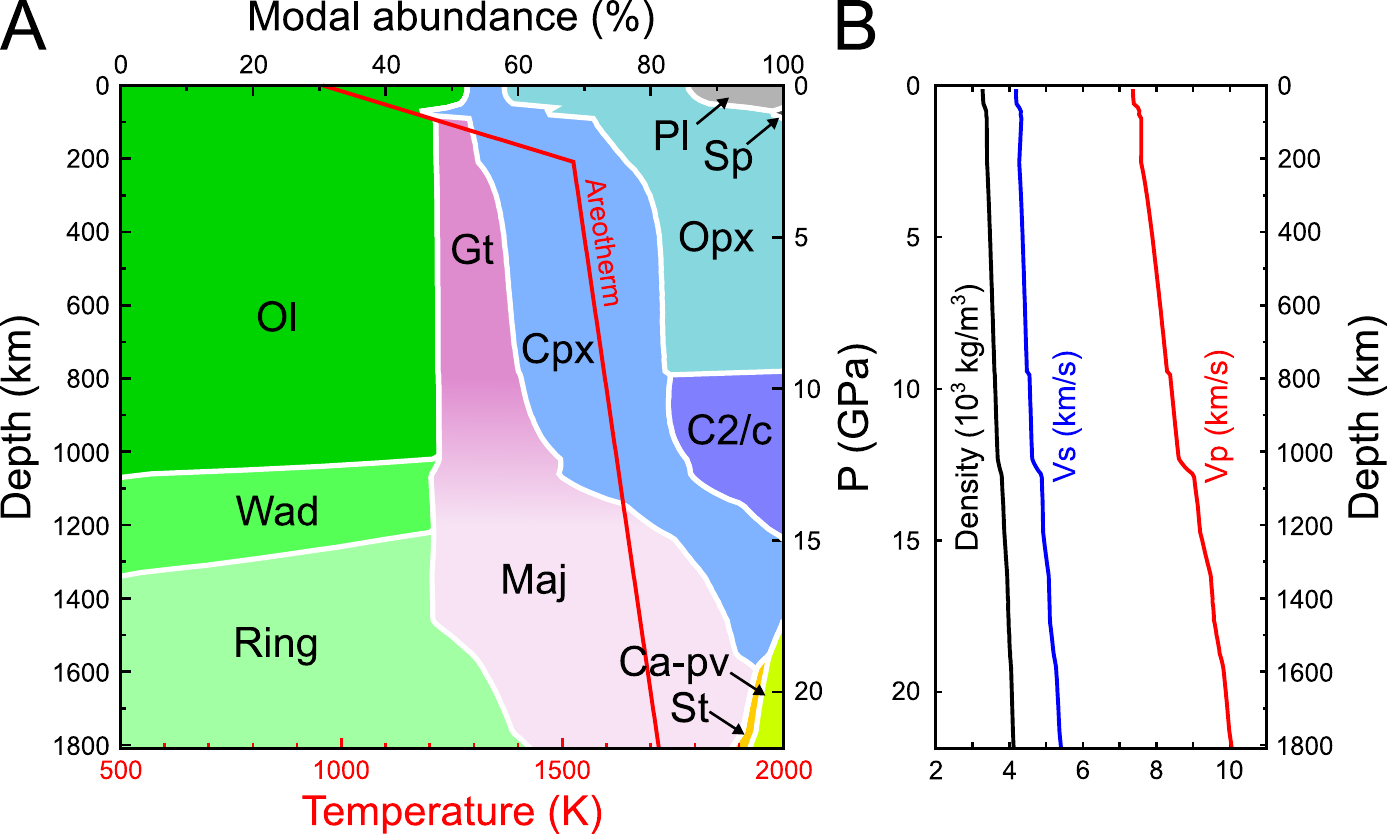}
		\caption{Mineralogy and physical properties of the Martian mantle. (A) Phase transitions in the Martian mantle. A red line shows an areotherm (Martian geotherm). (B) Depth versus P- and S-wave velocities and density from surface to the core-mantle boundary of Mars. Abbreviations: 
			Ol--olivine; 
			Wad--wadsleyite;
			Ring--ringwoodite;
			Gt--garnet;
			Maj--majorite;
			Cpx--clinopyroxene;
			Opx--orthopyroxene;
			C2/c--high-pressure clinopyroxene;
			Pl--plagioclase;
			Sp--spinel;
			Ca-pv--Ca-perovskite;
			St--stishovite.
		}
		\label{fig:phase_fig}
	\end{figure}
\end{landscape}

	To constrain the present-day Martian core composition, we modeled geophysical properties of Mars. Mineralogy, radial density distribution and seismic velocity profiles in the Martian mantle were computed using the Gibbs free energy minimization method that is employed in the thermodynamic modeling code Perple\_X version 6.8.6 \citep{connolly2009geodynamic}. Calculations for the BSM composition (i.e., primitive Martian mantle composition; \cref{tab:BSM_major}) were performed using thermodynamic parameters of \citet{stixrude2011thermodynamics} within a chemical system \ce{Na2O}--\ce{MgO}--\ce{Al2O3}--\ce{SiO2}--\ce{CaO}--\ce{FeO}. Temperature profile in the Martian mantle (areotherm) is estimated assuming an Earth-like profile \citep{katsura2010adiabatic}, mantle potential temperature of $ \sim $1,500 K \citep{baratoux2011thermal,baratoux2013petrological,putirka2016rates,filiberto2017geochemistry}, a lithosphere thickness of 200 km \citep{grott2013long} and conductive and adiabatic thermal gradients of 2.7 K/km and 0.12 K/km, respectively \citep{verhoeven2005interior}, which are consistent with the surface heat flux estimates \citep{parro2017present}.  \cref{fig:phase_fig} shows a result of the thermodynamic modeling. \bigskip
	
	We also estimated mineralogy and physical properties of the present-day Martian mantle (not the BSM), which is calculated using compositional models for the BSM (\cref{tab:BSM_major}) and the Martian crust \citep{taylor2009planetary}, and a crustal mass fraction of $ \sim $5 \% in the BSM. We observed small changes in the modal abundances of mineral species, but the density profiles are similar in the present-day and primitive mantle models. Thus, the use of the density profile obtained using the composition of primitive mantle or present-day mantle makes negligible changes in our discussion.	\bigskip
	
	Using the obtained radial density profile in the Martian mantle, we computed the Martian interior structure. Here we consider Mars as a spherically symmetric body divided into three layers (crust, mantle and core). The average crustal thickness of 50 km, which is consistent with geophysical and geochemical constraints \citep{zuber2000internal,mcgovern2002localized,mcgovern2004correction,wieczorek2004thickness,ruiz2009ancient}, is adopted in the modeling. We consider a crust of basaltic composition \citep[e.g.,][ and references therein]{,taylor2009planetary,mcsween2014mars} with a density of 3010 kg/m$ ^3 $, which is in agreement with gravity and topography studies \citep{mckenzie2002relationship,mcgovern2002localized,mcgovern2004correction,phillips2008mars}. The mass and moment of inertia (MOI) in Mars are expressed as a function of radial density distribution as:
	\begin{eqnarray}
	M = 4\pi \int_{0}^{a}   \rho(r) r^2 dr \label{eq:mass} \\
	C = \frac{8\pi}{3} \int_{0}^{a} \rho(r) r^4 dr, \label{eq:moi}
	\end{eqnarray}
	where $ M $ is mass, $ C $ is MOI, $ a $ is Mars' equatorial radius \citep[3389.5 km;][]{,seidelmann2002report}, $ \rho (r) $ is density and $ r $ is distance from the center of the planet. \bigskip
	
	Sulfur abundance in the Martian core has been of particular interests. Previous estimates vary between 3.5--25 wt\% S \citep[\cref{tab:core_comparison}; also see][]{,franz2019sulfur}. With the Martian mantle having 360 ppm S \citep{wang2017chalcophile}, the Martian volatility curve restricts the core to having $ \leq $7 wt\% S (\cref{fig:volatility_trend_S}) if its core mass fraction is 18 wt\% (see below). The low S composition for the Martian core is supported by metal-silicate partition coefficients for S \citep{rose2009effect,boujibar2014metal,wang2017chalcophile} and an experimental study of Fe isotopic fractionation under high pressure and temperature \citep{shahar2015sulfur}. If we adopt a higher S content for the BSM \citep[up to 2000 ppm;][]{gaillard2013geochemical,ding2015new}, the S content of the Martian core decreases. Thus, we conclude that the Martian core has $ \leq $7 wt\% S. \bigskip
	
	This estimate stands in contrast to widely adopted Martian core models that argue for $ > $10 wt\% S \citep{wanke1994chemistry,taylor2013bulk}. Such a high sulfur content for the bulk Mars would, in principle, treat S differently with respect to other moderately volatile elements. We find no justification for such a S enrichment (\cref{fig:volatility_trend_S}). \bigskip
	
	For the Earth's core, C, H, S, Si and O are proposed as candidate light elements that decrease its density \citep{birch1952elasticity,birch1964density}. Given the P-T-\textit{f}\ce{O2} condition of the Martian interior, we do not expect Si in the Martian core \citep{wade2005core,corgne2008metal}, which is supported by a lack of Si isotopic fractionation in Martian meteorites resulted from a metal-silicate segregation \citep{zambardi2013silicon,dauphas2015planetary}. We also exclude C as a candidate light element in the Martian core since the Martian meteorites lack C isotope fractionation unlike terrestrial rocks \citep{grady2004magmatic,wood2013carbon} and addition of C does not efficiently decrease the core density \citep{wood1993carbon,bertka1998implications}. In contrast, experimental studies indicate that O and H can be incorporated into the Martian core \citep{okuchi1997hydrogen,shibazaki2009hydrogen,tsuno2011effects} and their addition can decrease the core density \citep{badding1992high,bertka1998implications,zharkov1996internal,zharkov2005construction}. Moreover, having the Martian core-mantle boundary in contact with ringwoodite can contribute to the incorporation of H and O into the Martian core \citep{shibazaki2009hydrogen,tsuno2011effects,o2018suppressing}. Thus, we consider O and H, in addition to S, as candidates for light elements in the Martian core.  \bigskip
	
	We calculated density of the Martian core using elastic properties of liquid Fe \citep{anderson1994equation}, FeS \citep{nagamori1969density,kaiura1979densities,antonangeli2015toward,nishida2016towards} and FeO \citep{lee1974densities,komabayashi2014thermodynamics} and solid FeH \citep{umemoto2015liquid} at core pressures and temperatures and a third-order finite strain Birch-Murnaghan or Vinet equation of state \citep{stixrude2005thermodynamics}, following previous studies \citep{longhi1992bulk,sohl1997interior,bertka1998implications,verhoeven2005interior,zharkov2005construction,rivoldini2011geodesy,khan2018geophysical}. Given a lack of elastic properties of liquid FeH, we assumed that the properties of liquid FeH is similar to that of solid FeH. Temperature profile in the core is calculated based on the temperature of core-mantle boundary (\cref{fig:phase_fig}) and a convection in the score. \bigskip
	
\begin{figure}[p]
	\centering
	\includegraphics[width=1\linewidth]{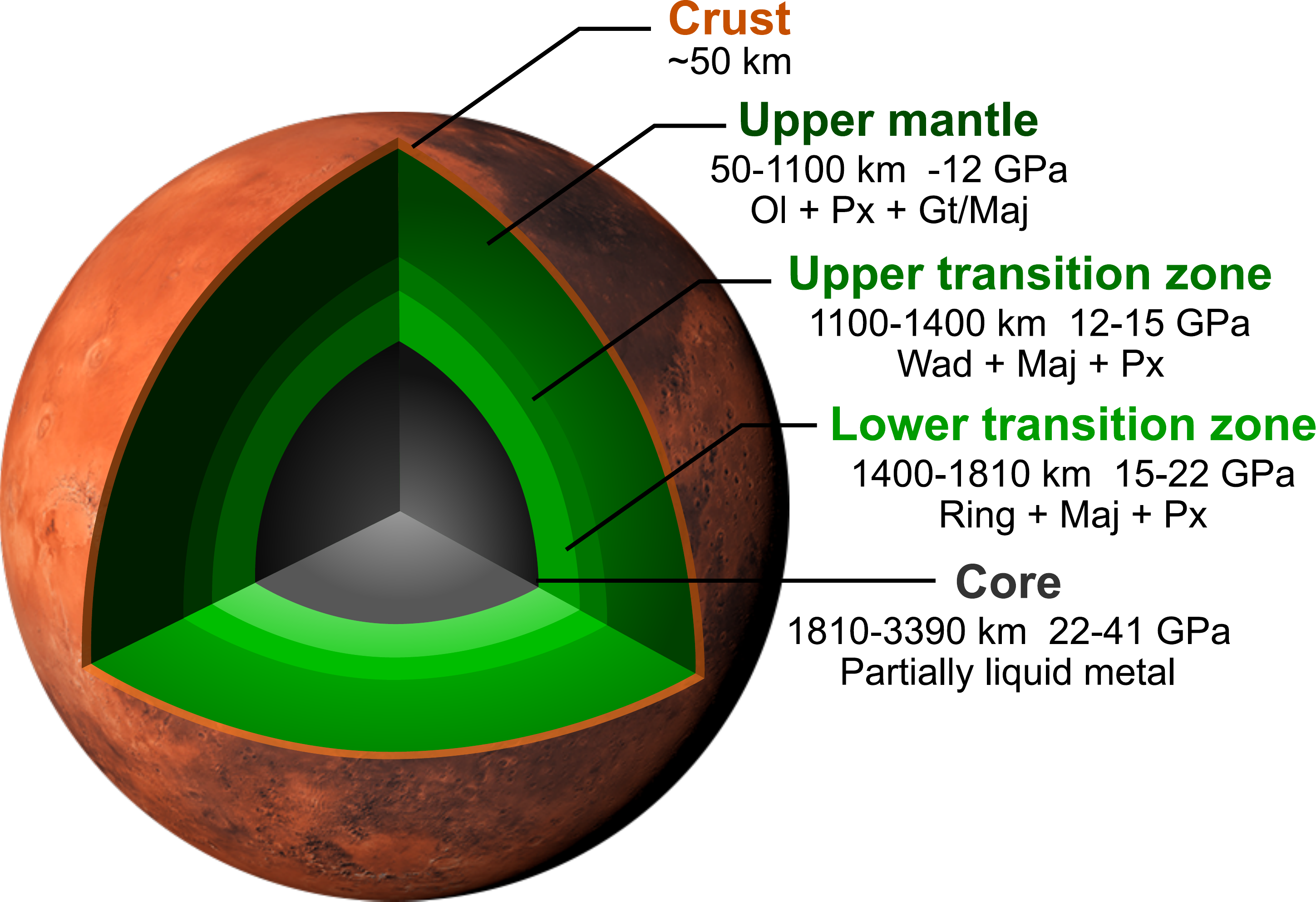}
	\caption{Interior structure of Mars. Abbreviations: Ol--olivine; Px--pyroxene; Gt--garnet; Wad--wadsleyite; Ring--ringwoodite; Maj--majorite.}
	\label{fig:pie_chart}
\end{figure}

	By fitting mass, density and MOI in the three layers and the bulk planet to geodetically constrained values (\cref{tab:phys_properties}), the density and composition of the metallic core is estimated. Finally we obtained a core composition with 6.6 wt\% S, 5.2 wt\% O and 0.9 wt\% H, and a mass fraction of 18\% as our best estimate. This core model yields a bulk planetary Fe/Si ratio of 1.36, which is lower than the CI value (1.74) but within the range of chondritic meteorite compositions \citep[1.0--1.8;][]{wasson1988compositions}. With a mass fraction of 18\%, the Martian core radius is 1580 km (i.e., 1810 km deep) and its mean density is $ \sim $6910 kg/m$ ^3 $ (\cref{fig:pie_chart}).  \bigskip
	
	Siderophile element abundances in the bulk Mars and its core were constrained by the BSM and a core model (\cref{tab:core}). This yields the Fe content in the core and bulk Mars to be 79.5 wt\% and 23.7 wt\%, respectively. Given chondritic meteorites (aside from the few examples of iron rich chondrites (e.g., CB)) show limited variation in Fe/Ni ($ \sim $17.4) and Ni/Co ($ \sim $20) values \citep{mcdonough2016composition}, we set the bulk Mars composition to these values. \bigskip
	
	Abundances of moderately volatile, lithophile elements in the BSM, which can be directly converted to the bulk Mars composition after correcting for the core mass fraction, define a robust depletion trend  (\cref{fig:volatility_trend}). The volatility trend provides a method to determine the rest of the element abundances, except for the atmophile elements, in the bulk Mars and its core (\cref{tab:BM,tab:core,tab:atomic_prop}). For the refractory siderophile and chalcophile elements (\cref{tab:elem_class}), the bulk Mars abundances are set at $ \sim $1.85 times CI abundance. \bigskip
	
	The possible conditions of Mars core formation (e.g., 10--17 GPa, 1900--2300 K, \textit{f}\ce{O2} = IW $ - $2 to $ - $1) are considered to be markedly different than that for the Earth's core   \citep[] {,righter2011moderately,rai2013core}. These findings (i.e, wholly lithophile character of Mn, Cr and V) are also in harmony with the relatively oxidized conditions for Mars' core formation as compared to the Earth \citep[e.g.,][]{,wadhwa2001redox,wadhwa2008redox,herd2002oxygen}. \bigskip

	\subsection{Comparison of the core models}
	\label{sec:comp_core_comparison}
	
	Although cosmochemical and geodetic insights constrain the composition and interior structure of Mars, we cannot determine a unique core model composition, given the available data. Thus, other Martian core models, distinct from that proposed in this study (\cref{tab:core}), are viable. The non-uniqueness can be readily viewed in \cref{fig:tradeoffs}, where we show trade-offs between core mass fraction, density and core radius, in order to constrain MOI, planetary density and mean mass. We tested three (end-member like) core compositions with (1) no H and O, (2) 9.5 wt\% O, and (3) 1.4 wt\% H, in addition to our preferred one (\cref{tab:core_model_comparison}). There is a non-uniqueness in core mass fraction -- core radius -- core density, with limits being $\sim$15 to 26$\%$, $\sim$1500 to 2000 km, and $\sim$5500 to 7500 kg/m$^3$, respectively. Composition and physical properties of core models were constrained by a combination of our BSM model, volatility trend, and geodetic properties of the planet. Importantly, as discussed in \cref{sec:comp_core_model}, the Martian volatility trend puts strong constraint on the S content in the Martian core (\cref{fig:volatility_trend_S}). \bigskip
	
	Core models assuming H- and O-free have the highest density ($ \sim $7460 kg/m$ ^3 $), given only S as the light element. A dense core forces down its mass fraction ($ \sim $17 wt\%) and low bulk Fe/Si (1.29) and Fe/Al (14.5) ratios, as compared to most chondritic meteorites \citep[1.0--1.8 and 15--25, respectively;][]{wasson1988compositions}. An exception, CV chondrite, has low Fe/Al ratio (13.4) due to its high abundance of refractory inclusions (i.e., high Al concentration). In contrast, however, the Fe/Si ratio in CV chondrites (1.36) is higher than that of Mars containing a core that is H- and O-free. Thus, to explain such a Mars' bulk composition requires a chemical fractionation process not recorded in chondritic meteorites. Similar arguments can be applied for the H-bearing, O-free core model.  \bigskip	
	
	Physical properties of an O-rich, H-free core model is similar to that of the O- and H-bearing model proposed as the best core model. However, solubility of oxygen in liquid iron does not support such a high O concentration in the Martian core \citep{rubie2004partitioning,tsuno2011effects}. \bigskip

	\section{Heat production in Mars}
	\label{sec:thermal_history}
	
\begin{figure}[p]
	\centering
	\includegraphics[width=1\linewidth]{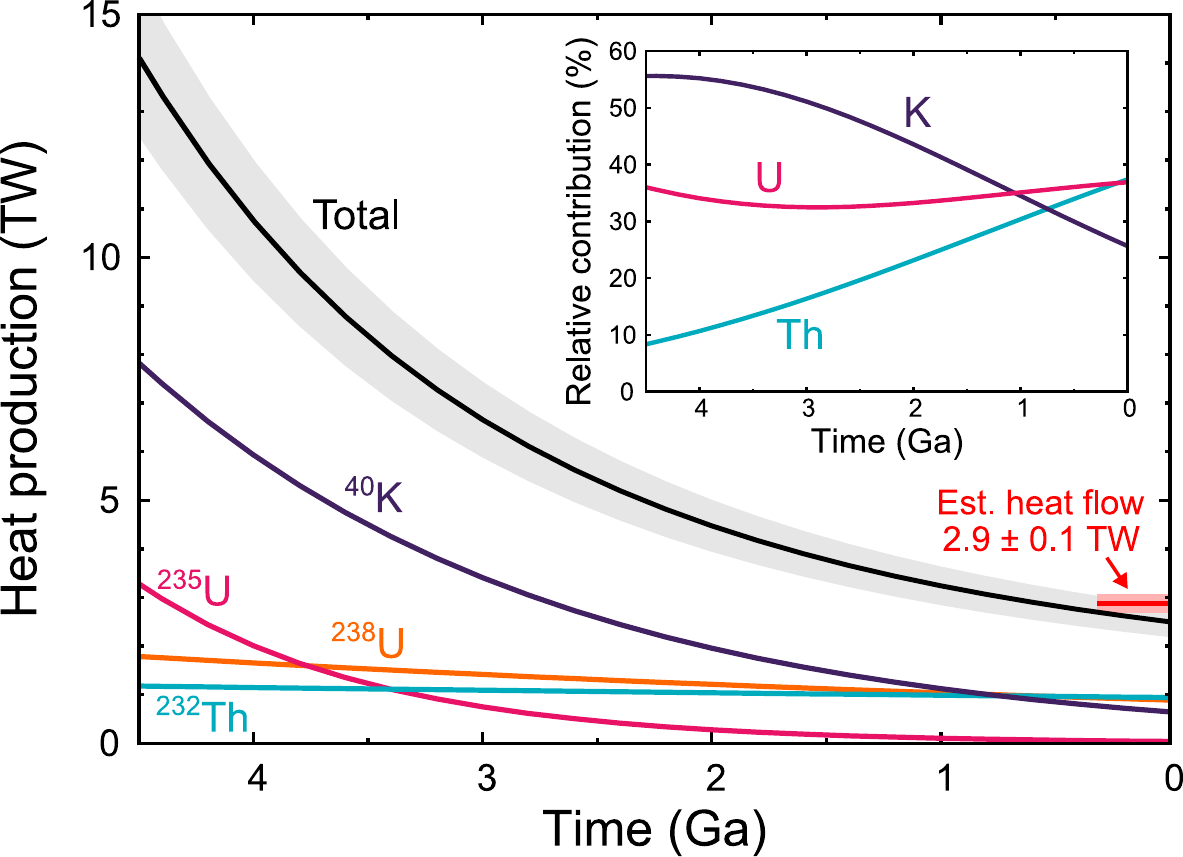}
	\caption{Radiogenic heat production in the Martian mantle through time. Grey band shows an uncertainty of the total radiogenic heat production. Present-day surface heat flow estimate (2.9 $ \pm $ 0.1 TW) is from \citet{parro2017present}. The inset shows relative contributions of heat-producing elements as a function of time.}
	\label{fig:HPE_rate}
\end{figure}

	Abundance of heat-producing elements (HPE: K, Th and U), which is constrained by the compositional modeling of the planets, is an important factor that influences thermal history of terrestrial planets. Our model predicts that the present-day HPE in the BSM produce 2.5 $ \pm $ 0.2 TW heat (\cref{tab:BSM_comparison}). In Mars, K is a dominant radiogenic heat source during its first 3.5 Gyr history (\cref{fig:HPE_rate}). Based on the BSM abundances of \ce{^{40}K}, \ce{^{232}Th}, \ce{^{235}U}, \ce{^{238}U} and \ce{^{87}Rb}, we estimate the Martian antineutrino (or areoneutrino) luminosity is (7.7 $ \pm $ 0.8) $ \times $ 10$ ^{24} $ $ \mathrm{\bar{\nu}_{e} /s} $ (i.e., (4.9 $ \pm $ 0.5) $ \times $ 10$ ^6 $ $ \mathrm{\bar{\nu}_{e} /cm^2/s} $).  \bigskip

	\section{Conclusions and implications for future works} 	\label{sec:implications}
	
	Compositional modeling of Mars reveals that the bulk silicate Mars (BSM) is enriched in refractory lithophile elements at 2.26 times higher than that in CI carbonaceous chondrites.  Moderately volatile elements are systematically depleted in Mars as a function of their volatility compared to the chondritic composition, but less so than in the Earth. The Martian core contains S, O and H as light elements, which is consistent with the volatility trend and occurrence of ringwoodite at the Martian core-mantle boundary. \bigskip
	
	The chemical compositions of Solar System bodies record accretion of solar nebular materials, core-mantle and mantle-crust differentiation and subsequent surface processes. The physicochemical similarities and differences between Mars and Earth provide insights into the origin and evolution of terrestrial planets, which are discussed elsewhere.  \bigskip
	
	%	The compositional model for Mars also provide some implications for the atmospheric evolution of the planet. For example, in the BSM having 360 ppm K (\cref{tab:BSM_all}), 2.7 $ \times $ 10$ ^{16} $ kg of \ce{^{40}Ar} has been produced during the 4.5 Gyr of Martian history (\cref{fig:Ar-40}). The amounts of \ce{^{40}Ar} in the present-day Martian atmosphere is $ \sim $10$ ^{14} $ kg \citep{lodders1998planetary,franz2017initial}, but those remaining in the solid reservoirs and having been lost to the space are model-dependent \citep{jakosky1994mars,hutchins1996evolution}. Future detailed modeling of evolution of Martian atmosphere combined with observational studies will improve our understandings of the evolution of Martian atmosphere.  \bigskip
	
	To constrain further the interior structure and composition of Mars direct evidence from the planet is needed. The best constraints would be provided by seismic determination of the depth of Martian core-mantle boundary, which would immediately define the core's mass fraction and density. NASA's ongoing Interior Exploration using Investigations, Geodesy and Heat Transport (InSight) mission will provide significant constraints on the Martian core composition \citep{smrekar2019pre}. In turn, our model for composition and interior structure of Mars can be tested by seismic and surface heat flux data from the InSight mission.  In addition, rock samples which will be returned from Phobos in JAXA’s planned Martian Moons eXploration (MMX) mission \citep{kuramoto2018martian} are keys to constrain further not only the origin of the Martian moons \citep{murchie2014value}, but also the composition of the Martian mantle, if Phobos has formed via giant impacts \citep{craddock2011phobos,citron2015formation,hyodo2017impact,canup2018origin}. \bigskip

	\subsection*{Acknowledgments}
	
	TY acknowledges supports from the Japanese Society for the Promotion of Science (JP18J20708), GP-EES Research Grant and DIARE Research Grant.  WFM gratefully acknowledges NSF support (EAR1650365). We acknowledge Eiji Ohtani and Attilio Rivoldini for their comments on early versions of this manuscript. We gratefully acknowledge Jeff Taylor and two anonymous referees for their constructive reviews that greatly improved this manuscript. We thank the Associate Editor James Day for his thoughtful comments and editorial effort.
	
	\subsection*{Author contributions}
	
	TY and WFM proposed and conceived various portions of this study and together calculated the compositional model of Mars. The manuscript was jointly written by TY and WFM and they read and approved the final manuscript.
	
	\subsection*{Competing interests}
	
	The authors declare no competing interests.
	
	\subsection*{Data and materials availability}
	
	Materials used in this study are provided as supplementary materials.
	
	\clearpage
	
	%\begin{landscape}
	%	\begin{figure}[h]
	%	\centering
	%	\includegraphics[width=1\linewidth]{lodders2003solar_cosmochem-table_20170513.pdf}
	%	\caption{Cosmochemical periodic table of elements \citep[modified after][]{,lodders2003solar}.}
	%	\label{fig:periodic_table}
	%\end{figure}
	%\end{landscape}
	%\clearpage
	
	\renewcommand{\arraystretch}{1.2}
	
\clearpage

\clearpage

\clearpage

\clearpage

\clearpage

%\begin{figure}[htbp]
%	\centering
%	\includegraphics[width=1\linewidth]{K_Th_GRS.jpg}
%	\caption{A log-log concentration plot of Na and Al in shergottites. (This figure is a draft. Modify units in the xy labels.)}
%	\label{fig:K_Th_GRS}
%\end{figure}

\clearpage

\clearpage

\begin{landscape}
	\begin{figure}[p]
		\centering
		\includegraphics[width=1\linewidth]{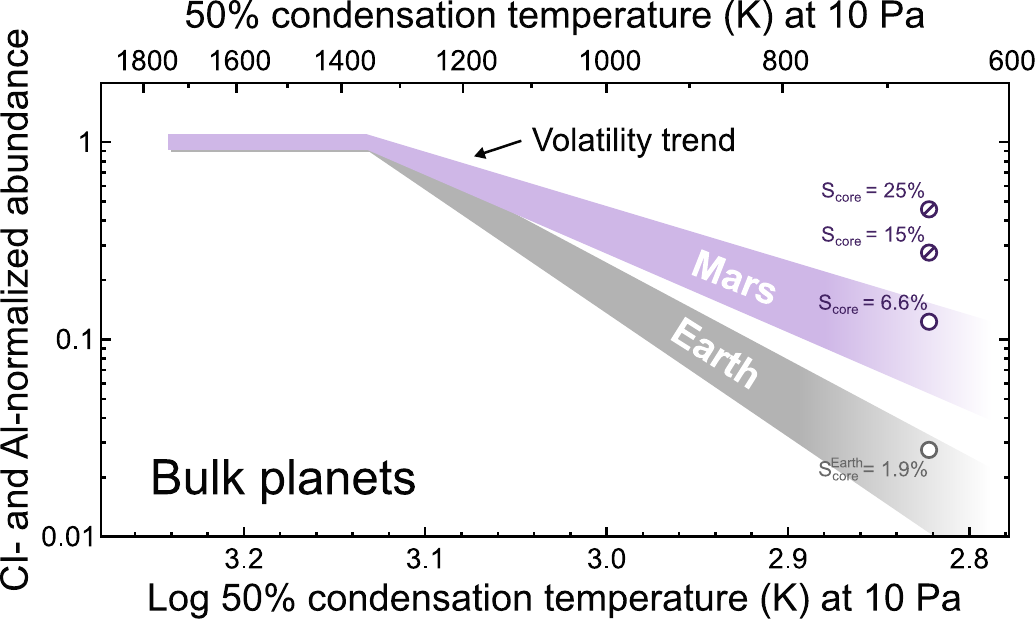}
		\caption{Volatility trends for Mars (this study) and Earth \citep{mcdonough1995composition} constrain S contents in the metallic cores. The S-rich Martian core models \citep[$ > $15 wt\%; e.g.,][; \cref{tab:core_comparison}]{,wanke1994chemistry,taylor2013bulk} are not consistent with the Martian volatility trend.}
		\label{fig:volatility_trend_S}
	\end{figure}
\end{landscape}
\clearpage

\clearpage

\clearpage

%	\begin{figure}[p]
%		\centering
%		\includegraphics[width=1\linewidth]{UC.pdf}
%		\caption{The Urey-Craig diagram. Redox condition of Mars is similar to L ordinary chondrites. The classification of chondrites into NC (non-carbonaceous; red) vs CC (carbonaceous; blue) groups follow that of \citet{warren2011stable,kruijer2017age}, where the NC group are derived from the inner Solar System and CC group the outer Solar System.}
%		\label{fig:UC}
%	\end{figure}
%	\clearpage

%	\begin{figure}[p]
%		\centering
%		\includegraphics[width=1\linewidth]{Ar-40_evolution.pdf}
%		\caption{Production of \ce{^{40}Ar} in the silicate Mars through time.}
%		\label{fig:Ar-40}
%	\end{figure}
%	\clearpage

\begin{landscape}
	\begin{table}[htbp]
		\centering
		\begin{threeparttable}
			\caption{Cosmochemical (refractory and volatile) and geochemical (lithophile, siderophile, chalcophile and atmophile) classification of elements. Elements are in the order of their 50\% condensation temperature (50\% Tc) in a gas of solar composition at 10 Pa \citep{lodders2003solar}. Elements with multiple geochemical affinity (e.g., lithophile and chalcophile behavior of Zn) are shown in both classifications.}
			\label{tab:elem_class}%
			\begin{tabular}{ccccc}
				\toprule
				& 50\% Tc (K) & Lithophile & Siderophile & Chalcophile \\
				\midrule
				Refractory & 1821--1355 & Zr, Hf, Sc, Y, Gd, Tb, Dy, &  Re, Os, W, Ir, Mo,   &  \\
				&       & Ho, Er, Tm, Lu, Th, Al, U, &  Ru, Pt, Rh    &  \\
				&       & Nd, Sm, Ti, Pr, La, Ta, Nb,  &   &  \\
				&       & Ca, Yb, Ce, Sr, Ba, Be, Eu &       &  \\
				&&&&\\
				Major component & 1355--1250 & Mg, Fe, Si, Cr & Ni, Co, Fe, Pd, Cr & Fe \\
				&&&&\\
				Moderately volatile & 1250--600 & Mn, Li, K, Na, Ga, B, & P, Mn, As, Au, Cu, Ag, Sb, & Zn, Te, Se, S, Cd \\
				&       & Rb, Cs, F, Zn & Ga, Ge, Bi, Pb, Te, Sn &  \\
				&&&&\\
				Volatile & 600--252 & Cl\tnotex{lab:Cl}, Br, I, Tl &       & In\tnotex{lab:In}, Tl, Hg \\
				\midrule
				&       &       &       &  \\
				\midrule
				&  50\% Tc (K)  & Atmophile &       &  \\
				\midrule
				Highly volatile & $ < $252  & O, N, Xe, Kr, Ar, C, Ne, He, H &       &  \\
				\bottomrule
			\end{tabular}%
			\begin{tablenotes}[flushleft]
				\item[a] Condensation temperature of Cl, Br and I may be similar \citep{clay2017halogens,wood2019condensation}.   \label{lab:Cl}
				\item[b] Condensation temperature of In might be $ \sim $800 K \citep{righter2017distribution}. \label{lab:In}
			\end{tablenotes}
		\end{threeparttable}
	\end{table}%
\end{landscape}
\clearpage

\begin{table}[htbp]
	\centering
	\scriptsize
	\begin{threeparttable}
		\caption{Composition of CI chondrites adopted in this study.}
		\label{tab:CI_comp}
		\begin{tabular}{ccccccccc}
			\toprule
			Element & Unit\tnotex{lab:ppm_abb}  & Value & Reference\tnotex{lab:ref_CI} &       & Element & Unit\tnotex{lab:ppm_abb}  & Value & Reference\tnotex{lab:ref_CI}  \\
			\midrule
			H     & \%    & 1.97  & P14   &       & Rh    & ppm   & 0.132 & P14 \\
			Li    & ppm   & 1.45  & P14   &       & Pd    & ppm   & 0.56  & P14 \\
			Be    & ppm   & 0.0219 & P14   &       & Ag    & ppm   & 0.201 & P14 \\
			B     & ppm   & 0.775 & P14   &       & Cd    & ppm   & 0.674 & P14 \\
			C     & ppm   & 34800 & P14   &       & In    & ppm   & 0.0778 & P14 \\
			N     & ppm   & 2950  & P14   &       & Sn    & ppm   & 1.63  & P14 \\
			O     & \%    & 45.9  & P14   &       & Sb    & ppm   & 0.145 & P14 \\
			F     & ppm   & 58.2  & P14   &       & Te    & ppm   & 2.28  & P14 \\
			Na    & ppm   & 4962  & P14   &       & I     & ppm   & 0.057 & C17 \\
			Mg    & \%    & 9.54  & P14   &       & Cs    & ppm   & 0.188 & P14 \\
			Al    & \%    & 0.840 & P14   &       & Ba    & ppm   & 2.42  & P14 \\
			Si    & \%    & 10.7  & P14   &       & La    & ppm   & 0.241 & P14 \\
			P     & ppm   & 985   & P14   &       & Ce    & ppm   & 0.619 & P14 \\
			S     & ppm   & 53500 & P14   &       & Pr    & ppm   & 0.0939 & P14 \\
			Cl    & ppm   & 115   & C17   &       & Nd    & ppm   & 0.474 & P14 \\
			K     & ppm   & 546   & P14   &       & Sm    & ppm   & 0.154 & P14 \\
			Ca    & \%    & 0.911 & P14   &       & Eu    & ppm   & 0.0588 & P14 \\
			Sc    & ppm   & 5.81  & P14   &       & Gd    & ppm   & 0.207 & P14 \\
			Ti    & ppm   & 447   & P14   &       & Tb    & ppm   & 0.0380 & P14 \\
			V     & ppm   & 54.6  & P14   &       & Dy    & ppm   & 0.256 & P14 \\
			Cr    & ppm   & 2623  & P14   &       & Ho    & ppm   & 0.0564 & P14 \\
			Mn    & ppm   & 1916  & P14   &       & Er    & ppm   & 0.166 & P14 \\
			Fe    & \%    & 18.66 & P14   &       & Tm    & ppm   & 0.0261 & P14 \\
			Co    & ppm   & 513   & P14   &       & Yb    & ppm   & 0.169 & P14 \\
			Ni    & ppm   & 10910 & P14   &       & Lu    & ppm   & 0.0250 & P14 \\
			Cu    & ppm   & 133   & P14   &       & Hf    & ppm   & 0.107 & P14 \\
			Zn    & ppm   & 309   & P14   &       & Ta    & ppm   & 0.0150 & P14 \\
			Ga    & ppm   & 9.62  & P14   &       & W     & ppm   & 0.0960 & P14 \\
			Ge    & ppm   & 32.6  & P14   &       & Re    & ppm   & 0.0381 & D16 \\
			As    & ppm   & 1.74  & P14   &       & Os    & ppm   & 0.461 & D16 \\
			Se    & ppm   & 20.3  & P14   &       & Ir    & ppm   & 0.431 & D16 \\
			Br    & ppm   & 0.189 & C17   &       & Pt    & ppm   & 0.874 & D16 \\
			Rb    & ppm   & 2.32  & P14   &       & Au    & ppm   & 0.175 & D16 \\
			Sr    & ppm   & 7.79  & P14   &       & Hg    & ppm   & 0.35  & P14 \\
			Y     & ppm   & 1.46  & P14   &       & Tl    & ppm   & 0.169 & W15 \\
			Zr    & ppm   & 3.63  & P14   &       & Pb    & ppm   & 2.62  & P14 \\
			Nb    & ppm   & 0.283 & P14   &       & Bi    & ppm   & 0.087 & W15 \\
			Mo    & ppm   & 0.87  & W15   &       & Th    & ppm   & 0.0300 & P14 \\
			Ru    & ppm   & 0.69  & P14   &       & U     & ppm   & 0.00796 & W18 \\
			\bottomrule
		\end{tabular}%
		\begin{tablenotes}[para]
			\item[a] ppm-- $ \mu $g/g. \label{lab:ppm_abb}
			\item[b] C17--\citet{clay2017halogens}; D16--\citet{day2016highly}; P14--\citet{palme2014solar}; W15--\citet{wang2015mass}; W18--\citet{wipperfurth2018earth}.	\label{lab:ref_CI}
		\end{tablenotes}
	\end{threeparttable}
\end{table}%

%\begin{landscape}
\begin{table}[p]
	\centering
	\small
	\begin{threeparttable}
		\caption{Comparison of compositional models of the bulk silicate Mars.}
		\label{tab:BSM_comparison}%
		\begin{tabular}{ccccccccc}
			\toprule
			wt\%  &    This study   &  WD94\tnotex{lab_WD94_BSM} & T13\tnotex{lab_T13_BSM} &  MA79\tnotex{lab_MA79_BSM} & OK92\tnotex{lab_OK92_BSM}  & LF97\tnotex{lab_LF97_BSM}  & S99\tnotex{lab_S99_BSM} & KC08\tnotex{lab_KC08_BSM} \\
			\midrule
			\ce{SiO2} & 45.5  & 44.4  & 43.7  & 41.6  & 43.0 & 45.4  & 47.5  & 44 \\
			\ce{TiO2} & 0.17   & 0.14  & 0.14  & 0.33  & 0.24  & 0.14  & 0.1   & - \\
			\ce{Al2O3} & 3.59   & 3.02  & 3.04  & 6.39  & 3.48  & 2.89  & 2.5   & 2.5 \\
			\ce{MnO} & 0.37   & 0.46  & 0.44  & 0.15  & 0.22  & 0.37  & 0.4   & - \\
			\ce{FeO} & 14.7  & 17.9  & 18.1  & 15.9 & 15.1 & 17.2  & 17.7  & 17 \\
			\ce{MgO} & 31.0  & 30.2  & 30.5  & 29.8 & 34.3 & 29.7  & 27.3  & 33 \\
			\ce{CaO} & 2.88   & 2.45  & 2.43  & 5.16  & 2.81  & 2.36  & 2.0   & 2.2 \\
			\ce{Na2O} & 0.59   & 0.5   & 0.53  & 0.10  & 0.46  & 0.98  & 1.2   & - \\
			\ce{K2O} & 0.043   & 0.037  & 0.037  & 0.009  &   -    & 0.11  & -     & - \\
			\ce{P2O5} & 0.17   & 0.16  & 0.19  & -     &   -    & 0.18  & -     & - \\
			\ce{NiO} & 0.046   & -     & 0.03  & -     &   -    & -     & -     & - \\
			\ce{Cr2O3} & 0.88   & 0.76  & 0.73  & 0.65  & 0.40   & 0.68  & 0.7   & - \\
			&       &       &       &       &       &       &       &  \\
			\ce{K} (ppm) & 360   & 305  & 309   & 76.5  & -  & 920  & -  & - \\
			\ce{Th} (ppb) & 68   & 56  & 58  & 125  &  -  & 55  &  -  & - \\
			\ce{U} (ppb) &  18 & 16  &  16 & 35  &  -  & 16  &  -  & - \\
			&       &       &       &       &       &       &       &  \\
			Total & 99.9  & 100   & 99.8  & 100   & 100   & 100   & 100   & 98.7 \\
			&       &       &       &       &       &       &       &  \\
			Mg\#  & 0.79    & 0.75    & 0.75    & 0.77    & 0.75    & 0.76    & 0.72    & 0.77 \\
			Mg/Si & 0.88  & 0.88  & 0.90  & 0.92  & 1.03  & 0.84  & 0.74  & 0.97 \\
			Al/Si & 0.09  & 0.08  & 0.08  & 0.17  & 0.09  & 0.07  & 0.06  & 0.06 \\
			RLE/CI\tnotex{lab_RLE} & 2.3 & 2.1 & 2.1 & 4.4 & 2.4 & 2.0 & 1.7 & 1.9 \\
			Fe/Si & 0.54  & 0.67  & 0.69  & 0.63  & 0.58  & 0.63  & 0.62  & 0.64 \\
			Fe/Al & 6.0   & 8.7   & 8.7   & 3.6   & 6.4   & 8.8   & 10.4  & 10.0 \\
			&       &       &       &       &       &       &       &  \\
			H$ \mathrm{_{BSM}} $\tnotex{lab:Habb} (TW) &   2.5  &  2.0   &  2.0   &  3.4    &   -    &   3.0  &  -    & -  \\
			H$ \mathrm{_{BSM}} $ (pW/kg) &   4.8  &  4.1     &  4.2    &  7.0    &   -    &    6.2  &  -    & -  \\
			\bottomrule
		\end{tabular}%
		\begin{tablenotes}[para]
			\item[a] \citet{wanke1994chemistry}. \label{lab_WD94_BSM}
			\item[b] \citet{taylor2013bulk}. \label{lab_T13_BSM}
			\item[c] \citet{morgan1979chemical}. \label{lab_MA79_BSM}
			\item[d] \citet{ohtani1992geochemical}. \label{lab_OK92_BSM}
			\item[e] \citet{lodders1997oxygen}. \label{lab_LF97_BSM}
			\item[f] \citet{sanloup1999simple} (EH45:H55 model). \label{lab_S99_BSM}
			\item[g] \citet{khan2008constraining}. \label{lab_KC08_BSM}
			\item[h] An average enrichment factor of major refractory lithophile elements (Ca, Al and Ti) compared to CI abundance (\cref{tab:CI_comp}). \label{lab_RLE}
			\item[i] Heat production in the bulk silicate Mars (BSM: mantle + crust). \label{lab:Habb}
		\end{tablenotes}
	\end{threeparttable}
\end{table}%
%\end{landscape}
\clearpage

\begin{table}[p]
	\centering
	\begin{threeparttable}
		\caption{Major element composition of the bulk silicate Mars (BSM). See text for the details of methods used to determine the BSM abundance of elements.}
		\label{tab:BSM_major}%
		\begin{tabular}{ccccc}
			\toprule
			& wt\%  & 1sd   & rsd\% & Method\tnotex{method} \\
			\midrule
			\ce{SiO2} & 45.5  & 1.8   & 4     & Mean SHE \\
			\ce{TiO2} & 0.17  & 0.02  & 10    & RLE \\
			\ce{Al2O3} & 3.59  & 0.36  & 10    & RLE \\
			\ce{MnO} & 0.37  & 0.07  & 18    & FeO/MnO in SNC \\
			\ce{FeO} & 14.7  & 1.0   & 7     & Mg\# = 0.79 $ \pm $ 0.02\tnotex{lab:Mgnum_1} \\
			\ce{MgO} & 31.0  & 2.0   & 6     & vs RLE in SHE \\
			\ce{CaO} & 2.88  & 0.29  & 10    & RLE \\
			\ce{Na2O} & 0.59  & 0.13  & 22    & Na/Al in SHE \\
			\ce{K2O} & 0.043  & 0.005  & 11    & K/Th (GRS)\tnotex{lab:GRS_1} \\
			\ce{P2O5} & 0.17  & 0.05  & 28    & P/Y in  SHE \\
			\ce{NiO} & 0.046  & 0.01  & 26   & Ni/Mg in SHE-P and SHE-Ol \\
			\ce{Cr2O3} & 0.88  & 0.15   & 17    & vs Al in SHE-P and SHE-Ol\\
			&       &       &       &  \\
			Total & 99.9  &       &       &  \\
			&       &       &       &  \\
			Mg\#  & 0.79   &       &       &  \\
			Mg/Si & 0.88  &       &       &  \\
			Al/Si & 0.09  &       &       &  \\
			RLE/CI & 2.26  &       &       &  \\
			Fe/Si & 0.54  &       &       &  \\
			Fe/Al & 6.0   &       &       &  \\
			\bottomrule
		\end{tabular}%
		\begin{tablenotes}[flushleft]
			\item[a] GRS--gamma-ray spectroscopy; RLE--refractory lithophile elements; SHE-P--poikilitic shergottites; SHE-Ol--olivine shergottites; SNC--shergottite, nakhlite and chassignite. \label{method}
			\item[b] \citet{borg2003petrogenetic,agee2004experimental,draper2005crystallization,minitti2006new,white2006experimental,collinet2015melting,mccoy2016experimentally_conf}. \label{lab:Mgnum_1}
			\item[c] \citet{taylor2006bulk,taylor2006variations}. \label{lab:GRS_1}
		\end{tablenotes}
	\end{threeparttable}
\end{table}%
\clearpage

\begin{landscape}
	\begin{table}[p]
		\centering
		\begin{threeparttable}
			\tiny
			\caption{Composition of the bulk silicate Mars. Concentrations are in ppm ($ \mu $g/g), otherwise noted. See text for the details of methods used to determine the BSM abundances of elements.}
			\label{tab:BSM_all}%
			\begin{tabular}{ccccccccc}
				\toprule
				Element &  BSM   & rsd\%\tnotex{lab:F4} &                         Method\tnotex{method_all}        &  & Element &   BSM   & rsd\% &           Method            \\
				\midrule
				H    &   16   &          U           &                           See text\tnotex{lab:H}                           &  &   Rh    & 0.0021  &   4   &                    Mean SNC                    \\
				Li    &  1.8   &          22          &                               vs RLE in SNC                                &  &   Pd    & 0.0074  &    70   &            vs MgO\tnotex{lab:TD18}             \\
				Be    &  0.05  &          10          &                                    RLE                                     &  &   Ag    &  0.02  &  U   &                  Sb/Ag in SHE                   \\
				B    &  0.84  &          63          &                          B/Ca in SHE-P and SHE-Ol                          &  &   Cd    &  0.020  &  50   &          Cd/Yb in SHE\tnotex{lab:Y15}          \\
				C    &   32   &          U           &                               Chondritic C/H                               &  &   In    &  0.010  &  50   &          In/Y in SHE\tnotex{lab:Y15}           \\
				N    &  1.6   &          U           &                               Chondritic N/H                               &  &   Sn    &  0.25   &  40   &          Sn/Sm in SHE\tnotex{lab:Y15}          \\
				O (\%)  &  43.2  &          8           & Major oxide stoichiometry, \ce{Fe^{3+}}/\ce{Fe^{tot}} = 0.1\tnotex{lab:Fe} &  &   Sb    &  0.02  &  50   &          Sb/Pr in SHE\tnotex{lab:Y15}          \\
				F    &   30   &          U          &                        Cl/F in SNC\tnotex{lab:F16}                         &  &   Te    & 0.0005  &  50   &    Cu/Te and Se/Te in SHE\tnotex{lab:WB17}     \\
				Na &  4380  &          22          &                               Na/Al in  SHE                                &  &    I    &  0.014  &   50   &                Chondritic I/Cl\tnotex{lab:C17}       \\
				Mg (\%) &  18.7  &          6           &                               vs RLE in  SHE                               &  &   Cs    &  0.07  &  41   &                 Cs/La in  SHE                  \\
				Al (\%) &  1.90  &          10          &                                    RLE                                     &  &   Ba    &   5.47   &  10   &                      RLE                       \\
				Si (\%) &  21.3  &          4           &                                  Mean SHE                                  &  &   La    &  0.546   &  10   &                      RLE                       \\
				P  & 740  &          28          &                                P/Y in  SHE                                 &  &   Ce    &   1.40   &  10   &                      RLE                       \\
				S  & 360  &          33          &                   S/Cu and S/Se in SHE\tnotex{lab:WB17}                    &  &   Pr    &  0.212   &  10   &                      RLE                       \\
				Cl    &   28   &          35          &                       Cl/La in  SHE\tnotex{lab:F16}                        &  &   Nd    &   1.07   &  10   &                      RLE                       \\
				K   & 360 &          11          &                         K/Th (GRS)\tnotex{lab:T06}                         &  &   Sm    &  0.347  &  10   &                      RLE                       \\
				Ca (\%) &  2.06   &          10          &                                    RLE                                     &  &   Eu    &  0.133   &  10   &                      RLE                       \\
				Sc    &   13.1   &          10          &                                    RLE                                     &  &   Gd    &  0.468   &  10   &                      RLE                       \\
				Ti  &  1010  &          10          &                                    RLE                                     &  &   Tb    &  0.0858   &  10   &                      RLE                       \\
				V    &  123   &          10          &                                    RLE                                     &  &   Dy    &  0.578   &  10   &                      RLE                       \\
				Cr  & 6000  &          17          &                         vs Al in SHE-P and SHE-Ol                          &  &   Ho    &  0.128   &  10   &                      RLE                       \\
				Mn &  2880  &          18          &                               FeO/MnO in SNC                               &  &   Er    &  0.374   &  10   &                      RLE                       \\
				Fe (\%) &  11.4  &          7           &                Mg\# = 0.79 $ \pm $ 0.02\tnotex{lab:Mgnum_2}                &  &   Tm    &  0.0590  &  10   &                      RLE                       \\
				Co    &   96   &          46          &                         Co/Ni in SHE-P and SHE-Ol                          &  &   Yb    &  0.381  &  10   &                      RLE                       \\
				Ni & 360  &          26          &                         Ni/Mg in SHE-P and SHE-Ol                          &  &   Lu    &  0.0566  &  10   &                      RLE                       \\
				Cu    &  2.6   &          23          &                               Cu/Ti in  SHE                                &  &   Hf    &  0.241   &  10   &                      RLE                       \\
				Zn    &   45   &          33          &                               vs Lu in  SHE                                &  &   Ta    &  0.0339  &  14   &                      RLE                       \\
				Ga    &  8.7   &          22         &                               Ga/Al in  SHE                                &  &    W    &  0.069   &  54   &                  W/Th in  SHE                  \\
				Ge    &  0.6   &          67          &                         vs MgO in SHE-P and SHE-Ol                         &  &   Re    & 0.0004  &   75    &            vs MgO\tnotex{lab:TD18}             \\
				As    &  0.03  &          76          &                        As/Ce in SHE\tnotex{lab:Y15}                        &  &   Os    &  0.004  &    50   &            vs MgO\tnotex{lab:TD18}             \\
				Se    &  0.10  &          27          &                       Cu/Se in SHE\tnotex{lab:WB17}                        &  &   Ir    & 0.0037  &   60    &            vs MgO\tnotex{lab:TD18}             \\
				Br    &  0.05  &          U          &                        Chondritic Cl/Br\tnotex{lab:C17}                     &  &   Pt    & 0.0096  &   80    &            vs MgO\tnotex{lab:TD18}             \\
				Rb    &  1.2  &          36         &       Initial \ce{^{87}Sr}/\ce{^{86}Sr} in SNC, vs La in  SHE       &  &   Au    & 0.002 &    80   &                     vs MgO                     \\
				Sr    &   17.6   &          10          &                                    RLE                                     &  &   Hg    &  0.007  &   U   &            Hg/Se in NAK and the BSE            \\
				Y    &  3.30   &          10          &                                    RLE                                     &  &   Tl    &  0.004  &  50   &                 Tl/Sm in  SHE                  \\
				Zr    &  8.20   &          10          &                                    RLE                                     &  &   Pb    &  0.255  &  18   & U-Pb isotope systematics in SNC\tnotex{lab:Pb} \\
				Nb    &  0.640  &          10         &                                    RLE                                     &  &   Bi    & 0.002  &  60   &                 Bi/Th in  SHE                  \\
				Mo    &  0.5   &          80          &                       Mo/Ce in SHE\tnotex{lab:RC11}                        &  &   Th    & 0.0678  &  10   &                      RLE                       \\
				Ru    & 0.0062 &      60      &                          vs MgO\tnotex{lab:TD18}                           &  &    U    & 0.0180  &  10   &                      RLE                       \\
				\bottomrule
			\end{tabular}%
			\begin{tablenotes}[flushleft,para]
				\item[a] U--$ \geq $2 factors of uncertainties. \label{lab:F4}
				\item[b] BSE--bulk silicate Earth; GRS--gamma-ray spectroscopy; NAK--nakhlite; SNC--shergottite, nakhlite and chassignite. RLE--refractory lithophile elements; SHE-P--poikilitic shergottites; SHE-Ol--olivine shergottites. \label{method_all}
				\item[c] \citet{mccubbin2016geologic,mccubbin2016heterogeneous,filiberto2019volatiles}. \label{lab:H}
				\item[d] \citet{schmidt2013primary,medard2014fe3+}. \label{lab:Fe}
				\item[e] \citet{filiberto2016constraints}; this study. \label{lab:F16}
				\item[f] \citet{wang2017chalcophile}. \label{lab:WB17}
				\item[g] \citet{taylor2006bulk,taylor2006variations}; this study. \label{lab:T06}
				\item[h] \citet{borg2003petrogenetic,agee2004experimental,draper2005crystallization,minitti2006new,white2006experimental,collinet2015melting,mccoy2016experimentally_conf}. \label{lab:Mgnum_2}
				\item[i] \citet{yang2015siderophile}. \label{lab:Y15}
				\item[j] \citet{clay2017halogens}. \label{lab:C17}
				\item[k] \citet{righter2011moderately}. \label{lab:RC11}
				\item[l] \citet{tait2018chondritic}. \label{lab:TD18}
				\item[m] $ \mu  $ (= \ce{^{238}U}/\ce{^{204}Pb}) = 3.6; \ce{^{208}Pb}/\ce{^{204}Pb} = 33; \ce{^{207}Pb}/\ce{^{204}Pb} =13; \ce{^{206}Pb}/\ce{^{204}Pb} = 14. \label{lab:Pb}
			\end{tablenotes}
		\end{threeparttable}
	\end{table}%
\end{landscape}
\clearpage

%\begin{landscape}
\begin{table}[p]
	\centering
	\small
	\begin{threeparttable}
		\caption{Comparison of compositional models of the Martian core.}
		\label{tab:core_comparison}%
		\begin{tabular}{ccccccccc}
			\toprule
			wt\%  &    This study   &  WD94\tnotex{lab_WD94_core} & T13\tnotex{lab_T13_core} &  MA79\tnotex{lab_MA79_core} & OK92\tnotex{lab_OK92_core}  & LF97\tnotex{lab_LF97_core}  & S99\tnotex{lab_S99_core} & KC08\tnotex{lab_KC08_core} \\
			\midrule
			Fe    & 79.5  & 77.8  & 78.6\tnotex{lab_FeNi}  & 88.1  & 78.4  & 81.1  & 76.6  & 75--78\tnotex{lab_FeNi} \\
			Ni    & 7.4  & 7.6   & -     & 8.0   & 7.6   & 7.7   & 7.2   & - \\
			Co    & 0.33   & 0.36  & -     & 0.37  & -     & 0.4   & -     & - \\
			S     & 6.6   & 14.24 & 21.4  & 3.5   & 14.0  & 10.6  & 16.2  & 22--25 \\
			P     & 0.33   & -     & -     & -     & -     & 0.2   & -     & - \\
			O     & 5.2   & -     & -     & -     & -     & -     & -     & - \\
			H     & 0.9   & -     & -     & -     & -     & -     & -     & - \\
			&&&&&&&&\\
			Total & 100  & 100   & 100   & 100   & 100   & 100   & 100   & - \\
			\bottomrule
		\end{tabular}%
		\begin{tablenotes}[para]
			\item[a] \citet{wanke1994chemistry}. \label{lab_WD94_core}
			\item[b] \citet{taylor2013bulk}. \label{lab_T13_core}
			\item[c] \citet{morgan1979chemical}. \label{lab_MA79_core}
			\item[d] \citet{ohtani1992geochemical}. \label{lab_OK92_core}
			\item[e] \citet{lodders1997oxygen}. \label{lab_LF97_core}
			\item[f] \citet{sanloup1999simple} (EH45:H55 model). \label{lab_S99_core}
			\item[g] \citet{khan2008constraining}. \label{lab_KC08_core}
			\item[h] Sum of Fe and Ni. \label{lab_FeNi}
		\end{tablenotes}
	\end{threeparttable}
\end{table}%
%\end{landscape}
\clearpage

\begin{landscape}
	\begin{table}[p]
		\centering
		\small
		\begin{threeparttable}
			\caption{Physical properties of Mars. Modeled values are in a normal font and reference values are in italic.}
			\label{tab:phys_properties}%
			\begin{tabularx}{480pt}{ccccccc}
				\toprule
				Observation & Unit & Crust & Mantle & Core  & Bulk planet  & Reference value \\
				\midrule
				%					\textbf{Mars} &&&&&&\\
				Mass & kg &  2.56$ \times $$10^{22} $ &5.01$\times $$ 10^{23} $ & 1.17$\times $$ 10^{23} $ &  6.419$\times$$10^{23} $ &  \textit{6.417(3)}$\times $\textit{10}$^{23} $\tnotex{lab_MOI} \\
				Mean density & kg/m$ ^3 $ & 3010  & 3640  &  6910  & 3936  & \textit{3935(1)}\tnotex{lab_density} \\
				Moment of inertia & -- & 7\%    & 89\%    & 4\%     & 0.3638 & \textit{0.3639(1)}\tnotex{lab_MOI} \\
				\multirow{2}[0]{*}{Heat production (K, Th, U)}  & TW & 1.3 & 1.3 & 0 & 2.5 & \textit{2.9(1)}\tnotex{lab_hf} \\
				& pW/kg & 49\tnotex{lab_HPE_crust}& 2.5\tnotex{lab_HPE_mantle} & 0 &  3.9\tnotex{lab_HPE_BSM}  & -- \\
				%	 &&&&&&\\
				%					\midrule
				%					&&&&&&\\
				%					\textbf{Earth} &&&&&&\\
				%					Mass  & kg & 3.12$ \times  $$10^{22}  $  & 4.00$\times $$10^{24} $  & 1.94$  \times $$ 10^{24} $  &  5.97$\times $$ 10^{24} $ &  \textit{5.97218(60)}$ \times  $\textit{10}$ ^{24} $\tnotex{lab_Earth_mass} \\
				%					Mean density & kg/m$ ^3 $  & 2800  & 4400  & 11870  & 5510 & \textit{5510}\tnotex{lab_PREM} \\
				%					Moment of inertia & -- & 1\%    & 88\%    & 11\%     & 0.3308 & \textit{0.3307}\tnotex{lab_Earth_mass}  \\
				%					\multirow{2}[0]{*}{Heat production (K, Th, U)}	& TW & \textit{7.3}\tnotex{lab_HPE_Earth}  & \textit{12.6}\tnotex{lab_HPE_Earth}  & 0 & \textit{19.9}\tnotex{lab_HPE_Earth}  & \textit{46}\tnotex{lab_hf_Earth} \\
				%					& pW/kg & 232  & 3.1  & 0 &  3.3 & --  \\	
				%					\midrule
				%					&&&&&&\\
				\bottomrule
			\end{tabularx}
			\begin{tablenotes}[para]
				\item[a] \citet{konopliv2016improved,khan2018geophysical}. \label{lab_MOI}
				\item[b] \citet{rivoldini2011geodesy}. \label{lab_density}
				\item[c] Surface heat loss estimated based on an average surface heat flux of 20 $ \pm $ 1 mW/m$ ^2 $ \citep{parro2017present}. \label{lab_hf}
				\item [d] Abundances of heat-producing elements (HPE) in the crust\label{lab_HPE_crust}: 3740 ppm K, 700 ppb Th and 180 ppb U \citep{taylor2009planetary}. 
				\item[e] HPE abundance in the mantle is calculated from mass-balance considerations: 190 ppm K, 36 ppb Th and 10 ppb U. \label{lab_HPE_mantle}
				\item[f] BSM abundances of HPE are 360 ppm K, 68 ppb Th and 18 ppb U (this study). \label{lab_HPE_BSM}
				%					\item[g] \citet{chambat2010flattening}. \label{lab_Earth_mass}
				%					\item[h] The preliminary Earth model \citep{dziewonski1981preliminary}. \label{lab_PREM}
				%					\item[i] \citet{huang2013reference}. \label{lab_HPE_Earth}
				%					\item[j] \citet{jaupart2015temperatures}. \label{lab_hf_Earth}
			\end{tablenotes}
		\end{threeparttable}
	\end{table}%
\end{landscape}
\clearpage

\begin{table}[p]
	\centering
	\begin{threeparttable}
		\caption{Composition of the Martian core. Concentrations are in ppm ($ \mu $g/g), otherwise noted.}
		\begin{tabular}{ccccc}
			\toprule
			Element & Martian core &       & Element & Martian core \\
			\midrule
			Fe (\%) & 79.5  &       & Os    & 5 \\
			Ni (\%) & 7.4   &       & Pd    & 5 \\
			O (\%) & 5.2   &       & Ir    & 5 \\
			S (\%) & 6.6   &       & Te    & 3 \\
			H (\%) & 0.9   &       & Pb    & 3.1 \\
			Co    & 3300  &       & Rh    & 1.3 \\
			P     & 3300  &       & Sn    & 1.3 \\
			Cu    & 560   &       & W     & 0.7 \\
			Zn    & 290   &       & Ag    & 0.7 \\
			Ge    & 90    &       & Au    & 0.7 \\
			Se    & 30    &       & Cd    & 0.7 \\
			Pt    & 9     &       & Re    & 0.4 \\
			As    & 8     &       & Sb    & 0.4 \\
			Mo    & 7     &       & Bi    & 0.1 \\
			Ru    & 7     &       & Tl    & 0.1 \\
			\bottomrule
		\end{tabular}%
		\label{tab:core}%
	\end{threeparttable}
\end{table}%
\clearpage

\begin{table}[htbp]
	\centering
	\scriptsize 
	\begin{threeparttable}
		\caption{Composition of the bulk Mars. Concentrations are in ppm ($ \mu $g/g), otherwise noted.}
		\begin{tabular}{ccccc}
			\toprule
			Element & Bulk Mars &       & Element & Bulk Mars \\
			\midrule
			H  & 1600 &       & Rh    & 0.24 \\
			Li    & 1.5  &       & Pd    & 0.84 \\
			Be    & 0.041  &       & Ag    & 0.14 \\
			B     & 0.69  &       & Cd    & 0.15 \\
			C     & 26 &       & In    & 0.01 \\
			N     & 1.4  &       & Sn    & 0.43 \\
			O (\%) & 36.3  &       & Sb    & 0.1 \\
			F     & 23  &       & Te    & 0.6 \\
			Na & 3600  &       & I     & 0.01 \\
			Mg (\%) & 15.3  &       & Cs    & 0.06 \\
			Al (\%) & 1.56  &       & Ba    & 4.5 \\
			Si (\%) & 17.4  &       & La    & 0.45 \\
			P  & 1200  &       & Ce    & 1.1 \\
			S & 12100  &       & Pr    & 0.17 \\
			Cl    & 23  &       & Nd    & 0.88 \\
			K & 300   &       & Sm    & 0.28 \\
			Ca (\%) & 1.69   &       & Eu    & 0.11 \\
			Sc    & 10.8  &       & Gd    & 0.38 \\
			Ti & 830   &       & Tb    & 0.070 \\
			V     & 100  &       & Dy    & 0.47 \\
			Cr  & 4900  &       & Ho    & 0.10 \\
			Mn & 2400  &       & Er    & 0.31 \\
			Fe (\%) & 23.7  &       & Tm    & 0.048 \\
			Co    & 680   &       & Yb    & 0.31 \\
			Ni  & 13600  &       & Lu    & 0.046 \\
			Cu    & 100    &       & Hf    & 0.20 \\
			Zn    & 89    &       & Ta    & 0.028 \\
			Ga    & 7.2   &       & W     & 0.18 \\
			Ge    & 16   &       & Re    & 0.07 \\
			As    & 1.4   &       & Os    & 0.9 \\
			Se    & 5.2   &       & Ir    & 0.9 \\
			Br    & 0.1  &       & Pt    & 1.7 \\
			Rb    & 0.96  &       & Au    & 0.1 \\
			Sr    & 14    &       & Hg    & 0.006 \\
			Y     & 2.7   &       & Tl    & 0.02 \\
			Zr    & 6.7   &       & Pb    & 0.8 \\
			Nb    & 0.52  &       & Bi    & 0.03 \\
			Mo    & 1.6   &       & Th    & 0.056 \\
			Ru    & 1.3   &       & U     & 0.015 \\
			\bottomrule
		\end{tabular}%
		\label{tab:BM}%
	\end{threeparttable}
\end{table}%

\begin{table}[p]
	\centering
	\begin{threeparttable}
		\caption{Compositional models for the bulk Mars, bulk silicate Mars (BSM) and core and atomic proportions for major elements.}
		\begin{tabular}{ccccccccc}
			\toprule
			wt\%  & Bulk  & BSM   & Core  &       & atomic\% & Bulk  & BSM   & Core \\
			\midrule
			O     & 36.3  & 43.2  & 5.2   &       & O     & 53  & 59 & 11 \\
			Mg    & 15.3  & 18.7  & 0     &       & Mg    & 15  & 17  & 0 \\
			Si    & 17.4  & 21.3  & 0     &       & Si    & 14  & 16  & 0 \\
			Fe    & 23.7  & 11.4  & 79.5  &       & Fe    & 10  & 4.4 & 48 \\
			Al    & 1.56   & 1.90   & 0     &       & Al    & 1.3 & 1.5 & 0 \\
			S     & 1.2   & 0.04  & 6.6   &       & S     & 0.9 & 0.02 & 6.9\\
			Ca    & 1.69   & 2.06   & 0     &       & Ca    & 1.0 & 1.1 & 0 \\
			Ni    & 1.4  & 0.04  & 7.4   &       & Ni    & 0.5 & 0.01 & 4.2 \\
			H    & 0.2  & 0.00  & 0.9   &       & H    & 3.8 & 0.03 & 30 \\
			\bottomrule
		\end{tabular}%
		\label{tab:atomic_prop}%
	\end{threeparttable}
\end{table}%
\clearpage

\begin{table}[p]
	\centering
	\small
	\begin{threeparttable}
		\caption{Comparison of chemical and physical properties of Martian core models.}
		\label{tab:core_model_comparison}%
		\begin{tabular}{cccccc}
			\toprule
			& H,O-bearing & H,O-free & H-free & O-free & Reference value  \\
			\midrule
			\multicolumn{6}{c}{Chemical composition} \\
			\midrule
			Core wt\%    &       &       &       &       &  \\
			Fe  & 79.5  & 81.6  & 76.5  & 83.3  &  \\
			Ni  & 7.4   & 7.8   & 7.1   & 7.8   &  \\
			Co  & 0.33  & 0.35  & 0.32  & 0.35  &  \\
			S & 6.6  & 10  & 6.10  & 7.04  &  \\
			P & 0.33  & 0.37  & 0.32  & 0.36  &  \\
			O & 5.2   & 0.0   & 9.5  & 0 &  \\
			H  & 0.9  & 0.0  & 0.0  & 1.4  &  \\
			Total & 100.2  & 100.1  & 99.9  & 100.3  &  \\
			&       &       &       &       &  \\
			Bulk Mars &       &       &       &       &  \\
			Fe/Si & 1.36  & 1.29  & 1.35  & 1.34  &  1.0--1.6\tnotex{lab_OC}  \\
			Fe/Al & 15.2  & 14.5  & 15.1  & 15.0 & 15.5--24.3\tnotex{lab_OC} \\ \midrule
			\multicolumn{6}{c}{Modeled physical properties\tnotex{lab_phys_prop_core}} \\
			\midrule
			$ \mathrm{M'_{core}} $ & 18\%  & 17\%  & 18\%  & 17\%  &  \\
			$ \mathrm{\rho_{core}} $ (kg/m$ ^3 $) & 6910  & 7460  & 6910  & 7180  &  \\
			$ \mathrm{r_{core}} $ (km) & 1580  & 1510  & 1600  & 1540  &  \\
			$ \mathrm{P_{CMB}} $ (GPa)   & 22    & 24    & 22    & 23  &  \\
			%				&       &       &       &       &  \\    
			%				$ \mathrm{M_{bulk}} $ (kg)   & 6.420$ \times $10$ ^{23} $ & 6.414$ \times $10$ ^{23} $ & 6.427$ \times $10$ ^{23} $ & 6.423$ \times $10$ ^{23} $ & 6.419$ \times $10$ ^{23} $\tnotex{lab_density_coretab} \\
			%				$ \mathrm{\rho_{bulk}} $  (kg/m$ ^3 $) & 3.935 & 3.932 & 3.939 & 3.937 & 3.935\tnotex{lab_density_coretab} \\
			%				MOI   & 0.3638 & 0.3644 & 0.3632 & 0.3627 & 0.3639\tnotex{lab_MOI_coretab} \\
			\bottomrule
		\end{tabular}%
		\begin{tablenotes}[para]
			\item[a] Compositional range of ordinary chondrites \citep{wasson1988compositions}. \label{lab_OC}
			\item[b] M'--mass fraction; $ \mathrm{\rho_{core}} $--density of the core; $ \mathrm{r_{core}} $--radius of the core; $ \mathrm{P_{CMB}} $--pressure at the core--mantle boundary. \label{lab_phys_prop_core}
			%				\item[c] \citet{rivoldini2011geodesy}. \label{lab_density_coretab}
			%				\item[d] \citet{konopliv2016improved}. \label{lab_MOI_coretab}
			%			\item[b] Mass fraction of the core. \label{lab_Mfrac}
			%			\item[c] Density of the core. \label{lab_rhocore}
			%			\item[d] Radius of the core. \label{lab_rcore}
			%			\item[e] Pressure at the core-mantle boundary. \label{lab_PCMB}
			%			\item[f]  Mass of the bulk Mars.	\label{lab_Mbulk}
		\end{tablenotes}
	\end{threeparttable}
\end{table}%
%\end{landscape}
\clearpage

\bibliographystyle{elsarticle-harv}
\bibliography{myrefs}

\begin{thebibliography}{212}
\expandafter\ifx\csname natexlab\endcsname\relax\def\natexlab#1{#1}\fi
\providecommand{\url}[1]{\texttt{#1}}
\providecommand{\href}[2]{#2}
\providecommand{\path}[1]{#1}
\providecommand{\DOIprefix}{doi:}
\providecommand{\ArXivprefix}{arXiv:}
\providecommand{\URLprefix}{URL: }
\providecommand{\Pubmedprefix}{pmid:}
\providecommand{\doi}[1]{\href{http://dx.doi.org/#1}{\path{#1}}}
\providecommand{\Pubmed}[1]{\href{pmid:#1}{\path{#1}}}
\providecommand{\bibinfo}[2]{#2}
\ifx\xfnm\relax \def\xfnm[#1]{\unskip,\space#1}\fi
%Type = Article
\bibitem[{Agee and Draper(2004)}]{agee2004experimental}
\bibinfo{author}{Agee, C.B.}, \bibinfo{author}{Draper, D.S.},
  \bibinfo{year}{2004}.
\newblock \bibinfo{title}{{Experimental constraints on the origin of Martian
  meteorites and the composition of the Martian mantle}}.
\newblock \bibinfo{journal}{Earth and Planetary Science Letters}
  \bibinfo{volume}{224}, \bibinfo{pages}{415--429}.
\newblock \DOIprefix\doi{10.1016/j.epsl.2004.05.022}.
%Type = Article
\bibitem[{Agostini et~al.(2018)Agostini, Altenm{\"u}ller, Appel, Atroshchenko,
  Bagdasarian, Basilico, Bellini, Benziger, Bick, Bonfini, Bravo, Caccianiga,
  Calaprice, Caminata, Caprioli, Carlini, Cavalcante, Chepurnov, Choi, Collica,
  D'Angelo, Davini, Derbin, Ding, Di~Ludovico, Di~Noto, Drachnev, Fomenko,
  Formozov, Franco, Gabriele, Galbiati, Ghiano, Giammarchi, Goretti, Gromov,
  Guffanti, Hagner, Houdy, Hungerford, Ianni, Ianni, Jany, Jeschke, Kobychev,
  Korablev, Korga, Kryn, Laubenstein, Litvinovich, Lombardi, Lombardi, Ludhova,
  Lukyanchenko, Lukyanchenko, Machulin, Manuzio, Marcocci, Martyn, Meroni,
  Meyer, Miramonti, Misiaszek, Muratova, Neumair, Oberauer, Opitz, Orekhov,
  Ortica, Pallavicini, Papp, Penek, Pilipenko, Pocar, Porcelli, Raikov,
  Ranucci, Razeto, Re, Redchuk, Romani, Roncin, Rossi, Sch{\"o}nert, Semenov,
  Skorokhvatov, Smirnov, Sotnikov, Stokes, Suvorov, Tartaglia, Testera, Thurn,
  Toropova, Unzhakov, Villante, Vishneva, Vogelaar, von Feilitzsch, Wang,
  Weinz, Wojcik, Wurm, Yokley, Zaimidoroga, Zavatarelli, Zuber and
  Zuzel}]{Borexino18}
\bibinfo{author}{Agostini, M.}, \bibinfo{author}{Altenm{\"u}ller, K.},
  \bibinfo{author}{Appel, S.}, \bibinfo{author}{Atroshchenko, V.},
  \bibinfo{author}{Bagdasarian, Z.}, \bibinfo{author}{Basilico, D.},
  \bibinfo{author}{Bellini, G.}, \bibinfo{author}{Benziger, J.},
  \bibinfo{author}{Bick, D.}, \bibinfo{author}{Bonfini, G.},
  \bibinfo{author}{Bravo, D.}, \bibinfo{author}{Caccianiga, B.},
  \bibinfo{author}{Calaprice, F.}, \bibinfo{author}{Caminata, A.},
  \bibinfo{author}{Caprioli, S.}, \bibinfo{author}{Carlini, M.},
  \bibinfo{author}{Cavalcante, P.}, \bibinfo{author}{Chepurnov, A.},
  \bibinfo{author}{Choi, K.}, \bibinfo{author}{Collica, L.},
  \bibinfo{author}{D'Angelo, D.}, \bibinfo{author}{Davini, S.},
  \bibinfo{author}{Derbin, A.}, \bibinfo{author}{Ding, X.F.},
  \bibinfo{author}{Di~Ludovico, A.}, \bibinfo{author}{Di~Noto, L.},
  \bibinfo{author}{Drachnev, I.}, \bibinfo{author}{Fomenko, K.},
  \bibinfo{author}{Formozov, A.}, \bibinfo{author}{Franco, D.},
  \bibinfo{author}{Gabriele, F.}, \bibinfo{author}{Galbiati, C.},
  \bibinfo{author}{Ghiano, C.}, \bibinfo{author}{Giammarchi, M.},
  \bibinfo{author}{Goretti, A.}, \bibinfo{author}{Gromov, M.},
  \bibinfo{author}{Guffanti, D.}, \bibinfo{author}{Hagner, C.},
  \bibinfo{author}{Houdy, T.}, \bibinfo{author}{Hungerford, E.},
  \bibinfo{author}{Ianni, A.}, \bibinfo{author}{Ianni, A.},
  \bibinfo{author}{Jany, A.}, \bibinfo{author}{Jeschke, D.},
  \bibinfo{author}{Kobychev, V.}, \bibinfo{author}{Korablev, D.},
  \bibinfo{author}{Korga, G.}, \bibinfo{author}{Kryn, D.},
  \bibinfo{author}{Laubenstein, M.}, \bibinfo{author}{Litvinovich, E.},
  \bibinfo{author}{Lombardi, F.}, \bibinfo{author}{Lombardi, P.},
  \bibinfo{author}{Ludhova, L.}, \bibinfo{author}{Lukyanchenko, G.},
  \bibinfo{author}{Lukyanchenko, L.}, \bibinfo{author}{Machulin, I.},
  \bibinfo{author}{Manuzio, G.}, \bibinfo{author}{Marcocci, S.},
  \bibinfo{author}{Martyn, J.}, \bibinfo{author}{Meroni, E.},
  \bibinfo{author}{Meyer, M.}, \bibinfo{author}{Miramonti, L.},
  \bibinfo{author}{Misiaszek, M.}, \bibinfo{author}{Muratova, V.},
  \bibinfo{author}{Neumair, B.}, \bibinfo{author}{Oberauer, L.},
  \bibinfo{author}{Opitz, B.}, \bibinfo{author}{Orekhov, V.},
  \bibinfo{author}{Ortica, F.}, \bibinfo{author}{Pallavicini, M.},
  \bibinfo{author}{Papp, L.}, \bibinfo{author}{Penek, {\"O}.},
  \bibinfo{author}{Pilipenko, N.}, \bibinfo{author}{Pocar, A.},
  \bibinfo{author}{Porcelli, A.}, \bibinfo{author}{Raikov, G.},
  \bibinfo{author}{Ranucci, G.}, \bibinfo{author}{Razeto, A.},
  \bibinfo{author}{Re, A.}, \bibinfo{author}{Redchuk, M.},
  \bibinfo{author}{Romani, A.}, \bibinfo{author}{Roncin, R.},
  \bibinfo{author}{Rossi, N.}, \bibinfo{author}{Sch{\"o}nert, S.},
  \bibinfo{author}{Semenov, D.}, \bibinfo{author}{Skorokhvatov, M.},
  \bibinfo{author}{Smirnov, O.}, \bibinfo{author}{Sotnikov, A.},
  \bibinfo{author}{Stokes, L.F.F.}, \bibinfo{author}{Suvorov, Y.},
  \bibinfo{author}{Tartaglia, R.}, \bibinfo{author}{Testera, G.},
  \bibinfo{author}{Thurn, J.}, \bibinfo{author}{Toropova, M.},
  \bibinfo{author}{Unzhakov, E.}, \bibinfo{author}{Villante, F.L.},
  \bibinfo{author}{Vishneva, A.}, \bibinfo{author}{Vogelaar, R.B.},
  \bibinfo{author}{von Feilitzsch, F.}, \bibinfo{author}{Wang, H.},
  \bibinfo{author}{Weinz, S.}, \bibinfo{author}{Wojcik, M.},
  \bibinfo{author}{Wurm, M.}, \bibinfo{author}{Yokley, Z.},
  \bibinfo{author}{Zaimidoroga, O.}, \bibinfo{author}{Zavatarelli, S.},
  \bibinfo{author}{Zuber, K.}, \bibinfo{author}{Zuzel, G.},
  \bibinfo{year}{2018}.
\newblock \bibinfo{title}{Comprehensive measurement of \textit{pp}-chain solar
  neutrinos}.
\newblock \bibinfo{journal}{Nature} \bibinfo{volume}{562},
  \bibinfo{pages}{505--510}.
\newblock \DOIprefix\doi{10.1038/s41586-018-0624-y}.
%Type = Article
\bibitem[{Alexander et~al.(2012)Alexander, Bowden, Fogel, Howard, Herd and
  Nittler}]{alexander2012provenances}
\bibinfo{author}{Alexander, C.M.O'D.}, \bibinfo{author}{Bowden, R.},
  \bibinfo{author}{Fogel, M.L.}, \bibinfo{author}{Howard, K.T.},
  \bibinfo{author}{Herd, C.D.K.}, \bibinfo{author}{Nittler, L.R.},
  \bibinfo{year}{2012}.
\newblock \bibinfo{title}{{The provenances of asteroids, and their
  contributions to the volatile inventories of the terrestrial planets}}.
\newblock \bibinfo{journal}{Science} \bibinfo{volume}{337},
  \bibinfo{pages}{721--723}.
\newblock \DOIprefix\doi{10.1126/science.1223474}.
%Type = Article
\bibitem[{Anderson and Ahrens(1994)}]{anderson1994equation}
\bibinfo{author}{Anderson, W.W.}, \bibinfo{author}{Ahrens, T.J.},
  \bibinfo{year}{1994}.
\newblock \bibinfo{title}{{An equation of state for liquid iron and
  implications for the Earth's core}}.
\newblock \bibinfo{journal}{Journal of Geophysical Research: Solid Earth}
  \bibinfo{volume}{99}, \bibinfo{pages}{4273--4284}.
\newblock \DOIprefix\doi{10.1029/93JB03158}.
%Type = Article
\bibitem[{Antonangeli et~al.(2015)Antonangeli, Morard, Schmerr, Komabayashi,
  Krisch, Fiquet and Fei}]{antonangeli2015toward}
\bibinfo{author}{Antonangeli, D.}, \bibinfo{author}{Morard, G.},
  \bibinfo{author}{Schmerr, N.C.}, \bibinfo{author}{Komabayashi, T.},
  \bibinfo{author}{Krisch, M.}, \bibinfo{author}{Fiquet, G.},
  \bibinfo{author}{Fei, Y.}, \bibinfo{year}{2015}.
\newblock \bibinfo{title}{{Toward a mineral physics reference model for the
  Moon's core}}.
\newblock \bibinfo{journal}{Proceedings of the National Academy of Sciences}
  \bibinfo{volume}{112}, \bibinfo{pages}{3916--3919}.
\newblock \DOIprefix\doi{10.1073/pnas.1417490112}.
%Type = Article
\bibitem[{Arndt and Jenner(1986)}]{arndt1986crustally}
\bibinfo{author}{Arndt, N.T.}, \bibinfo{author}{Jenner, G.A.},
  \bibinfo{year}{1986}.
\newblock \bibinfo{title}{{Crustally contaminated komatiites and basalts from
  Kambalda, Western Australia}}.
\newblock \bibinfo{journal}{Chemical Geology} \bibinfo{volume}{56},
  \bibinfo{pages}{229--255}.
\newblock \DOIprefix\doi{10.1016/0009-2541(86)90006-9}.
%Type = Article
\bibitem[{Asplund et~al.(2009)Asplund, Grevesse, Sauval and
  Scott}]{asplund2009chemical}
\bibinfo{author}{Asplund, M.}, \bibinfo{author}{Grevesse, N.},
  \bibinfo{author}{Sauval, A.J.}, \bibinfo{author}{Scott, P.},
  \bibinfo{year}{2009}.
\newblock \bibinfo{title}{{The chemical composition of the Sun}}.
\newblock \bibinfo{journal}{Annual Review of Astronomy and Astrophysics}
  \bibinfo{volume}{47}, \bibinfo{pages}{481--522}.
\newblock \DOIprefix\doi{10.1146/annurev.astro.46.060407.145222}.
%Type = Incollection
\bibitem[{Badding et~al.(1992)Badding, Mao and Hemley}]{badding1992high}
\bibinfo{author}{Badding, J.V.}, \bibinfo{author}{Mao, H.K.},
  \bibinfo{author}{Hemley, R.J.}, \bibinfo{year}{1992}.
\newblock \bibinfo{title}{{High-pressure crystal structure and equation of
  state of iron hydride: Implications for the Earth's core}}, in:
  \bibinfo{editor}{Syono, Y.}, \bibinfo{editor}{Manghnani, M.H.} (Eds.),
  \bibinfo{booktitle}{High Pressure Research: Application to Earth and
  Planetary Sciences}. \bibinfo{publisher}{American Geophysical Union}.
  volume~\bibinfo{volume}{67}. chapter~\bibinfo{chapter}{5}, pp.
  \bibinfo{pages}{363--371}.
\newblock \DOIprefix\doi{10.1029/GM067p0363}.
%Type = Article
\bibitem[{Bailey et~al.(2015)Bailey, Nagayama, Loisel, Rochau, Blancard,
  Colgan, Cosse, Faussurier, Fontes, Gilleron, Golovkin, Hansen, Iglesias,
  Kilcrease, MacFarlane, Mancini, Nahar, Orban, Pain, Pradhan, Sherrill and
  Wilson}]{bailey2015higher}
\bibinfo{author}{Bailey, J.E.}, \bibinfo{author}{Nagayama, T.},
  \bibinfo{author}{Loisel, G.P.}, \bibinfo{author}{Rochau, G.A.},
  \bibinfo{author}{Blancard, C.}, \bibinfo{author}{Colgan, J.},
  \bibinfo{author}{Cosse, P.}, \bibinfo{author}{Faussurier, G.},
  \bibinfo{author}{Fontes, C.J.}, \bibinfo{author}{Gilleron, F.},
  \bibinfo{author}{Golovkin, I.}, \bibinfo{author}{Hansen, S.B.},
  \bibinfo{author}{Iglesias, C.A.}, \bibinfo{author}{Kilcrease, D.P.},
  \bibinfo{author}{MacFarlane, J.J.}, \bibinfo{author}{Mancini, R.C.},
  \bibinfo{author}{Nahar, S.N.}, \bibinfo{author}{Orban, C.},
  \bibinfo{author}{Pain, J.C.}, \bibinfo{author}{Pradhan, A.K.},
  \bibinfo{author}{Sherrill, M.}, \bibinfo{author}{Wilson, B.G.},
  \bibinfo{year}{2015}.
\newblock \bibinfo{title}{{A higher-than-predicted measurement of iron opacity
  at solar interior temperatures}}.
\newblock \bibinfo{journal}{Nature} \bibinfo{volume}{517},
  \bibinfo{pages}{56--59}.
\newblock \DOIprefix\doi{10.1038/nature14048}.
%Type = Article
\bibitem[{Baratoux et~al.(2013)Baratoux, Toplis, Monnereau and
  Sautter}]{baratoux2013petrological}
\bibinfo{author}{Baratoux, D.}, \bibinfo{author}{Toplis, M.},
  \bibinfo{author}{Monnereau, M.}, \bibinfo{author}{Sautter, V.},
  \bibinfo{year}{2013}.
\newblock \bibinfo{title}{{The petrological expression of early Mars
  volcanism}}.
\newblock \bibinfo{journal}{Journal of Geophysical Research: Planets}
  \bibinfo{volume}{118}, \bibinfo{pages}{59--64}.
\newblock \DOIprefix\doi{10.1029/2012JE004234}.
%Type = Article
\bibitem[{Baratoux et~al.(2011)Baratoux, Toplis, Monnereau and
  Gasnault}]{baratoux2011thermal}
\bibinfo{author}{Baratoux, D.}, \bibinfo{author}{Toplis, M.J.},
  \bibinfo{author}{Monnereau, M.}, \bibinfo{author}{Gasnault, O.},
  \bibinfo{year}{2011}.
\newblock \bibinfo{title}{{Thermal history of Mars inferred from orbital
  geochemistry of volcanic provinces}}.
\newblock \bibinfo{journal}{Nature} \bibinfo{volume}{472},
  \bibinfo{pages}{338--341}.
\newblock \DOIprefix\doi{10.1038/nature09903}.
%Type = Article
\bibitem[{Barrat et~al.(2016)Barrat, Greenwood, Keil, Rouget, Boesenberg, Zanda
  and Franchi}]{barrat2016origin}
\bibinfo{author}{Barrat, J.A.}, \bibinfo{author}{Greenwood, R.C.},
  \bibinfo{author}{Keil, K.}, \bibinfo{author}{Rouget, M.L.},
  \bibinfo{author}{Boesenberg, J.S.}, \bibinfo{author}{Zanda, B.},
  \bibinfo{author}{Franchi, I.A.}, \bibinfo{year}{2016}.
\newblock \bibinfo{title}{{The origin of aubrites: Evidence from lithophile
  trace element abundances and oxygen isotope compositions}}.
\newblock \bibinfo{journal}{Geochimica et Cosmochimica Acta}
  \bibinfo{volume}{192}, \bibinfo{pages}{29--48}.
\newblock \DOIprefix\doi{10.1016/j.gca.2016.07.025}.
%Type = Article
\bibitem[{Basu and Antia(2008)}]{basu2008helioseismology}
\bibinfo{author}{Basu, S.}, \bibinfo{author}{Antia, H.M.},
  \bibinfo{year}{2008}.
\newblock \bibinfo{title}{{Helioseismology and solar abundances}}.
\newblock \bibinfo{journal}{Physics Reports} \bibinfo{volume}{457},
  \bibinfo{pages}{217--283}.
\newblock \DOIprefix\doi{10.1016/j.physrep.2007.12.002}.
%Type = Article
\bibitem[{Bellucci et~al.(2018)Bellucci, Nemchin, Whitehouse, Snape, Bland,
  Benedix and Roszjar}]{bellucci2018pb}
\bibinfo{author}{Bellucci, J.J.}, \bibinfo{author}{Nemchin, A.A.},
  \bibinfo{author}{Whitehouse, M.J.}, \bibinfo{author}{Snape, J.F.},
  \bibinfo{author}{Bland, P.}, \bibinfo{author}{Benedix, G.K.},
  \bibinfo{author}{Roszjar, J.}, \bibinfo{year}{2018}.
\newblock \bibinfo{title}{{Pb evolution in the Martian mantle}}.
\newblock \bibinfo{journal}{Earth and Planetary Science Letters}
  \bibinfo{volume}{485}, \bibinfo{pages}{79--87}.
\newblock \DOIprefix\doi{10.1016/j.epsl.2017.12.039}.
%Type = Incollection
\bibitem[{Bergemann and Serenelli(2014)}]{bergemann2014solar}
\bibinfo{author}{Bergemann, M.}, \bibinfo{author}{Serenelli, A.},
  \bibinfo{year}{2014}.
\newblock \bibinfo{title}{{Solar abundance problem}}, in:
  \bibinfo{editor}{Niemczura, E.}, \bibinfo{editor}{Smalley, B.},
  \bibinfo{editor}{Pych, W.} (Eds.), \bibinfo{booktitle}{Determination of
  Atmospheric Parameters of B-, A-, F-and G-Type Stars}.
  \bibinfo{publisher}{Springer}, \bibinfo{address}{Cham}, pp.
  \bibinfo{pages}{245--258}.
\newblock \DOIprefix\doi{10.1007/978-3-319-06956-2_21}.
%Type = Article
\bibitem[{Bertka and Fei(1997)}]{bertka1997mineralogy}
\bibinfo{author}{Bertka, C.M.}, \bibinfo{author}{Fei, Y.},
  \bibinfo{year}{1997}.
\newblock \bibinfo{title}{{Mineralogy of the Martian interior up to core-mantle
  boundary pressures}}.
\newblock \bibinfo{journal}{Journal of Geophysical Research: Solid Earth}
  \bibinfo{volume}{102}, \bibinfo{pages}{5251--5264}.
\newblock \DOIprefix\doi{10.1029/96JB03270}.
%Type = Article
\bibitem[{Bertka and Fei(1998a)}]{bertka1998density}
\bibinfo{author}{Bertka, C.M.}, \bibinfo{author}{Fei, Y.},
  \bibinfo{year}{1998}a.
\newblock \bibinfo{title}{{Density profile of an SNC model Martian interior and
  the moment-of-inertia factor of Mars}}.
\newblock \bibinfo{journal}{Earth and Planetary Science Letters}
  \bibinfo{volume}{157}, \bibinfo{pages}{79--88}.
\newblock \DOIprefix\doi{10.1016/S0012-821X(98)00030-2}.
%Type = Article
\bibitem[{Bertka and Fei(1998b)}]{bertka1998implications}
\bibinfo{author}{Bertka, C.M.}, \bibinfo{author}{Fei, Y.},
  \bibinfo{year}{1998}b.
\newblock \bibinfo{title}{{Implications of Mars Pathfinder data for the
  accretion history of the terrestrial planets}}.
\newblock \bibinfo{journal}{Science} \bibinfo{volume}{281},
  \bibinfo{pages}{1838--1840}.
\newblock \DOIprefix\doi{10.1126/science.281.5384.1838}.
%Type = Article
\bibitem[{Birch(1952)}]{birch1952elasticity}
\bibinfo{author}{Birch, F.}, \bibinfo{year}{1952}.
\newblock \bibinfo{title}{{Elasticity and constitution of the Earth's
  interior}}.
\newblock \bibinfo{journal}{Journal of Geophysical Research}
  \bibinfo{volume}{57}, \bibinfo{pages}{227--286}.
\newblock \DOIprefix\doi{10.1029/JZ057i002p00227}.
%Type = Article
\bibitem[{Birch(1964)}]{birch1964density}
\bibinfo{author}{Birch, F.}, \bibinfo{year}{1964}.
\newblock \bibinfo{title}{{Density and composition of mantle and core}}.
\newblock \bibinfo{journal}{Journal of Geophysical Research}
  \bibinfo{volume}{69}, \bibinfo{pages}{4377--4388}.
\newblock \DOIprefix\doi{10.1029/JZ069i020p04377}.
%Type = Article
\bibitem[{Blichert-Toft et~al.(1999)Blichert-Toft, Gleason, T{\'e}louk and
  Albar{\`e}de}]{blichert1999lu}
\bibinfo{author}{Blichert-Toft, J.}, \bibinfo{author}{Gleason, J.D.},
  \bibinfo{author}{T{\'e}louk, P.}, \bibinfo{author}{Albar{\`e}de, F.},
  \bibinfo{year}{1999}.
\newblock \bibinfo{title}{{The Lu--Hf isotope geochemistry of shergottites and
  the evolution of the Martian mantle--crust system}}.
\newblock \bibinfo{journal}{Earth and Planetary Science Letters}
  \bibinfo{volume}{173}, \bibinfo{pages}{25--39}.
\newblock \DOIprefix\doi{10.1016/S0012-821X(99)00222-8}.
%Type = Article
\bibitem[{Bogard et~al.(2001)Bogard, Clayton, Marti, Owen and
  Turner}]{bogard2001martian}
\bibinfo{author}{Bogard, D.D.}, \bibinfo{author}{Clayton, R.N.},
  \bibinfo{author}{Marti, K.}, \bibinfo{author}{Owen, T.},
  \bibinfo{author}{Turner, G.}, \bibinfo{year}{2001}.
\newblock \bibinfo{title}{{Martian volatiles: isotopic composition, origin, and
  evolution}}.
\newblock \bibinfo{journal}{Space Science Reviews} \bibinfo{volume}{96},
  \bibinfo{pages}{425--458}.
\newblock \DOIprefix\doi{10.1023/A:1011974028370}.
%Type = Article
\bibitem[{Borg and Draper(2003)}]{borg2003petrogenetic}
\bibinfo{author}{Borg, L.E.}, \bibinfo{author}{Draper, D.S.},
  \bibinfo{year}{2003}.
\newblock \bibinfo{title}{{A petrogenetic model for the origin and
  compositional variation of the Martian basaltic meteorites}}.
\newblock \bibinfo{journal}{Meteoritics \& Planetary Science}
  \bibinfo{volume}{38}, \bibinfo{pages}{1713--1731}.
\newblock \DOIprefix\doi{10.1111/j.1945-5100.2003.tb00011.x}.
%Type = Article
\bibitem[{Borg et~al.(2005)Borg, Edmunson and Asmerom}]{borg2005constraints}
\bibinfo{author}{Borg, L.E.}, \bibinfo{author}{Edmunson, J.E.},
  \bibinfo{author}{Asmerom, Y.}, \bibinfo{year}{2005}.
\newblock \bibinfo{title}{{Constraints on the U-Pb isotopic systematics of Mars
  inferred from a combined U-Pb, Rb-Sr, and Sm-Nd isotopic study of the Martian
  meteorite Zagami}}.
\newblock \bibinfo{journal}{Geochimica et Cosmochimica Acta}
  \bibinfo{volume}{69}, \bibinfo{pages}{5819--5830}.
\newblock \DOIprefix\doi{10.1016/j.gca.2005.08.007}.
%Type = Article
\bibitem[{Boujibar et~al.(2014)Boujibar, Andrault, Bouhifd, Bolfan-Casanova,
  Devidal and Trcera}]{boujibar2014metal}
\bibinfo{author}{Boujibar, A.}, \bibinfo{author}{Andrault, D.},
  \bibinfo{author}{Bouhifd, M.A.}, \bibinfo{author}{Bolfan-Casanova, N.},
  \bibinfo{author}{Devidal, J.L.}, \bibinfo{author}{Trcera, N.},
  \bibinfo{year}{2014}.
\newblock \bibinfo{title}{{Metal--silicate partitioning of sulphur, new
  experimental and thermodynamic constraints on planetary accretion}}.
\newblock \bibinfo{journal}{Earth and Planetary Science Letters}
  \bibinfo{volume}{391}, \bibinfo{pages}{42--54}.
\newblock \DOIprefix\doi{10.1016/j.epsl.2014.01.021}.
%Type = Article
\bibitem[{Bouvier et~al.(2009)Bouvier, Blichert-Toft and
  Albarede}]{bouvier2009martian}
\bibinfo{author}{Bouvier, A.}, \bibinfo{author}{Blichert-Toft, J.},
  \bibinfo{author}{Albarede, F.}, \bibinfo{year}{2009}.
\newblock \bibinfo{title}{{Martian meteorite chronology and the evolution of
  the interior of Mars}}.
\newblock \bibinfo{journal}{Earth and Planetary Science Letters}
  \bibinfo{volume}{280}, \bibinfo{pages}{285--295}.
\newblock \DOIprefix\doi{10.1016/j.epsl.2009.01.042}.
%Type = Article
\bibitem[{Bouvier et~al.(2005)Bouvier, Blichert-Toft, Vervoort and
  Albarede}]{bouvier2005age}
\bibinfo{author}{Bouvier, A.}, \bibinfo{author}{Blichert-Toft, J.},
  \bibinfo{author}{Vervoort, J.D.}, \bibinfo{author}{Albarede, F.},
  \bibinfo{year}{2005}.
\newblock \bibinfo{title}{{The age of SNC meteorites and the antiquity of the
  Martian surface}}.
\newblock \bibinfo{journal}{Earth and Planetary Science Letters}
  \bibinfo{volume}{240}, \bibinfo{pages}{221--233}.
\newblock \DOIprefix\doi{10.1016/j.epsl.2005.09.007}.
%Type = Article
\bibitem[{Bouvier et~al.(2008a)Bouvier, Blichert-Toft, Vervoort, Gillet and
  Albar{\`e}de}]{bouvier2008case}
\bibinfo{author}{Bouvier, A.}, \bibinfo{author}{Blichert-Toft, J.},
  \bibinfo{author}{Vervoort, J.D.}, \bibinfo{author}{Gillet, P.},
  \bibinfo{author}{Albar{\`e}de, F.}, \bibinfo{year}{2008}a.
\newblock \bibinfo{title}{{The case for old basaltic shergottites}}.
\newblock \bibinfo{journal}{Earth and Planetary Science Letters}
  \bibinfo{volume}{266}, \bibinfo{pages}{105--124}.
\newblock \DOIprefix\doi{10.1016/j.epsl.2007.11.006}.
%Type = Article
\bibitem[{Bouvier et~al.(2008b)Bouvier, Vervoort and Patchett}]{bouvier2008lu}
\bibinfo{author}{Bouvier, A.}, \bibinfo{author}{Vervoort, J.D.},
  \bibinfo{author}{Patchett, P.J.}, \bibinfo{year}{2008}b.
\newblock \bibinfo{title}{{The Lu--Hf and Sm--Nd isotopic composition of CHUR:
  Constraints from unequilibrated chondrites and implications for the bulk
  composition of terrestrial planets}}.
\newblock \bibinfo{journal}{Earth and Planetary Science Letters}
  \bibinfo{volume}{273}, \bibinfo{pages}{48--57}.
\newblock \DOIprefix\doi{10.1016/j.epsl.2008.06.010}.
%Type = Article
\bibitem[{Bouvier et~al.(2018)Bouvier, Costa, Connelly, Jensen, Wielandt,
  Storey, Nemchin, Whitehouse, Snape, Bellucci, {Moynier}, {Agranier},
  {Gueguen}, {Sch{\"o}nb{\"a}chler} and {Bizzarro}}]{bouvier2018evidence}
\bibinfo{author}{Bouvier, L.C.}, \bibinfo{author}{Costa, M.M.},
  \bibinfo{author}{Connelly, J.N.}, \bibinfo{author}{Jensen, N.K.},
  \bibinfo{author}{Wielandt, D.}, \bibinfo{author}{Storey, M.},
  \bibinfo{author}{Nemchin, A.A.}, \bibinfo{author}{Whitehouse, M.J.},
  \bibinfo{author}{Snape, J.F.}, \bibinfo{author}{Bellucci, J.J.},
  \bibinfo{author}{{Moynier}, F.}, \bibinfo{author}{{Agranier}, A.},
  \bibinfo{author}{{Gueguen}, B.}, \bibinfo{author}{{Sch{\"o}nb{\"a}chler},
  M.}, \bibinfo{author}{{Bizzarro}, M.}, \bibinfo{year}{2018}.
\newblock \bibinfo{title}{{Evidence for extremely rapid magma ocean
  crystallization and crust formation on Mars}}.
\newblock \bibinfo{journal}{Nature} \bibinfo{volume}{558},
  \bibinfo{pages}{586--589}.
\newblock \DOIprefix\doi{10.1038/s41586-018-0222-z}.
%Type = Article
\bibitem[{Boynton et~al.(2007)Boynton, Taylor, Evans, Reedy, Starr, Janes,
  Kerry, Drake, Kim, Williams, Crombie, Dohm, Baker, Metzger, Karunatillake,
  Keller, Newsom, Arnold, Brückner, Englert, Gasnault, Sprague, Mitrofanov,
  Squyres, Trombka, d'Uston, Wänke and Hamara}]{boynton2007concentration}
\bibinfo{author}{Boynton, W.V.}, \bibinfo{author}{Taylor, G.J.},
  \bibinfo{author}{Evans, L.G.}, \bibinfo{author}{Reedy, R.C.},
  \bibinfo{author}{Starr, R.}, \bibinfo{author}{Janes, D.M.},
  \bibinfo{author}{Kerry, K.E.}, \bibinfo{author}{Drake, D.M.},
  \bibinfo{author}{Kim, K.J.}, \bibinfo{author}{Williams, R.M.S.},
  \bibinfo{author}{Crombie, M.K.}, \bibinfo{author}{Dohm, J.M.},
  \bibinfo{author}{Baker, V.}, \bibinfo{author}{Metzger, A.E.},
  \bibinfo{author}{Karunatillake, S.}, \bibinfo{author}{Keller, J.M.},
  \bibinfo{author}{Newsom, H.E.}, \bibinfo{author}{Arnold, J.R.},
  \bibinfo{author}{Brückner, J.}, \bibinfo{author}{Englert, P.A.J.},
  \bibinfo{author}{Gasnault, O.}, \bibinfo{author}{Sprague, A.L.},
  \bibinfo{author}{Mitrofanov, I.}, \bibinfo{author}{Squyres, S.W.},
  \bibinfo{author}{Trombka, J.I.}, \bibinfo{author}{d'Uston, L.},
  \bibinfo{author}{Wänke, H.}, \bibinfo{author}{Hamara, D.K.},
  \bibinfo{year}{2007}.
\newblock \bibinfo{title}{{Concentration of H, Si, Cl, K, Fe, and Th in the
  low-and mid-latitude regions of Mars}}.
\newblock \bibinfo{journal}{Journal of Geophysical Research: Planets}
  \bibinfo{volume}{112}, \bibinfo{pages}{E12S99}.
\newblock \DOIprefix\doi{10.1029/2007JE002887}.
%Type = Incollection
\bibitem[{Boynton et~al.(2008)Boynton, Taylor, Karunatillake, Reedy and
  Keller}]{boynton2008elemental}
\bibinfo{author}{Boynton, W.V.}, \bibinfo{author}{Taylor, G.J.},
  \bibinfo{author}{Karunatillake, S.}, \bibinfo{author}{Reedy, R.C.},
  \bibinfo{author}{Keller, J.M.}, \bibinfo{year}{2008}.
\newblock \bibinfo{title}{{Elemental abundances determined via the Mars Odyssey
  GRS}}, in: \bibinfo{editor}{Bell~III, J.} (Ed.), \bibinfo{booktitle}{The
  Martian Surface: Composition, Mineralogy, and Physical Properties}.
  \bibinfo{publisher}{Cambridge University Press},
  \bibinfo{address}{Cambridge}, pp. \bibinfo{pages}{105--124}.
\newblock \DOIprefix\doi{10.1017/CBO9780511536076.006}.
%Type = Article
\bibitem[{Brandon et~al.(2012)Brandon, Puchtel, Walker, Day, Irving and
  Taylor}]{brandon2012evolution}
\bibinfo{author}{Brandon, A.D.}, \bibinfo{author}{Puchtel, I.S.},
  \bibinfo{author}{Walker, R.J.}, \bibinfo{author}{Day, J.M.D.},
  \bibinfo{author}{Irving, A.J.}, \bibinfo{author}{Taylor, L.A.},
  \bibinfo{year}{2012}.
\newblock \bibinfo{title}{{Evolution of the martian mantle inferred from the
  \ce{^{187}Re}--\ce{^{187}Os} isotope and highly siderophile element abundance
  systematics of shergottite meteorites}}.
\newblock \bibinfo{journal}{Geochimica et Cosmochimica Acta}
  \bibinfo{volume}{76}, \bibinfo{pages}{206--235}.
\newblock \DOIprefix\doi{10.1016/j.gca.2011.09.047}.
%Type = Article
\bibitem[{Burbine and O'Brien(2004)}]{burbine2004determining}
\bibinfo{author}{Burbine, T.H.}, \bibinfo{author}{O'Brien, K.M.},
  \bibinfo{year}{2004}.
\newblock \bibinfo{title}{{Determining the possible building blocks of the
  Earth and Mars}}.
\newblock \bibinfo{journal}{Meteoritics \& Planetary Science}
  \bibinfo{volume}{39}, \bibinfo{pages}{667--681}.
\newblock \DOIprefix\doi{10.1111/j.1945-5100.2004.tb00110.x}.
%Type = Article
\bibitem[{Canup and Salmon(2018)}]{canup2018origin}
\bibinfo{author}{Canup, R.}, \bibinfo{author}{Salmon, J.},
  \bibinfo{year}{2018}.
\newblock \bibinfo{title}{{Origin of Phobos and Deimos by the impact of a
  Vesta-to-Ceres sized body with Mars}}.
\newblock \bibinfo{journal}{Science Advances} \bibinfo{volume}{4},
  \bibinfo{pages}{eaar6887}.
\newblock \DOIprefix\doi{10.1126/sciadv.aar6887}.
%Type = Article
\bibitem[{Chen and Wasserburg(1986)}]{chen1986formation}
\bibinfo{author}{Chen, J.H.}, \bibinfo{author}{Wasserburg, G.J.},
  \bibinfo{year}{1986}.
\newblock \bibinfo{title}{{Formation ages and evolution of Shergotty and its
  parent planet from U-Th-Pb systematics}}.
\newblock \bibinfo{journal}{Geochimica et Cosmochimica Acta}
  \bibinfo{volume}{50}, \bibinfo{pages}{955--968}.
\newblock \DOIprefix\doi{10.1016/0016-7037(86)90376-5}.
%Type = Article
\bibitem[{Citron et~al.(2015)Citron, Genda and Ida}]{citron2015formation}
\bibinfo{author}{Citron, R.I.}, \bibinfo{author}{Genda, H.},
  \bibinfo{author}{Ida, S.}, \bibinfo{year}{2015}.
\newblock \bibinfo{title}{{Formation of Phobos and Deimos via a giant impact}}.
\newblock \bibinfo{journal}{Icarus} \bibinfo{volume}{252},
  \bibinfo{pages}{334--338}.
\newblock \DOIprefix\doi{10.1016/j.icarus.2015.02.011}.
%Type = Article
\bibitem[{Clay et~al.(2017)Clay, Burgess, Busemann, Ruzi{\'e}-Hamilton,
  Joachim, Day and Ballentine}]{clay2017halogens}
\bibinfo{author}{Clay, P.L.}, \bibinfo{author}{Burgess, R.},
  \bibinfo{author}{Busemann, H.}, \bibinfo{author}{Ruzi{\'e}-Hamilton, L.},
  \bibinfo{author}{Joachim, B.}, \bibinfo{author}{Day, J.M.D.},
  \bibinfo{author}{Ballentine, C.J.}, \bibinfo{year}{2017}.
\newblock \bibinfo{title}{{Halogens in chondritic meteorites and terrestrial
  accretion}}.
\newblock \bibinfo{journal}{Nature} \bibinfo{volume}{551},
  \bibinfo{pages}{614--618}.
\newblock \DOIprefix\doi{10.1038/nature24625}.
%Type = Article
\bibitem[{Collinet et~al.(2015)Collinet, M{\'e}dard, Charlier, Vander~Auwera
  and Grove}]{collinet2015melting}
\bibinfo{author}{Collinet, M.}, \bibinfo{author}{M{\'e}dard, E.},
  \bibinfo{author}{Charlier, B.}, \bibinfo{author}{Vander~Auwera, J.},
  \bibinfo{author}{Grove, T.L.}, \bibinfo{year}{2015}.
\newblock \bibinfo{title}{{Melting of the primitive Martian mantle at 0.5--2.2
  GPa and the origin of basalts and alkaline rocks on Mars}}.
\newblock \bibinfo{journal}{Earth and Planetary Science Letters}
  \bibinfo{volume}{427}, \bibinfo{pages}{83--94}.
\newblock \DOIprefix\doi{10.1016/j.epsl.2015.06.056}.
%Type = Article
\bibitem[{Connolly(2009)}]{connolly2009geodynamic}
\bibinfo{author}{Connolly, J.A.D.}, \bibinfo{year}{2009}.
\newblock \bibinfo{title}{{The geodynamic equation of state: what and how}}.
\newblock \bibinfo{journal}{Geochemistry, Geophysics, Geosystems}
  \bibinfo{volume}{10}, \bibinfo{pages}{Q10014}.
\newblock \DOIprefix\doi{10.1029/2009GC002540}.
%Type = Article
\bibitem[{Corgne et~al.(2008)Corgne, Keshav, Wood, McDonough and
  Fei}]{corgne2008metal}
\bibinfo{author}{Corgne, A.}, \bibinfo{author}{Keshav, S.},
  \bibinfo{author}{Wood, B.J.}, \bibinfo{author}{McDonough, W.F.},
  \bibinfo{author}{Fei, Y.}, \bibinfo{year}{2008}.
\newblock \bibinfo{title}{{Metal--silicate partitioning and constraints on core
  composition and oxygen fugacity during Earth accretion}}.
\newblock \bibinfo{journal}{Geochimica et Cosmochimica Acta}
  \bibinfo{volume}{72}, \bibinfo{pages}{574--589}.
\newblock \DOIprefix\doi{10.1016/j.gca.2007.10.006}.
%Type = Article
\bibitem[{Coryell et~al.(1963)Coryell, Chase and
  Winchester}]{coryell1963procedure}
\bibinfo{author}{Coryell, C.D.}, \bibinfo{author}{Chase, J.W.},
  \bibinfo{author}{Winchester, J.W.}, \bibinfo{year}{1963}.
\newblock \bibinfo{title}{{A procedure for geochemical interpretation of
  terrestrial rare-earth abundance patterns}}.
\newblock \bibinfo{journal}{Journal of Geophysical Research}
  \bibinfo{volume}{68}, \bibinfo{pages}{559--566}.
\newblock \DOIprefix\doi{10.1029/JZ068i002p00559}.
%Type = Article
\bibitem[{Craddock(2011)}]{craddock2011phobos}
\bibinfo{author}{Craddock, R.A.}, \bibinfo{year}{2011}.
\newblock \bibinfo{title}{{Are Phobos and Deimos the result of a giant
  impact?}}
\newblock \bibinfo{journal}{Icarus} \bibinfo{volume}{211},
  \bibinfo{pages}{1150--1161}.
\newblock \DOIprefix\doi{10.1016/j.icarus.2010.10.023}.
%Type = Article
\bibitem[{Curtis et~al.(1980)Curtis, Gladney and Jurney}]{curtis1980revision}
\bibinfo{author}{Curtis, D.}, \bibinfo{author}{Gladney, E.},
  \bibinfo{author}{Jurney, E.}, \bibinfo{year}{1980}.
\newblock \bibinfo{title}{{A revision of the meteorite based cosmic abundance
  of boron}}.
\newblock \bibinfo{journal}{Geochimica et Cosmochimica Acta}
  \bibinfo{volume}{44}, \bibinfo{pages}{1945--1953}.
\newblock \DOIprefix\doi{10.1016/0016-7037(80)90194-5}.
%Type = Incollection
\bibitem[{Dauphas and Morbidelli(2014)}]{dauphas2014chemical}
\bibinfo{author}{Dauphas, N.}, \bibinfo{author}{Morbidelli, A.},
  \bibinfo{year}{2014}.
\newblock \bibinfo{title}{{Geochemical and Planetary Dynamical Views on the
  Origin of Earth's Atmosphere and Oceans}}, in: \bibinfo{editor}{Holland,
  H.D.}, \bibinfo{editor}{Turekian, K.K.} (Eds.), \bibinfo{booktitle}{Treatise
  on Geochemistry (Second Edition)}. \bibinfo{publisher}{Elsevier},
  \bibinfo{address}{Oxford}. volume~\bibinfo{volume}{6}, pp.
  \bibinfo{pages}{1--35}.
\newblock \DOIprefix\doi{10.1016/B978-0-08-095975-7.01301-2}.
%Type = Article
\bibitem[{Dauphas et~al.(2015)Dauphas, Poitrasson, Burkhardt, Kobayashi and
  Kurosawa}]{dauphas2015planetary}
\bibinfo{author}{Dauphas, N.}, \bibinfo{author}{Poitrasson, F.},
  \bibinfo{author}{Burkhardt, C.}, \bibinfo{author}{Kobayashi, H.},
  \bibinfo{author}{Kurosawa, K.}, \bibinfo{year}{2015}.
\newblock \bibinfo{title}{{Planetary and meteoritic Mg/Si and
  $\delta$\ce{^{30}}Si variations inherited from solar nebula chemistry}}.
\newblock \bibinfo{journal}{Earth and Planetary Science Letters}
  \bibinfo{volume}{427}, \bibinfo{pages}{236--248}.
\newblock \DOIprefix\doi{10.1016/j.epsl.2015.07.008}.
%Type = Article
\bibitem[{Dauphas and Pourmand(2011)}]{dauphas2011hf}
\bibinfo{author}{Dauphas, N.}, \bibinfo{author}{Pourmand, A.},
  \bibinfo{year}{2011}.
\newblock \bibinfo{title}{{Hf--W--Th evidence for rapid growth of Mars and its
  status as a planetary embryo}}.
\newblock \bibinfo{journal}{Nature} \bibinfo{volume}{473},
  \bibinfo{pages}{489--492}.
\newblock \DOIprefix\doi{10.1038/nature10077}.
%Type = Article
\bibitem[{Dauphas and Pourmand(2015)}]{dauphas2015thulium}
\bibinfo{author}{Dauphas, N.}, \bibinfo{author}{Pourmand, A.},
  \bibinfo{year}{2015}.
\newblock \bibinfo{title}{{Thulium anomalies and rare earth element patterns in
  meteorites and Earth: Nebular fractionation and the nugget effect}}.
\newblock \bibinfo{journal}{Geochimica et Cosmochimica Acta}
  \bibinfo{volume}{163}, \bibinfo{pages}{234--261}.
\newblock \DOIprefix\doi{10.1016/j.gca.2015.03.037}.
%Type = Article
\bibitem[{Dauphas and Schauble(2016)}]{dauphas2016mass}
\bibinfo{author}{Dauphas, N.}, \bibinfo{author}{Schauble, E.A.},
  \bibinfo{year}{2016}.
\newblock \bibinfo{title}{{Mass fractionation laws, mass-independent effects,
  and isotopic anomalies}}.
\newblock \bibinfo{journal}{Annual Review of Earth and Planetary Sciences}
  \bibinfo{volume}{44}, \bibinfo{pages}{709--783}.
\newblock \DOIprefix\doi{10.1146/annurev-earth-060115-012157}.
%Type = Article
\bibitem[{Davis et~al.(2013)Davis, Humayun, Hirschmann and
  Cooper}]{davis2013experimentally}
\bibinfo{author}{Davis, F.A.}, \bibinfo{author}{Humayun, M.},
  \bibinfo{author}{Hirschmann, M.M.}, \bibinfo{author}{Cooper, R.S.},
  \bibinfo{year}{2013}.
\newblock \bibinfo{title}{{Experimentally determined mineral/melt partitioning
  of first-row transition elements (FRTE) during partial melting of peridotite
  at 3 GPa}}.
\newblock \bibinfo{journal}{Geochimica et Cosmochimica Acta}
  \bibinfo{volume}{104}, \bibinfo{pages}{232--260}.
\newblock \DOIprefix\doi{10.1016/j.gca.2012.11.009}.
%Type = Article
\bibitem[{Day et~al.(2016)Day, Brandon and Walker}]{day2016highly}
\bibinfo{author}{Day, J.M.D.}, \bibinfo{author}{Brandon, A.D.},
  \bibinfo{author}{Walker, R.J.}, \bibinfo{year}{2016}.
\newblock \bibinfo{title}{{Highly siderophile elements in Earth, Mars, the
  Moon, and asteroids}}.
\newblock \bibinfo{journal}{Reviews in Mineralogy and Geochemistry}
  \bibinfo{volume}{81}, \bibinfo{pages}{161--238}.
\newblock \DOIprefix\doi{10.2138/rmg.2016.81.04}.
%Type = Article
\bibitem[{Day et~al.(2018)Day, Tait, Udry, Moynier, Liu and
  Neal}]{day2018martian}
\bibinfo{author}{Day, J.M.D.}, \bibinfo{author}{Tait, K.T.},
  \bibinfo{author}{Udry, A.}, \bibinfo{author}{Moynier, F.},
  \bibinfo{author}{Liu, Y.}, \bibinfo{author}{Neal, C.R.},
  \bibinfo{year}{2018}.
\newblock \bibinfo{title}{{Martian magmatism from plume metasomatized mantle}}.
\newblock \bibinfo{journal}{Nature Communications} \bibinfo{volume}{9},
  \bibinfo{pages}{4799}.
\newblock \DOIprefix\doi{10.1038/s41467-018-07191-0}.
%Type = Article
\bibitem[{Desch et~al.(2018)Desch, Kalyaan and Alexander}]{desch2017effect}
\bibinfo{author}{Desch, S.J.}, \bibinfo{author}{Kalyaan, A.},
  \bibinfo{author}{Alexander, C.M.O'D.}, \bibinfo{year}{2018}.
\newblock \bibinfo{title}{{The effect of Jupiter's formation on the
  distribution of refractory elements and inclusions in meteorites}}.
\newblock \bibinfo{journal}{The Astrophysical Journal Supplement Series}
  \bibinfo{volume}{238}, \bibinfo{pages}{11}.
\newblock \DOIprefix\doi{10.3847/1538-4365/aad95f}.
%Type = Article
\bibitem[{Ding et~al.(2015)Ding, Dasgupta, Lee and Wadhwa}]{ding2015new}
\bibinfo{author}{Ding, S.}, \bibinfo{author}{Dasgupta, R.},
  \bibinfo{author}{Lee, C.T.A.}, \bibinfo{author}{Wadhwa, M.},
  \bibinfo{year}{2015}.
\newblock \bibinfo{title}{{New bulk sulfur measurements of Martian meteorites
  and modeling the fate of sulfur during melting and
  crystallization--Implications for sulfur transfer from Martian mantle to
  crust--atmosphere system}}.
\newblock \bibinfo{journal}{Earth and Planetary Science Letters}
  \bibinfo{volume}{409}, \bibinfo{pages}{157--167}.
\newblock \DOIprefix\doi{10.1016/j.epsl.2014.10.046}.
%Type = Inproceedings
\bibitem[{Draper et~al.(2005)Draper, Borg and Agee}]{draper2005crystallization}
\bibinfo{author}{Draper, D.S.}, \bibinfo{author}{Borg, L.E.},
  \bibinfo{author}{Agee, C.B.}, \bibinfo{year}{2005}.
\newblock \bibinfo{title}{{Crystallization of a martian magma ocean and the
  formation of Shergottite source regions: A less Fe-rich Mars?}}, in:
  \bibinfo{booktitle}{Lunar and Planetary Science Conference}, p.
  \bibinfo{pages}{1429}.
%Type = Inproceedings
\bibitem[{Dreibus and W{\"a}nke(1984)}]{dreibus1984accretion}
\bibinfo{author}{Dreibus, G.}, \bibinfo{author}{W{\"a}nke, H.},
  \bibinfo{year}{1984}.
\newblock \bibinfo{title}{{Accretion of the Earth and the inner planets}}, in:
  \bibinfo{booktitle}{Proceedings of the 27th International Geological
  Congress}, \bibinfo{organization}{VNU Science Press}. pp.
  \bibinfo{pages}{1--20}.
%Type = Article
\bibitem[{Dreibus and W{\"a}nke(1985)}]{dreibus1985mars}
\bibinfo{author}{Dreibus, G.}, \bibinfo{author}{W{\"a}nke, H.},
  \bibinfo{year}{1985}.
\newblock \bibinfo{title}{{Mars, a volatile-rich planet}}.
\newblock \bibinfo{journal}{Meteoritics} \bibinfo{volume}{20},
  \bibinfo{pages}{367--381}.
%Type = Article
\bibitem[{Dreibus and W{\"a}nke(1987)}]{dreibus1987volatiles}
\bibinfo{author}{Dreibus, G.}, \bibinfo{author}{W{\"a}nke, H.},
  \bibinfo{year}{1987}.
\newblock \bibinfo{title}{{Volatiles on Earth and Mars: A comparison}}.
\newblock \bibinfo{journal}{Icarus} \bibinfo{volume}{71},
  \bibinfo{pages}{225--240}.
\newblock \DOIprefix\doi{10.1016/0019-1035(87)90148-5}.
%Type = Article
\bibitem[{Ehmann and Lovering(1967)}]{ehmann1967abundance}
\bibinfo{author}{Ehmann, W.D.}, \bibinfo{author}{Lovering, J.F.},
  \bibinfo{year}{1967}.
\newblock \bibinfo{title}{{The abundance of mercury in meteorites and rocks by
  neutron activation analysis}}.
\newblock \bibinfo{journal}{Geochimica et Cosmochimica Acta}
  \bibinfo{volume}{31}, \bibinfo{pages}{357--376}.
\newblock \DOIprefix\doi{10.1016/0016-7037(67)90047-6}.
%Type = Article
\bibitem[{Filiberto(2017)}]{filiberto2017geochemistry}
\bibinfo{author}{Filiberto, J.}, \bibinfo{year}{2017}.
\newblock \bibinfo{title}{{Geochemistry of Martian basalts with constraints on
  magma genesis}}.
\newblock \bibinfo{journal}{Chemical Geology} \bibinfo{volume}{466},
  \bibinfo{pages}{1--14}.
\newblock \DOIprefix\doi{10.1016/j.chemgeo.2017.06.009}.
%Type = Article
\bibitem[{Filiberto and Dasgupta(2011)}]{filiberto2011fe2+}
\bibinfo{author}{Filiberto, J.}, \bibinfo{author}{Dasgupta, R.},
  \bibinfo{year}{2011}.
\newblock \bibinfo{title}{{\ce{Fe^{2+}}--Mg partitioning between olivine and
  basaltic melts: Applications to genesis of olivine-phyric shergottites and
  conditions of melting in the Martian interior}}.
\newblock \bibinfo{journal}{Earth and Planetary Science Letters}
  \bibinfo{volume}{304}, \bibinfo{pages}{527--537}.
\newblock \DOIprefix\doi{10.1016/j.epsl.2011.02.029}.
%Type = Article
\bibitem[{Filiberto et~al.(2016)Filiberto, Gross and
  McCubbin}]{filiberto2016constraints}
\bibinfo{author}{Filiberto, J.}, \bibinfo{author}{Gross, J.},
  \bibinfo{author}{McCubbin, F.M.}, \bibinfo{year}{2016}.
\newblock \bibinfo{title}{{Constraints on the water, chlorine, and fluorine
  content of the Martian mantle}}.
\newblock \bibinfo{journal}{Meteoritics \& Planetary Science}
  \bibinfo{volume}{51}, \bibinfo{pages}{2023--2035}.
\newblock \DOIprefix\doi{10.1111/maps.12624}.
%Type = Incollection
\bibitem[{Filiberto et~al.(2019)Filiberto, McCubbin and
  Taylor}]{filiberto2019volatiles}
\bibinfo{author}{Filiberto, J.}, \bibinfo{author}{McCubbin, F.M.},
  \bibinfo{author}{Taylor, G.J.}, \bibinfo{year}{2019}.
\newblock \bibinfo{title}{{Volatiles in Martian Magmas and the Interior: Inputs
  of Volatiles Into the Crust and Atmosphere}}, in: \bibinfo{editor}{Filiberto,
  J.}, \bibinfo{editor}{Schwenzer, S.P.} (Eds.), \bibinfo{booktitle}{Volatiles
  in the Martian Crust}. \bibinfo{publisher}{Elsevier}, pp.
  \bibinfo{pages}{13--33}.
\newblock \DOIprefix\doi{10.1016/B978-0-12-804191-8.00002-7}.
%Type = Article
\bibitem[{Filiberto and Treiman(2009)}]{filiberto2009martian}
\bibinfo{author}{Filiberto, J.}, \bibinfo{author}{Treiman, A.H.},
  \bibinfo{year}{2009}.
\newblock \bibinfo{title}{{Martian magmas contained abundant chlorine, but
  little water}}.
\newblock \bibinfo{journal}{Geology} \bibinfo{volume}{37},
  \bibinfo{pages}{1087--1090}.
\newblock \DOIprefix\doi{10.1130/G30488A.1}.
%Type = Article
\bibitem[{Foley et~al.(2005)Foley, Wadhwa, Borg, Janney, Hines and
  Grove}]{foley2005early}
\bibinfo{author}{Foley, C.N.}, \bibinfo{author}{Wadhwa, M.},
  \bibinfo{author}{Borg, L.}, \bibinfo{author}{Janney, P.},
  \bibinfo{author}{Hines, R.}, \bibinfo{author}{Grove, T.},
  \bibinfo{year}{2005}.
\newblock \bibinfo{title}{{The early differentiation history of Mars from
  \ce{^{182}W}-\ce{^{142}Nd} isotope systematics in the SNC meteorites}}.
\newblock \bibinfo{journal}{Geochimica et Cosmochimica Acta}
  \bibinfo{volume}{69}, \bibinfo{pages}{4557--4571}.
\newblock \DOIprefix\doi{10.1016/j.gca.2005.05.009}.
%Type = Incollection
\bibitem[{Franz et~al.(2019)Franz, King and Gaillard}]{franz2019sulfur}
\bibinfo{author}{Franz, H.B.}, \bibinfo{author}{King, P.L.},
  \bibinfo{author}{Gaillard, F.}, \bibinfo{year}{2019}.
\newblock \bibinfo{title}{{Sulfur on Mars from the Atmosphere to the Core}},
  in: \bibinfo{editor}{Filiberto, J.}, \bibinfo{editor}{Schwenzer, S.P.}
  (Eds.), \bibinfo{booktitle}{Volatiles in the Martian Crust}.
  \bibinfo{publisher}{Elsevier}, pp. \bibinfo{pages}{119--183}.
\newblock \DOIprefix\doi{10.1016/B978-0-12-804191-8.00006-4}.
%Type = Article
\bibitem[{Gaffney et~al.(2007)Gaffney, Borg and Connelly}]{gaffney2007uranium}
\bibinfo{author}{Gaffney, A.M.}, \bibinfo{author}{Borg, L.E.},
  \bibinfo{author}{Connelly, J.N.}, \bibinfo{year}{2007}.
\newblock \bibinfo{title}{{Uranium--lead isotope systematics of Mars inferred
  from the basaltic shergottite QUE 94201}}.
\newblock \bibinfo{journal}{Geochimica et Cosmochimica Acta}
  \bibinfo{volume}{71}, \bibinfo{pages}{5016--5031}.
\newblock \DOIprefix\doi{10.1016/j.gca.2007.08.009}.
%Type = Article
\bibitem[{Gaillard et~al.(2013)Gaillard, Michalski, Berger, McLennan and
  Scaillet}]{gaillard2013geochemical}
\bibinfo{author}{Gaillard, F.}, \bibinfo{author}{Michalski, J.},
  \bibinfo{author}{Berger, G.}, \bibinfo{author}{McLennan, S.M.},
  \bibinfo{author}{Scaillet, B.}, \bibinfo{year}{2013}.
\newblock \bibinfo{title}{{Geochemical reservoirs and timing of sulfur cycling
  on Mars}}.
\newblock \bibinfo{journal}{Space Science Reviews} \bibinfo{volume}{174},
  \bibinfo{pages}{251--300}.
\newblock \DOIprefix\doi{10.1007/s11214-012-9947-4}.
%Type = Article
\bibitem[{Grady et~al.(2004)Grady, Verchovsky and Wright}]{grady2004magmatic}
\bibinfo{author}{Grady, M.M.}, \bibinfo{author}{Verchovsky, A.B.},
  \bibinfo{author}{Wright, I.}, \bibinfo{year}{2004}.
\newblock \bibinfo{title}{{Magmatic carbon in Martian meteorites: attempts to
  constrain the carbon cycle on Mars}}.
\newblock \bibinfo{journal}{International Journal of Astrobiology}
  \bibinfo{volume}{3}, \bibinfo{pages}{117--124}.
\newblock \DOIprefix\doi{10.1017/S1473550404002071}.
%Type = Article
\bibitem[{Grevesse et~al.(2015)Grevesse, Scott, Asplund and
  Sauval}]{grevesse2015elemental}
\bibinfo{author}{Grevesse, N.}, \bibinfo{author}{Scott, P.},
  \bibinfo{author}{Asplund, M.}, \bibinfo{author}{Sauval, A.J.},
  \bibinfo{year}{2015}.
\newblock \bibinfo{title}{{The elemental composition of the Sun-III. The heavy
  elements Cu to Th}}.
\newblock \bibinfo{journal}{Astronomy \& Astrophysics} \bibinfo{volume}{573},
  \bibinfo{pages}{A27}.
\newblock \DOIprefix\doi{10.1051/0004-6361/201424111}.
%Type = Article
\bibitem[{Grott et~al.(2013)Grott, Baratoux, Hauber, Sautter, Mustard,
  Gasnault, Ruff, Karato, Debaille, Knapmeyer, Sohl, {Van Hoolst}, Breuer,
  Morschhauser and Toplis}]{grott2013long}
\bibinfo{author}{Grott, M.}, \bibinfo{author}{Baratoux, D.},
  \bibinfo{author}{Hauber, E.}, \bibinfo{author}{Sautter, V.},
  \bibinfo{author}{Mustard, J.}, \bibinfo{author}{Gasnault, O.},
  \bibinfo{author}{Ruff, S.W.}, \bibinfo{author}{Karato, S.I.},
  \bibinfo{author}{Debaille, V.}, \bibinfo{author}{Knapmeyer, M.},
  \bibinfo{author}{Sohl, F.}, \bibinfo{author}{{Van Hoolst}, T.},
  \bibinfo{author}{Breuer, D.}, \bibinfo{author}{Morschhauser, A.},
  \bibinfo{author}{Toplis, M.J.}, \bibinfo{year}{2013}.
\newblock \bibinfo{title}{{Long-term evolution of the Martian crust-mantle
  system}}.
\newblock \bibinfo{journal}{Space Science Reviews} \bibinfo{volume}{174},
  \bibinfo{pages}{49--111}.
\newblock \DOIprefix\doi{10.1007/s11214-012-9948-3}.
%Type = Article
\bibitem[{Halliday et~al.(2001)Halliday, W{\"a}nke, Birck and
  Clayton}]{halliday2001accretion}
\bibinfo{author}{Halliday, A.N.}, \bibinfo{author}{W{\"a}nke, H.},
  \bibinfo{author}{Birck, J.L.}, \bibinfo{author}{Clayton, R.N.},
  \bibinfo{year}{2001}.
\newblock \bibinfo{title}{{The accretion, composition and early differentiation
  of Mars}}.
\newblock \bibinfo{journal}{Space Science Reviews} \bibinfo{volume}{96},
  \bibinfo{pages}{197--230}.
\newblock \DOIprefix\doi{10.1023/A:1011997206080}.
%Type = Article
\bibitem[{Haxton et~al.(2013)Haxton, Hamish~Robertson and Serenelli}]{Haxton13}
\bibinfo{author}{Haxton, W.C.}, \bibinfo{author}{Hamish~Robertson, R.G.},
  \bibinfo{author}{Serenelli, A.M.}, \bibinfo{year}{2013}.
\newblock \bibinfo{title}{{Solar neutrinos: status and prospects}}.
\newblock \bibinfo{journal}{Annual Review of Astronomy and Astrophysics}
  \bibinfo{volume}{51}, \bibinfo{pages}{21--61}.
\newblock \DOIprefix\doi{10.1146/annurev-astro-081811-125539}.
%Type = Article
\bibitem[{Herd et~al.(2002)Herd, Borg, Jones and Papike}]{herd2002oxygen}
\bibinfo{author}{Herd, C.D.K.}, \bibinfo{author}{Borg, L.E.},
  \bibinfo{author}{Jones, J.H.}, \bibinfo{author}{Papike, J.J.},
  \bibinfo{year}{2002}.
\newblock \bibinfo{title}{{Oxygen fugacity and geochemical variations in the
  martian basalts: Implications for martian basalt petrogenesis and the
  oxidation state of the upper mantle of Mars}}.
\newblock \bibinfo{journal}{Geochimica et Cosmochimica Acta}
  \bibinfo{volume}{66}, \bibinfo{pages}{2025--2036}.
\newblock \DOIprefix\doi{10.1016/S0016-7037(02)00828-1}.
%Type = Article
\bibitem[{Herzberg and Zhang(1996)}]{herzberg1996melting}
\bibinfo{author}{Herzberg, C.}, \bibinfo{author}{Zhang, J.},
  \bibinfo{year}{1996}.
\newblock \bibinfo{title}{{Melting experiments on anhydrous peridotite KLB-1:
  Compositions of magmas in the upper mantle and transition zone}}.
\newblock \bibinfo{journal}{Journal of Geophysical Research: Solid Earth}
  \bibinfo{volume}{101}, \bibinfo{pages}{8271--8295}.
\newblock \DOIprefix\doi{10.1029/96JB00170}.
%Type = Article
\bibitem[{Hui et~al.(2011)Hui, Peslier, Lapen, Shafer, Brandon and
  Irving}]{hui2011petrogenesis}
\bibinfo{author}{Hui, H.}, \bibinfo{author}{Peslier, A.H.},
  \bibinfo{author}{Lapen, T.J.}, \bibinfo{author}{Shafer, J.T.},
  \bibinfo{author}{Brandon, A.D.}, \bibinfo{author}{Irving, A.J.},
  \bibinfo{year}{2011}.
\newblock \bibinfo{title}{{Petrogenesis of basaltic shergottite Northwest
  Africa 5298: Closed-system crystallization of an oxidized mafic melt}}.
\newblock \bibinfo{journal}{Meteoritics \& Planetary Science}
  \bibinfo{volume}{46}, \bibinfo{pages}{1313--1328}.
\newblock \DOIprefix\doi{10.1111/j.1945-5100.2011.01231.x}.
%Type = Article
\bibitem[{Hyodo et~al.(2017)Hyodo, Genda, Charnoz and
  Rosenblatt}]{hyodo2017impact}
\bibinfo{author}{Hyodo, R.}, \bibinfo{author}{Genda, H.},
  \bibinfo{author}{Charnoz, S.}, \bibinfo{author}{Rosenblatt, P.},
  \bibinfo{year}{2017}.
\newblock \bibinfo{title}{{On the impact origin of Phobos and Deimos. I.
  Thermodynamic and physical aspects}}.
\newblock \bibinfo{journal}{The Astrophysical Journal} \bibinfo{volume}{845},
  \bibinfo{pages}{125}.
\newblock \DOIprefix\doi{10.3847/1538-4357/aa81c4}.
%Type = Article
\bibitem[{Johansen et~al.(2015)Johansen, Mac~Low, Lacerda and
  Bizzarro}]{johansen2015growth}
\bibinfo{author}{Johansen, A.}, \bibinfo{author}{Mac~Low, M.M.},
  \bibinfo{author}{Lacerda, P.}, \bibinfo{author}{Bizzarro, M.},
  \bibinfo{year}{2015}.
\newblock \bibinfo{title}{{Growth of asteroids, planetary embryos, and Kuiper
  belt objects by chondrule accretion}}.
\newblock \bibinfo{journal}{Science Advances} \bibinfo{volume}{1},
  \bibinfo{pages}{e1500109}.
\newblock \DOIprefix\doi{10.1126/sciadv.1500109}.
%Type = Article
\bibitem[{Jones(2003)}]{jones2003constraints}
\bibinfo{author}{Jones, J.H.}, \bibinfo{year}{2003}.
\newblock \bibinfo{title}{{Constraints on the structure of the martian interior
  determined from the chemical and isotopic systematics of SNC meteorites}}.
\newblock \bibinfo{journal}{Meteoritics \& Planetary Science}
  \bibinfo{volume}{38}, \bibinfo{pages}{1807--1814}.
\newblock \DOIprefix\doi{10.1111/j.1945-5100.2003.tb00016.x}.
%Type = Article
\bibitem[{Kaiura and Toguri(1979)}]{kaiura1979densities}
\bibinfo{author}{Kaiura, G.H.}, \bibinfo{author}{Toguri, J.M.},
  \bibinfo{year}{1979}.
\newblock \bibinfo{title}{{Densities of the molten FeS, FeS--\ce{Cu_2S} and
  Fe--S--O systems--utilizing a bottom-balance Archimedean technique}}.
\newblock \bibinfo{journal}{Canadian Metallurgical Quarterly}
  \bibinfo{volume}{18}, \bibinfo{pages}{155--164}.
\newblock \DOIprefix\doi{10.1179/cmq.1979.18.2.155}.
%Type = Article
\bibitem[{Katsura et~al.(2010)Katsura, Yoneda, Yamazaki, Yoshino and
  Ito}]{katsura2010adiabatic}
\bibinfo{author}{Katsura, T.}, \bibinfo{author}{Yoneda, A.},
  \bibinfo{author}{Yamazaki, D.}, \bibinfo{author}{Yoshino, T.},
  \bibinfo{author}{Ito, E.}, \bibinfo{year}{2010}.
\newblock \bibinfo{title}{{Adiabatic temperature profile in the mantle}}.
\newblock \bibinfo{journal}{Physics of the Earth and Planetary Interiors}
  \bibinfo{volume}{183}, \bibinfo{pages}{212--218}.
\newblock \DOIprefix\doi{10.1016/j.pepi.2010.07.001}.
%Type = Article
\bibitem[{Keller et~al.(2006)Keller, Boynton, Karunatillake, Baker, Dohm,
  Evans, Finch, Hahn, Hamara, Janes, Kerry, Newsom, Reedy, Sprague, Squyres,
  Starr, Taylor and Williams}]{keller2006equatorial}
\bibinfo{author}{Keller, J.M.}, \bibinfo{author}{Boynton, W.V.},
  \bibinfo{author}{Karunatillake, S.}, \bibinfo{author}{Baker, V.R.},
  \bibinfo{author}{Dohm, J.M.}, \bibinfo{author}{Evans, L.G.},
  \bibinfo{author}{Finch, M.J.}, \bibinfo{author}{Hahn, B.C.},
  \bibinfo{author}{Hamara, D.K.}, \bibinfo{author}{Janes, D.M.},
  \bibinfo{author}{Kerry, K.E.}, \bibinfo{author}{Newsom, H.E.},
  \bibinfo{author}{Reedy, R.C.}, \bibinfo{author}{Sprague, A.L.},
  \bibinfo{author}{Squyres, S.W.}, \bibinfo{author}{Starr, R.D.},
  \bibinfo{author}{Taylor, G.J.}, \bibinfo{author}{Williams, R.M.},
  \bibinfo{year}{2006}.
\newblock \bibinfo{title}{{Equatorial and midlatitude distribution of chlorine
  measured by Mars Odyssey GRS}}.
\newblock \bibinfo{journal}{Journal of Geophysical Research: Planets}
  \bibinfo{volume}{111}, \bibinfo{pages}{E03S08}.
\newblock \DOIprefix\doi{10.1029/2006JE002679}.
%Type = Article
\bibitem[{Khan and Connolly(2008)}]{khan2008constraining}
\bibinfo{author}{Khan, A.}, \bibinfo{author}{Connolly, J.A.D.},
  \bibinfo{year}{2008}.
\newblock \bibinfo{title}{{Constraining the composition and thermal state of
  Mars from inversion of geophysical data}}.
\newblock \bibinfo{journal}{Journal of Geophysical Research: Planets}
  \bibinfo{volume}{113}, \bibinfo{pages}{E07003}.
\newblock \DOIprefix\doi{10.1029/2007JE002996}.
%Type = Article
\bibitem[{Khan et~al.(2018)Khan, Liebske, Rozel, Rivoldini, Nimmo, Connolly,
  Plesa and Giardini}]{khan2018geophysical}
\bibinfo{author}{Khan, A.}, \bibinfo{author}{Liebske, C.},
  \bibinfo{author}{Rozel, A.}, \bibinfo{author}{Rivoldini, A.},
  \bibinfo{author}{Nimmo, F.}, \bibinfo{author}{Connolly, J.A.D.},
  \bibinfo{author}{Plesa, A.C.}, \bibinfo{author}{Giardini, D.},
  \bibinfo{year}{2018}.
\newblock \bibinfo{title}{{A geophysical perspective on the bulk composition of
  Mars}}.
\newblock \bibinfo{journal}{Journal of Geophysical Research: Planets}
  \bibinfo{volume}{123}, \bibinfo{pages}{575--611}.
\newblock \DOIprefix\doi{10.1002/2017JE005371}.
%Type = Article
\bibitem[{Komabayashi(2014)}]{komabayashi2014thermodynamics}
\bibinfo{author}{Komabayashi, T.}, \bibinfo{year}{2014}.
\newblock \bibinfo{title}{{Thermodynamics of melting relations in the system
  Fe-FeO at high pressure: Implications for oxygen in the Earth's core}}.
\newblock \bibinfo{journal}{Journal of Geophysical Research: Solid Earth}
  \bibinfo{volume}{119}, \bibinfo{pages}{4164--4177}.
\newblock \DOIprefix\doi{10.1002/2014JB010980}.
%Type = Article
\bibitem[{Konopliv et~al.(2016)Konopliv, Park and
  Folkner}]{konopliv2016improved}
\bibinfo{author}{Konopliv, A.S.}, \bibinfo{author}{Park, R.S.},
  \bibinfo{author}{Folkner, W.M.}, \bibinfo{year}{2016}.
\newblock \bibinfo{title}{{An improved JPL Mars gravity field and orientation
  from Mars orbiter and lander tracking data}}.
\newblock \bibinfo{journal}{Icarus} \bibinfo{volume}{274},
  \bibinfo{pages}{253--260}.
\newblock \DOIprefix\doi{10.1016/j.icarus.2016.02.052}.
%Type = Article
\bibitem[{Kruijer et~al.(2017a)Kruijer, Burkhardt, Budde and
  Kleine}]{kruijer2017age}
\bibinfo{author}{Kruijer, T.S.}, \bibinfo{author}{Burkhardt, C.},
  \bibinfo{author}{Budde, G.}, \bibinfo{author}{Kleine, T.},
  \bibinfo{year}{2017}a.
\newblock \bibinfo{title}{{Age of Jupiter inferred from the distinct genetics
  and formation times of meteorites}}.
\newblock \bibinfo{journal}{Proceedings of the National Academy of Sciences}
  \bibinfo{volume}{114}, \bibinfo{pages}{6712--6716}.
\newblock \DOIprefix\doi{10.1073/pnas.1704461114}.
%Type = Article
\bibitem[{Kruijer et~al.(2017b)Kruijer, Kleine, Borg, Brennecka, Irving,
  Bischoff and Agee}]{kruijer2017early}
\bibinfo{author}{Kruijer, T.S.}, \bibinfo{author}{Kleine, T.},
  \bibinfo{author}{Borg, L.E.}, \bibinfo{author}{Brennecka, G.A.},
  \bibinfo{author}{Irving, A.J.}, \bibinfo{author}{Bischoff, A.},
  \bibinfo{author}{Agee, C.B.}, \bibinfo{year}{2017}b.
\newblock \bibinfo{title}{{The early differentiation of Mars inferred from
  Hf--W chronometry}}.
\newblock \bibinfo{journal}{Earth and Planetary Science Letters}
  \bibinfo{volume}{474}, \bibinfo{pages}{345--354}.
\newblock \DOIprefix\doi{10.1016/j.epsl.2017.06.047}.
%Type = Inproceedings
\bibitem[{Kuramoto et~al.(2018)Kuramoto, Kawakatsu, Fujimoto, Bibring, Genda,
  Imamura, Kameda, Lawrence, Matsumoto, Miyamoto, Morota, Nagaoka, Nakamura,
  Ogawa, Otake, Ozaki, Sasaki, Senshu, Tachibana, Terada, Usui, J, Watanabe and
  {MMX study team}}]{kuramoto2018martian}
\bibinfo{author}{Kuramoto, K.}, \bibinfo{author}{Kawakatsu, Y.},
  \bibinfo{author}{Fujimoto, M.}, \bibinfo{author}{Bibring, J.P.},
  \bibinfo{author}{Genda, H.}, \bibinfo{author}{Imamura, T.},
  \bibinfo{author}{Kameda, S.}, \bibinfo{author}{Lawrence, D.},
  \bibinfo{author}{Matsumoto, K.}, \bibinfo{author}{Miyamoto, H.},
  \bibinfo{author}{Morota, T.}, \bibinfo{author}{Nagaoka, H.},
  \bibinfo{author}{Nakamura, T.}, \bibinfo{author}{Ogawa, K.},
  \bibinfo{author}{Otake, H.}, \bibinfo{author}{Ozaki, M.},
  \bibinfo{author}{Sasaki, S.}, \bibinfo{author}{Senshu, H.},
  \bibinfo{author}{Tachibana, S.}, \bibinfo{author}{Terada, N.},
  \bibinfo{author}{Usui, T.}, \bibinfo{author}{J, W.},
  \bibinfo{author}{Watanabe, S.}, \bibinfo{author}{{MMX study team}},
  \bibinfo{year}{2018}.
\newblock \bibinfo{title}{{Martian Moons eXploration (MMX): an overview of its
  science}}, in: \bibinfo{booktitle}{European Planetary Science Congress}, pp.
  \bibinfo{pages}{EPSC2018--1036}.
%Type = Incollection
\bibitem[{Larimer and Wasson(1988)}]{larimer1988refractory}
\bibinfo{author}{Larimer, J.W.}, \bibinfo{author}{Wasson, J.T.},
  \bibinfo{year}{1988}.
\newblock \bibinfo{title}{{Refractory lithophile elements}}, in:
  \bibinfo{editor}{Kerridge, J.F.}, \bibinfo{editor}{Matthews, M.S.} (Eds.),
  \bibinfo{booktitle}{Meteorites and the Early Solar System}.
  \bibinfo{publisher}{The University of Arizona Press},
  \bibinfo{address}{Tucson}, pp. \bibinfo{pages}{394--415}.
%Type = Article
\bibitem[{Lauretta et~al.(1999)Lauretta, Devouard and
  Buseck}]{lauretta1999cosmochemical}
\bibinfo{author}{Lauretta, D.S.}, \bibinfo{author}{Devouard, B.},
  \bibinfo{author}{Buseck, P.R.}, \bibinfo{year}{1999}.
\newblock \bibinfo{title}{{The cosmochemical behavior of mercury}}.
\newblock \bibinfo{journal}{Earth and Planetary Science Letters}
  \bibinfo{volume}{171}, \bibinfo{pages}{35--47}.
\newblock \DOIprefix\doi{10.1016/S0012-821X(99)00129-6}.
%Type = Article
\bibitem[{Le~Roux et~al.(2011)Le~Roux, Dasgupta and Lee}]{le2011mineralogical}
\bibinfo{author}{Le~Roux, V.}, \bibinfo{author}{Dasgupta, R.},
  \bibinfo{author}{Lee, C.T.A.}, \bibinfo{year}{2011}.
\newblock \bibinfo{title}{{Mineralogical heterogeneities in the Earth's mantle:
  constraints from Mn, Co, Ni and Zn partitioning during partial melting}}.
\newblock \bibinfo{journal}{Earth and Planetary Science Letters}
  \bibinfo{volume}{307}, \bibinfo{pages}{395--408}.
\newblock \DOIprefix\doi{10.1016/j.epsl.2011.05.014}.
%Type = Article
\bibitem[{Lee and Gaskell(1974)}]{lee1974densities}
\bibinfo{author}{Lee, Y.E.}, \bibinfo{author}{Gaskell, D.R.},
  \bibinfo{year}{1974}.
\newblock \bibinfo{title}{{The densities and structures of melts in the system
  CaO-“FeO”-\ce{SiO2}}}.
\newblock \bibinfo{journal}{Metallurgical Transactions} \bibinfo{volume}{5},
  \bibinfo{pages}{853--860}.
\newblock \DOIprefix\doi{10.1007/BF02643138}.
%Type = Article
\bibitem[{Levison et~al.(2015)Levison, Kretke, Walsh and
  Bottke}]{levison2015growing}
\bibinfo{author}{Levison, H.F.}, \bibinfo{author}{Kretke, K.A.},
  \bibinfo{author}{Walsh, K.J.}, \bibinfo{author}{Bottke, W.F.},
  \bibinfo{year}{2015}.
\newblock \bibinfo{title}{{Growing the terrestrial planets from the gradual
  accumulation of submeter-sized objects}}.
\newblock \bibinfo{journal}{Proceedings of the National Academy of Sciences}
  \bibinfo{volume}{112}, \bibinfo{pages}{14180--14185}.
\newblock \DOIprefix\doi{10.1073/pnas.1513364112}.
%Type = Article
\bibitem[{Lodders(2003)}]{lodders2003solar}
\bibinfo{author}{Lodders, K.}, \bibinfo{year}{2003}.
\newblock \bibinfo{title}{{Solar system abundances and condensation
  temperatures of the elements}}.
\newblock \bibinfo{journal}{Astrophysical Journal} \bibinfo{volume}{591},
  \bibinfo{pages}{1220--1247}.
\newblock \DOIprefix\doi{10.1086/375492}.
%Type = Article
\bibitem[{Lodders and Fegley(1997)}]{lodders1997oxygen}
\bibinfo{author}{Lodders, K.}, \bibinfo{author}{Fegley, Jr., B.},
  \bibinfo{year}{1997}.
\newblock \bibinfo{title}{{An oxygen isotope model for the composition of
  Mars}}.
\newblock \bibinfo{journal}{Icarus} \bibinfo{volume}{126},
  \bibinfo{pages}{373--394}.
\newblock \DOIprefix\doi{10.1006/icar.1996.5653}.
%Type = Incollection
\bibitem[{Longhi et~al.(1992)Longhi, Knittle, Holloway and
  W{\"a}nke}]{longhi1992bulk}
\bibinfo{author}{Longhi, J.}, \bibinfo{author}{Knittle, E.},
  \bibinfo{author}{Holloway, J.R.}, \bibinfo{author}{W{\"a}nke, H.},
  \bibinfo{year}{1992}.
\newblock \bibinfo{title}{{The bulk composition, mineralogy and internal
  structure of Mars}}, in: \bibinfo{editor}{Kieffer, H.H.},
  \bibinfo{editor}{Jakosky, B.M.}, \bibinfo{editor}{Snyder, C.W.},
  \bibinfo{editor}{Matthews, M.S.} (Eds.), \bibinfo{booktitle}{Mars}.
  \bibinfo{publisher}{University of Arizona Press}, \bibinfo{address}{Tuscon},
  pp. \bibinfo{pages}{184--208}.
%Type = Article
\bibitem[{Masuda(1957)}]{masuda1957simple}
\bibinfo{author}{Masuda, A.}, \bibinfo{year}{1957}.
\newblock \bibinfo{title}{{Simple regularity in the variation of relative
  abundances of rare earth elements}}.
\newblock \bibinfo{journal}{The Journal of Earth Sciences, Nagoya University}
  \bibinfo{volume}{5}, \bibinfo{pages}{125--134}.
%Type = Article
\bibitem[{Mathew and Marti(2001)}]{mathew2001early}
\bibinfo{author}{Mathew, K.J.}, \bibinfo{author}{Marti, K.},
  \bibinfo{year}{2001}.
\newblock \bibinfo{title}{{Early evolution of Martian volatiles: Nitrogen and
  noble gas components in ALH84001 and Chassigny}}.
\newblock \bibinfo{journal}{Journal of Geophysical Research: Planets}
  \bibinfo{volume}{106}, \bibinfo{pages}{1401--1422}.
\newblock \DOIprefix\doi{10.1029/2000JE001255}.
%Type = Inproceedings
\bibitem[{McCoy et~al.(2016)McCoy, Chartrand, Caroebter, Gross and
  Filiberto}]{mccoy2016experimentally_conf}
\bibinfo{author}{McCoy, C.L.}, \bibinfo{author}{Chartrand, Z.},
  \bibinfo{author}{Caroebter, P.K.}, \bibinfo{author}{Gross, J.},
  \bibinfo{author}{Filiberto, J.}, \bibinfo{year}{2016}.
\newblock \bibinfo{title}{{Experimentally melting a Mg\#80 Martian mantle at
  0.5 to 1.5 GPa: Implications for basalt genesis}}, in:
  \bibinfo{booktitle}{The Geological Society of America Annual Meeting},
  p.~\bibinfo{pages}{7}.
\newblock \DOIprefix\doi{10.1130/abs/2016AM-282245}.
%Type = Article
\bibitem[{McCubbin et~al.(2016a)McCubbin, Boyce, Nov{\'{a}}k-Szab{\'{o}},
  Santos, Tart{\`{e}}se, Muttik, Domokos, Vazquez, Keller, Moser, Jerolmack,
  Shearer, Steele, Elardo, Rahman, Anand, Delhaye and
  Agee}]{mccubbin2016geologic}
\bibinfo{author}{McCubbin, F.M.}, \bibinfo{author}{Boyce, J.W.},
  \bibinfo{author}{Nov{\'{a}}k-Szab{\'{o}}, T.}, \bibinfo{author}{Santos,
  A.R.}, \bibinfo{author}{Tart{\`{e}}se, R.}, \bibinfo{author}{Muttik, N.},
  \bibinfo{author}{Domokos, G.}, \bibinfo{author}{Vazquez, J.},
  \bibinfo{author}{Keller, L.P.}, \bibinfo{author}{Moser, D.E.},
  \bibinfo{author}{Jerolmack, D.J.}, \bibinfo{author}{Shearer, C.K.},
  \bibinfo{author}{Steele, A.}, \bibinfo{author}{Elardo, S.M.},
  \bibinfo{author}{Rahman, Z.}, \bibinfo{author}{Anand, M.},
  \bibinfo{author}{Delhaye, T.}, \bibinfo{author}{Agee, C.B.},
  \bibinfo{year}{2016}a.
\newblock \bibinfo{title}{{Geologic history of Martian regolith breccia
  Northwest Africa 7034: Evidence for hydrothermal activity and lithologic
  diversity in the Martian crust}}.
\newblock \bibinfo{journal}{Journal of Geophysical Research: Planets}
  \bibinfo{volume}{121}, \bibinfo{pages}{2120--2149}.
\newblock \DOIprefix\doi{10.1002/2016JE005143}.
%Type = Article
\bibitem[{McCubbin et~al.(2016b)McCubbin, Boyce, Srinivasan, Santos, Elardo,
  Filiberto, Steele and Shearer}]{mccubbin2016heterogeneous}
\bibinfo{author}{McCubbin, F.M.}, \bibinfo{author}{Boyce, J.W.},
  \bibinfo{author}{Srinivasan, P.}, \bibinfo{author}{Santos, A.R.},
  \bibinfo{author}{Elardo, S.M.}, \bibinfo{author}{Filiberto, J.},
  \bibinfo{author}{Steele, A.}, \bibinfo{author}{Shearer, C.K.},
  \bibinfo{year}{2016}b.
\newblock \bibinfo{title}{{Heterogeneous distribution of \ce{H2O} in the
  Martian interior: Implications for the abundance of \ce{H2O} in depleted and
  enriched mantle sources}}.
\newblock \bibinfo{journal}{Meteoritics \& Planetary Science}
  \bibinfo{volume}{51}, \bibinfo{pages}{2036--2060}.
\newblock \DOIprefix\doi{10.1111/maps.12639}.
%Type = Article
\bibitem[{McCubbin et~al.(2013)McCubbin, Elardo, Shearer, Smirnov, Hauri and
  Draper}]{mccubbin2013petrogenetic}
\bibinfo{author}{McCubbin, F.M.}, \bibinfo{author}{Elardo, S.M.},
  \bibinfo{author}{Shearer, Jr., C.K.}, \bibinfo{author}{Smirnov, A.},
  \bibinfo{author}{Hauri, E.H.}, \bibinfo{author}{Draper, D.S.},
  \bibinfo{year}{2013}.
\newblock \bibinfo{title}{{A petrogenetic model for the comagmatic origin of
  chassignites and nakhlites: Inferences from chlorine-rich minerals,
  petrology, and geochemistry}}.
\newblock \bibinfo{journal}{Meteoritics \& Planetary Science}
  \bibinfo{volume}{48}, \bibinfo{pages}{819--853}.
\newblock \DOIprefix\doi{10.1111/maps.12095}.
%Type = Article
\bibitem[{McCubbin et~al.(2010)McCubbin, Smirnov, Nekvasil, Wang, Hauri and
  Lindsley}]{mccubbin2010hydrous}
\bibinfo{author}{McCubbin, F.M.}, \bibinfo{author}{Smirnov, A.},
  \bibinfo{author}{Nekvasil, H.}, \bibinfo{author}{Wang, J.},
  \bibinfo{author}{Hauri, E.}, \bibinfo{author}{Lindsley, D.H.},
  \bibinfo{year}{2010}.
\newblock \bibinfo{title}{{Hydrous magmatism on Mars: A source of water for the
  surface and subsurface during the Amazonian}}.
\newblock \bibinfo{journal}{Earth and Planetary Science Letters}
  \bibinfo{volume}{292}, \bibinfo{pages}{132--138}.
\newblock \DOIprefix\doi{10.1016/j.epsl.2010.01.028}.
%Type = Article
\bibitem[{McDonough(1990)}]{mcdonough1990comment}
\bibinfo{author}{McDonough, W.F.}, \bibinfo{year}{1990}.
\newblock \bibinfo{title}{{Comment on “Abundance and distribution of gallium
  in some spinel and garnet lherzolites” by D.B. McKay and R.H. Mitchell}}.
\newblock \bibinfo{journal}{Geochimica et Cosmochimica Acta}
  \bibinfo{volume}{54}, \bibinfo{pages}{471--473}.
\newblock \DOIprefix\doi{10.1016/0016-7037(90)90335-I}.
%Type = Incollection
\bibitem[{McDonough(2014)}]{mcdonough2014compositional}
\bibinfo{author}{McDonough, W.F.}, \bibinfo{year}{2014}.
\newblock \bibinfo{title}{{Compositional model for the Earth's core}}, in:
  \bibinfo{editor}{Holland, H.D.}, \bibinfo{editor}{Turekian, K.K.} (Eds.),
  \bibinfo{booktitle}{Treatise on Geochemistry (Second Edition)}.
  \bibinfo{publisher}{Elsevier}, \bibinfo{address}{Oxford}.
  volume~\bibinfo{volume}{3}, pp. \bibinfo{pages}{559--577}.
\newblock \DOIprefix\doi{10.1016/B978-0-08-095975-7.00215-1}.
%Type = Incollection
\bibitem[{McDonough(2016)}]{mcdonough2016composition}
\bibinfo{author}{McDonough, W.F.}, \bibinfo{year}{2016}.
\newblock \bibinfo{title}{{The composition of the lower mantle and core}}, in:
  \bibinfo{editor}{Terasaki, H.}, \bibinfo{editor}{Fischer, R.A.} (Eds.),
  \bibinfo{booktitle}{Deep Earth}. \bibinfo{publisher}{American Geophysical
  Union (AGU)}. chapter~\bibinfo{chapter}{12}, pp. \bibinfo{pages}{145--159}.
\newblock \DOIprefix\doi{10.1002/9781118992487.ch12}.
%Type = Article
\bibitem[{McDonough and Sun(1995)}]{mcdonough1995composition}
\bibinfo{author}{McDonough, W.F.}, \bibinfo{author}{Sun, S.s.},
  \bibinfo{year}{1995}.
\newblock \bibinfo{title}{{The composition of the Earth}}.
\newblock \bibinfo{journal}{Chemical Geology} \bibinfo{volume}{120},
  \bibinfo{pages}{223--253}.
\newblock \DOIprefix\doi{10.1016/0009-2541(94)00140-4}.
%Type = Article
\bibitem[{McGovern et~al.(2002)McGovern, Solomon, Smith, Zuber, Simons,
  Wieczorek, Phillips, Neumann, Aharonson and Head}]{mcgovern2002localized}
\bibinfo{author}{McGovern, P.J.}, \bibinfo{author}{Solomon, S.C.},
  \bibinfo{author}{Smith, D.E.}, \bibinfo{author}{Zuber, M.T.},
  \bibinfo{author}{Simons, M.}, \bibinfo{author}{Wieczorek, M.A.},
  \bibinfo{author}{Phillips, R.J.}, \bibinfo{author}{Neumann, G.A.},
  \bibinfo{author}{Aharonson, O.}, \bibinfo{author}{Head, J.W.},
  \bibinfo{year}{2002}.
\newblock \bibinfo{title}{{Localized gravity/topography admittance and
  correlation spectra on Mars: Implications for regional and global
  evolution}}.
\newblock \bibinfo{journal}{Journal of Geophysical Research: Planets}
  \bibinfo{volume}{107}, \bibinfo{pages}{19--1--19--25}.
\newblock \DOIprefix\doi{10.1029/2002JE001854}.
%Type = Article
\bibitem[{McGovern et~al.(2004)McGovern, Solomon, Smith, Zuber, Simons,
  Wieczorek, Phillips, Neumann, Aharonson and Head}]{mcgovern2004correction}
\bibinfo{author}{McGovern, P.J.}, \bibinfo{author}{Solomon, S.C.},
  \bibinfo{author}{Smith, D.E.}, \bibinfo{author}{Zuber, M.T.},
  \bibinfo{author}{Simons, M.}, \bibinfo{author}{Wieczorek, M.A.},
  \bibinfo{author}{Phillips, R.J.}, \bibinfo{author}{Neumann, G.A.},
  \bibinfo{author}{Aharonson, O.}, \bibinfo{author}{Head, J.W.},
  \bibinfo{year}{2004}.
\newblock \bibinfo{title}{{Correction to “Localized gravity/topography
  admittance and correlation spectra on Mars: Implications for regional and
  global evolution”}}.
\newblock \bibinfo{journal}{Journal of Geophysical Research: Planets}
  \bibinfo{volume}{109}, \bibinfo{pages}{E07007}.
\newblock \DOIprefix\doi{10.1029/2004JE002286}.
%Type = Article
\bibitem[{McKenzie et~al.(2002)McKenzie, Barnett and
  Yuan}]{mckenzie2002relationship}
\bibinfo{author}{McKenzie, D.}, \bibinfo{author}{Barnett, D.N.},
  \bibinfo{author}{Yuan, D.N.}, \bibinfo{year}{2002}.
\newblock \bibinfo{title}{{The relationship between Martian gravity and
  topography}}.
\newblock \bibinfo{journal}{Earth and Planetary Science Letters}
  \bibinfo{volume}{195}, \bibinfo{pages}{1--16}.
\newblock \DOIprefix\doi{10.1016/S0012-821X(01)00555-6}.
%Type = Article
\bibitem[{McLennan(2003)}]{mclennan2003large}
\bibinfo{author}{McLennan, S.M.}, \bibinfo{year}{2003}.
\newblock \bibinfo{title}{{Large-ion lithophile element fractionation during
  the early differentiation of Mars and the composition of the Martian
  primitive mantle}}.
\newblock \bibinfo{journal}{Meteoritics \& Planetary Science}
  \bibinfo{volume}{38}, \bibinfo{pages}{895--904}.
\newblock \DOIprefix\doi{10.1111/j.1945-5100.2003.tb00286.x}.
%Type = Incollection
\bibitem[{McSween(2008)}]{mcsween2008martian}
\bibinfo{author}{McSween, Jr., H.Y.}, \bibinfo{year}{2008}.
\newblock \bibinfo{title}{{Martian meteorites as crustal samples}}, in:
  \bibinfo{editor}{Bell~III, J.} (Ed.), \bibinfo{booktitle}{The Martian
  Surface: Composition, Mineralogy, and Physical Properties}.
  \bibinfo{publisher}{Cambridge University Press}, pp.
  \bibinfo{pages}{383--395}.
\newblock \DOIprefix\doi{10.1017/CBO9780511536076.018}.
%Type = Article
\bibitem[{McSween(2015)}]{mcsween2015petrology}
\bibinfo{author}{McSween, Jr., H.Y.}, \bibinfo{year}{2015}.
\newblock \bibinfo{title}{{Petrology on Mars}}.
\newblock \bibinfo{journal}{American Mineralogist} \bibinfo{volume}{100},
  \bibinfo{pages}{2380--2395}.
\newblock \DOIprefix\doi{10.2138/am-2015-5257}.
%Type = Incollection
\bibitem[{McSween and McLennan(2014)}]{mcsween2014mars}
\bibinfo{author}{McSween, Jr., H.Y.}, \bibinfo{author}{McLennan, S.M.},
  \bibinfo{year}{2014}.
\newblock \bibinfo{title}{Mars}, in: \bibinfo{editor}{Holland, H.D.},
  \bibinfo{editor}{Turekian, K.K.} (Eds.), \bibinfo{booktitle}{Treatise on
  Geochemistry (Second Edition)}. \bibinfo{publisher}{Elsevier},
  \bibinfo{address}{Oxford}. volume~\bibinfo{volume}{2}, pp.
  \bibinfo{pages}{251--300}.
\newblock \DOIprefix\doi{10.1016/B978-0-08-095975-7.00125-X}.
%Type = Article
\bibitem[{McSween et~al.(2009)McSween, Taylor and Wyatt}]{mcsween2009elemental}
\bibinfo{author}{McSween, Jr., H.Y.}, \bibinfo{author}{Taylor, G.J.},
  \bibinfo{author}{Wyatt, M.B.}, \bibinfo{year}{2009}.
\newblock \bibinfo{title}{{Elemental composition of the Martian crust}}.
\newblock \bibinfo{journal}{Science} \bibinfo{volume}{324},
  \bibinfo{pages}{736--739}.
\newblock \DOIprefix\doi{10.1126/science.1165871}.
%Type = Inproceedings
\bibitem[{Medard et~al.(2014)Medard, Martin, Collinet, Righter, Grove, Newville
  and Lanzirotti}]{medard2014fe3+}
\bibinfo{author}{Medard, E.}, \bibinfo{author}{Martin, A.M.},
  \bibinfo{author}{Collinet, M.}, \bibinfo{author}{Righter, K.},
  \bibinfo{author}{Grove, T.L.}, \bibinfo{author}{Newville, M.},
  \bibinfo{author}{Lanzirotti, A.}, \bibinfo{year}{2014}.
\newblock \bibinfo{title}{{\ce{Fe^{3+}} partitioning during basalt
  differentiation on Mars: insights into the oxygen fugacity of the shergottite
  mantle source(s)}}, in: \bibinfo{booktitle}{AGU Fall Meeting Abstracts}, pp.
  \bibinfo{pages}{V52B--03}.
%Type = Article
\bibitem[{Meier et~al.(2016)Meier, Cloquet and Marty}]{meier2016mercury}
\bibinfo{author}{Meier, M.M.}, \bibinfo{author}{Cloquet, C.},
  \bibinfo{author}{Marty, B.}, \bibinfo{year}{2016}.
\newblock \bibinfo{title}{{Mercury (Hg) in meteorites: Variations in abundance,
  thermal release profile, mass-dependent and mass-independent isotopic
  fractionation}}.
\newblock \bibinfo{journal}{Geochimica et Cosmochimica Acta}
  \bibinfo{volume}{182}, \bibinfo{pages}{55--72}.
\newblock \DOIprefix\doi{10.1016/j.gca.2016.03.007}.
%Type = Inproceedings
\bibitem[{Minitti et~al.(2006)Minitti, Fei and Bertka}]{minitti2006new}
\bibinfo{author}{Minitti, M.E.}, \bibinfo{author}{Fei, Y.},
  \bibinfo{author}{Bertka, C.M.}, \bibinfo{year}{2006}.
\newblock \bibinfo{title}{{New, geophysically-constrained martian mantle
  compositions}}, in: \bibinfo{booktitle}{Workshop on Early Planetary
  Differentiation}, pp. \bibinfo{pages}{72--73}.
%Type = Article
\bibitem[{Misawa et~al.(1997)Misawa, Nakamura, Premo and
  Tatsumoto}]{misawa1997u}
\bibinfo{author}{Misawa, K.}, \bibinfo{author}{Nakamura, N.},
  \bibinfo{author}{Premo, W.R.}, \bibinfo{author}{Tatsumoto, M.},
  \bibinfo{year}{1997}.
\newblock \bibinfo{title}{{U-Th-Pb isotopic systematics of lherzolitic
  shergottite Yamato-793605}}.
\newblock \bibinfo{journal}{Antarctic Meteorite Research} \bibinfo{volume}{10},
  \bibinfo{pages}{95--108}.
%Type = Article
\bibitem[{Morgan and Anders(1979)}]{morgan1979chemical}
\bibinfo{author}{Morgan, J.W.}, \bibinfo{author}{Anders, E.},
  \bibinfo{year}{1979}.
\newblock \bibinfo{title}{{Chemical composition of Mars}}.
\newblock \bibinfo{journal}{Geochimica et Cosmochimica Acta}
  \bibinfo{volume}{43}, \bibinfo{pages}{1601--1610}.
\newblock \DOIprefix\doi{10.1016/0016-7037(79)90180-7}.
%Type = Article
\bibitem[{Moriwaki et~al.(2017)Moriwaki, Usui, Simon, Jones, Yokoyama and
  Tobita}]{moriwaki2017coupled}
\bibinfo{author}{Moriwaki, R.}, \bibinfo{author}{Usui, T.},
  \bibinfo{author}{Simon, J.I.}, \bibinfo{author}{Jones, J.H.},
  \bibinfo{author}{Yokoyama, T.}, \bibinfo{author}{Tobita, M.},
  \bibinfo{year}{2017}.
\newblock \bibinfo{title}{{Coupled Pb isotopic and trace element systematics of
  the Tissint meteorite: Geochemical signatures of the depleted shergottite
  source mantle}}.
\newblock \bibinfo{journal}{Earth and Planetary Science Letters}
  \bibinfo{volume}{474}, \bibinfo{pages}{180--189}.
\newblock \DOIprefix\doi{10.1016/j.epsl.2017.06.044}.
%Type = Article
\bibitem[{Murchie et~al.(2014)Murchie, Britt and Pieters}]{murchie2014value}
\bibinfo{author}{Murchie, S.L.}, \bibinfo{author}{Britt, D.T.},
  \bibinfo{author}{Pieters, C.M.}, \bibinfo{year}{2014}.
\newblock \bibinfo{title}{{The value of Phobos sample return}}.
\newblock \bibinfo{journal}{Planetary and Space Science} \bibinfo{volume}{102},
  \bibinfo{pages}{176--182}.
\newblock \DOIprefix\doi{10.1016/j.pss.2014.04.014}.
%Type = Article
\bibitem[{Musselwhite et~al.(2006)Musselwhite, Dalton, Kiefer and
  Treiman}]{white2006experimental}
\bibinfo{author}{Musselwhite, D.S.}, \bibinfo{author}{Dalton, H.A.},
  \bibinfo{author}{Kiefer, W.S.}, \bibinfo{author}{Treiman, A.H.},
  \bibinfo{year}{2006}.
\newblock \bibinfo{title}{{Experimental petrology of the basaltic shergottite
  Yamato-980459: Implications for the thermal structure of the Martian
  mantle}}.
\newblock \bibinfo{journal}{Meteoritics \& Planetary Science}
  \bibinfo{volume}{41}, \bibinfo{pages}{1271--1290}.
\newblock \DOIprefix\doi{10.1111/j.1945-5100.2006.tb00521.x}.
%Type = Article
\bibitem[{Nagamori(1969)}]{nagamori1969density}
\bibinfo{author}{Nagamori, M.}, \bibinfo{year}{1969}.
\newblock \bibinfo{title}{{Density of molten Ag-S, Cu-S, Fe-S, and Ni-S
  systems}}.
\newblock \bibinfo{journal}{Transactions of the Metallurgical Society of the
  American Institute of Mechanical Engineers} \bibinfo{volume}{245},
  \bibinfo{pages}{1897--1902}.
%Type = Article
\bibitem[{Nagayama et~al.(2019)Nagayama, Bailey, Loisel, Dunham, Rochau,
  Blancard, Colgan, Coss\'e, Faussurier, Fontes, Gilleron, Hansen, Iglesias,
  Golovkin, Kilcrease, MacFarlane, Mancini, More, Orban, Pain, Sherrill and
  Wilson}]{nagayama2019systematic}
\bibinfo{author}{Nagayama, T.}, \bibinfo{author}{Bailey, J.E.},
  \bibinfo{author}{Loisel, G.P.}, \bibinfo{author}{Dunham, G.S.},
  \bibinfo{author}{Rochau, G.A.}, \bibinfo{author}{Blancard, C.},
  \bibinfo{author}{Colgan, J.}, \bibinfo{author}{Coss\'e, P.},
  \bibinfo{author}{Faussurier, G.}, \bibinfo{author}{Fontes, C.J.},
  \bibinfo{author}{Gilleron, F.}, \bibinfo{author}{Hansen, S.B.},
  \bibinfo{author}{Iglesias, C.A.}, \bibinfo{author}{Golovkin, I.E.},
  \bibinfo{author}{Kilcrease, D.P.}, \bibinfo{author}{MacFarlane, J.J.},
  \bibinfo{author}{Mancini, R.C.}, \bibinfo{author}{More, R.M.},
  \bibinfo{author}{Orban, C.}, \bibinfo{author}{Pain, J.C.},
  \bibinfo{author}{Sherrill, M.E.}, \bibinfo{author}{Wilson, B.G.},
  \bibinfo{year}{2019}.
\newblock \bibinfo{title}{{Systematic study of $L$-shell opacity at stellar
  interior temperatures}}.
\newblock \bibinfo{journal}{Physical Review Letters} \bibinfo{volume}{122},
  \bibinfo{pages}{235001}.
\newblock \DOIprefix\doi{10.1103/PhysRevLett.122.235001}.
%Type = Article
\bibitem[{Nakamura et~al.(1982)Nakamura, Unruh, Tatsumoto and
  Hutchison}]{nakamura1982origin}
\bibinfo{author}{Nakamura, N.}, \bibinfo{author}{Unruh, D.M.},
  \bibinfo{author}{Tatsumoto, M.}, \bibinfo{author}{Hutchison, R.},
  \bibinfo{year}{1982}.
\newblock \bibinfo{title}{{Origin and evolution of the Nakhla meteorite
  inferred from the Sm-Nd and U-Pb systematics and REE, Ba, Sr, Rb and K
  abundances}}.
\newblock \bibinfo{journal}{Geochimica et Cosmochimica Acta}
  \bibinfo{volume}{46}, \bibinfo{pages}{1555--1573}.
\newblock \DOIprefix\doi{10.1016/0016-7037(82)90314-3}.
%Type = Article
\bibitem[{Nishida et~al.(2016)Nishida, Suzuki, Terasaki, Shibazaki, Higo,
  Kuwabara, Shimoyama, Sakurai, Ushioda, Takahashi, Kikegawa, Wakabayashi and
  Funamori}]{nishida2016towards}
\bibinfo{author}{Nishida, K.}, \bibinfo{author}{Suzuki, A.},
  \bibinfo{author}{Terasaki, H.}, \bibinfo{author}{Shibazaki, Y.},
  \bibinfo{author}{Higo, Y.}, \bibinfo{author}{Kuwabara, S.},
  \bibinfo{author}{Shimoyama, Y.}, \bibinfo{author}{Sakurai, M.},
  \bibinfo{author}{Ushioda, M.}, \bibinfo{author}{Takahashi, E.},
  \bibinfo{author}{Kikegawa, T.}, \bibinfo{author}{Wakabayashi, D.},
  \bibinfo{author}{Funamori, N.}, \bibinfo{year}{2016}.
\newblock \bibinfo{title}{{Towards a consensus on the pressure and composition
  dependence of sound velocity in the liquid Fe--S system}}.
\newblock \bibinfo{journal}{Physics of the Earth and Planetary Interiors}
  \bibinfo{volume}{257}, \bibinfo{pages}{230--239}.
\newblock \DOIprefix\doi{10.1016/j.pepi.2016.06.009}.
%Type = Article
\bibitem[{Noll et~al.(1996)Noll, Newsom, Leeman and Ryan}]{noll1996role}
\bibinfo{author}{Noll, Jr., P.D.}, \bibinfo{author}{Newsom, H.E.},
  \bibinfo{author}{Leeman, W.P.}, \bibinfo{author}{Ryan, J.G.},
  \bibinfo{year}{1996}.
\newblock \bibinfo{title}{{The role of hydrothermal fluids in the production of
  subduction zone magmas: evidence from siderophile and chalcophile trace
  elements and boron}}.
\newblock \bibinfo{journal}{Geochimica et Cosmochimica Acta}
  \bibinfo{volume}{60}, \bibinfo{pages}{587--611}.
\newblock \DOIprefix\doi{10.1016/0016-7037(95)00405-X}.
%Type = Article
\bibitem[{Norman(1999)}]{norman1999composition}
\bibinfo{author}{Norman, M.D.}, \bibinfo{year}{1999}.
\newblock \bibinfo{title}{{The composition and thickness of the crust of Mars
  estimated from rare earth elements and neodymium-isotopic compositions of
  Martian meteorites}}.
\newblock \bibinfo{journal}{Meteoritics \& Planetary Science}
  \bibinfo{volume}{34}, \bibinfo{pages}{439--449}.
\newblock \DOIprefix\doi{10.1111/j.1945-5100.1999.tb01352.x}.
%Type = Article
\bibitem[{Ohtani and Kamaya(1992)}]{ohtani1992geochemical}
\bibinfo{author}{Ohtani, E.}, \bibinfo{author}{Kamaya, N.},
  \bibinfo{year}{1992}.
\newblock \bibinfo{title}{{The geochemical model of Mars: An estimation from
  the high pressure experiments}}.
\newblock \bibinfo{journal}{Geophysical Research Letters} \bibinfo{volume}{19},
  \bibinfo{pages}{2239--2242}.
\newblock \DOIprefix\doi{10.1029/92GL02369}.
%Type = Article
\bibitem[{Okuchi(1997)}]{okuchi1997hydrogen}
\bibinfo{author}{Okuchi, T.}, \bibinfo{year}{1997}.
\newblock \bibinfo{title}{{Hydrogen partitioning into molten iron at high
  pressure: implications for Earth's core}}.
\newblock \bibinfo{journal}{Science} \bibinfo{volume}{278},
  \bibinfo{pages}{1781--1784}.
\newblock \DOIprefix\doi{10.1126/science.278.5344.1781}.
%Type = Inproceedings
\bibitem[{O'Rourke and Shim(2018)}]{o2018suppressing}
\bibinfo{author}{O'Rourke, J.G.}, \bibinfo{author}{Shim, S.H.},
  \bibinfo{year}{2018}.
\newblock \bibinfo{title}{{Suppressing the Martian dynamo with ongoing
  hydrogenation of the core by hydrated mantle minerals}}, in:
  \bibinfo{booktitle}{Lunar and Planetary Science Conference}, p.
  \bibinfo{pages}{2390}.
%Type = Article
\bibitem[{Ott(1988)}]{ott1988noble}
\bibinfo{author}{Ott, U.}, \bibinfo{year}{1988}.
\newblock \bibinfo{title}{{Noble gases in SNC meteorites: Shergotty, Nakhla,
  Chassigny}}.
\newblock \bibinfo{journal}{Geochimica et Cosmochimica Acta}
  \bibinfo{volume}{52}, \bibinfo{pages}{1937--1948}.
\newblock \DOIprefix\doi{10.1016/0016-7037(88)90017-8}.
%Type = Incollection
\bibitem[{Ott et~al.(2019)Ott, Swindle and Schwenzer}]{ott2019noble}
\bibinfo{author}{Ott, U.}, \bibinfo{author}{Swindle, T.D.},
  \bibinfo{author}{Schwenzer, S.P.}, \bibinfo{year}{2019}.
\newblock \bibinfo{title}{{Noble Gases in Martian Meteorites: Budget, Sources,
  Sinks, and Processes}}, in: \bibinfo{editor}{Filiberto, J.},
  \bibinfo{editor}{Schwenzer, S.P.} (Eds.), \bibinfo{booktitle}{Volatiles in
  the Martian Crust}. \bibinfo{publisher}{Elsevier}, pp.
  \bibinfo{pages}{35--70}.
\newblock \DOIprefix\doi{10.1016/B978-0-12-804191-8.00003-9}.
%Type = Incollection
\bibitem[{Palme et~al.(2014)Palme, Lodders and Jones}]{palme2014solar}
\bibinfo{author}{Palme, H.}, \bibinfo{author}{Lodders, K.},
  \bibinfo{author}{Jones, A.}, \bibinfo{year}{2014}.
\newblock \bibinfo{title}{Solar system abundances of the elements}, in:
  \bibinfo{editor}{Holland, H.D.}, \bibinfo{editor}{Turekian, K.K.} (Eds.),
  \bibinfo{booktitle}{Treatise on Geochemistry (Second Edition)}.
  \bibinfo{publisher}{Elsevier}, \bibinfo{address}{Oxford}.
  volume~\bibinfo{volume}{2}, pp. \bibinfo{pages}{15--36}.
\newblock \DOIprefix\doi{10.1016/B978-0-08-095975-7.00118-2}.
%Type = Incollection
\bibitem[{Palme and O'Neill(2014)}]{palme2014cosmochemical}
\bibinfo{author}{Palme, H.}, \bibinfo{author}{O'Neill, H.S.C.},
  \bibinfo{year}{2014}.
\newblock \bibinfo{title}{Cosmochemical estimates of mantle composition}, in:
  \bibinfo{editor}{Holland, H.D.}, \bibinfo{editor}{Turekian, K.K.} (Eds.),
  \bibinfo{booktitle}{Treatise on Geochemistry (Second Edition)}.
  \bibinfo{publisher}{Elsevier}, \bibinfo{address}{Oxford}.
  volume~\bibinfo{volume}{3}, pp. \bibinfo{pages}{1--39}.
\newblock \DOIprefix\doi{10.1016/B978-0-08-095975-7.00201-1}.
%Type = Article
\bibitem[{Parro et~al.(2017)Parro, Jim{\'e}nez-D{\'\i}az, Mansilla and
  Ruiz}]{parro2017present}
\bibinfo{author}{Parro, L.M.}, \bibinfo{author}{Jim{\'e}nez-D{\'\i}az, A.},
  \bibinfo{author}{Mansilla, F.}, \bibinfo{author}{Ruiz, J.},
  \bibinfo{year}{2017}.
\newblock \bibinfo{title}{{Present-day heat flow model of Mars}}.
\newblock \bibinfo{journal}{Scientific Reports} \bibinfo{volume}{7},
  \bibinfo{pages}{45629}.
\newblock \DOIprefix\doi{10.1038/srep45629}.
%Type = Article
\bibitem[{Peplowski et~al.(2011)Peplowski, Evans, Hauck, McCoy, Boynton,
  Gillis-Davis, Ebel, Goldsten, Hamara, Lawrence, McNutt, Nittler, Solomon,
  Rhodes, Sprague, Starr and Stockstill-Cahill}]{peplowski2011radioactive}
\bibinfo{author}{Peplowski, P.N.}, \bibinfo{author}{Evans, L.G.},
  \bibinfo{author}{Hauck, S.A.}, \bibinfo{author}{McCoy, T.J.},
  \bibinfo{author}{Boynton, W.V.}, \bibinfo{author}{Gillis-Davis, J.J.},
  \bibinfo{author}{Ebel, D.S.}, \bibinfo{author}{Goldsten, J.O.},
  \bibinfo{author}{Hamara, D.K.}, \bibinfo{author}{Lawrence, D.J.},
  \bibinfo{author}{McNutt, R.L.}, \bibinfo{author}{Nittler, L.R.},
  \bibinfo{author}{Solomon, S.C.}, \bibinfo{author}{Rhodes, E.A.},
  \bibinfo{author}{Sprague, A.L.}, \bibinfo{author}{Starr, R.D.},
  \bibinfo{author}{Stockstill-Cahill, K.R.}, \bibinfo{year}{2011}.
\newblock \bibinfo{title}{{Radioactive elements on Mercury’s surface from
  MESSENGER: Implications for the planet’s formation and evolution}}.
\newblock \bibinfo{journal}{Science} \bibinfo{volume}{333},
  \bibinfo{pages}{1850--1852}.
\newblock \DOIprefix\doi{10.1126/science.1211576}.
%Type = Article
\bibitem[{Phillips et~al.(2008)Phillips, Zuber, Smrekar, Mellon, Head, Tanaka,
  Putzig, Milkovich, Campbell, Plaut, Safaeinili, Seu, Biccari, Carter,
  Picardi, Orosei, {Surdas Mohit}, Heggy, Zurek, Egan, Giacomoni, Russo,
  Cutigni, Pettinelli, Holt, Leuschen and Marinangeli}]{phillips2008mars}
\bibinfo{author}{Phillips, R.J.}, \bibinfo{author}{Zuber, M.T.},
  \bibinfo{author}{Smrekar, S.E.}, \bibinfo{author}{Mellon, M.T.},
  \bibinfo{author}{Head, J.W.}, \bibinfo{author}{Tanaka, K.L.},
  \bibinfo{author}{Putzig, N.E.}, \bibinfo{author}{Milkovich, S.M.},
  \bibinfo{author}{Campbell, B.A.}, \bibinfo{author}{Plaut, J.J.},
  \bibinfo{author}{Safaeinili, A.}, \bibinfo{author}{Seu, R.},
  \bibinfo{author}{Biccari, D.}, \bibinfo{author}{Carter, L.M.},
  \bibinfo{author}{Picardi, G.}, \bibinfo{author}{Orosei, R.},
  \bibinfo{author}{{Surdas Mohit}, P.}, \bibinfo{author}{Heggy, E.},
  \bibinfo{author}{Zurek, R.W.}, \bibinfo{author}{Egan, A.F.},
  \bibinfo{author}{Giacomoni, E.}, \bibinfo{author}{Russo, F.},
  \bibinfo{author}{Cutigni, M.}, \bibinfo{author}{Pettinelli, E.},
  \bibinfo{author}{Holt, J.W.}, \bibinfo{author}{Leuschen, C.J.},
  \bibinfo{author}{Marinangeli, L.}, \bibinfo{year}{2008}.
\newblock \bibinfo{title}{{Mars north polar deposits: Stratigraphy, age, and
  geodynamical response}}.
\newblock \bibinfo{journal}{Science} \bibinfo{volume}{320},
  \bibinfo{pages}{1182--1185}.
\newblock \DOIprefix\doi{10.1126/science.1157546}.
%Type = Article
\bibitem[{Prettyman et~al.(2015)Prettyman, Yamashita, Reedy, McSween,
  Mittlefehldt, Hendricks and Toplis}]{prettyman2015concentrations}
\bibinfo{author}{Prettyman, T.H.}, \bibinfo{author}{Yamashita, N.},
  \bibinfo{author}{Reedy, R.C.}, \bibinfo{author}{McSween, Jr., H.Y.},
  \bibinfo{author}{Mittlefehldt, D.W.}, \bibinfo{author}{Hendricks, J.S.},
  \bibinfo{author}{Toplis, M.J.}, \bibinfo{year}{2015}.
\newblock \bibinfo{title}{{Concentrations of potassium and thorium within
  Vesta’s regolith}}.
\newblock \bibinfo{journal}{Icarus} \bibinfo{volume}{259},
  \bibinfo{pages}{39--52}.
\newblock \DOIprefix\doi{10.1016/j.icarus.2015.05.035}.
%Type = Article
\bibitem[{Putirka(2016)}]{putirka2016rates}
\bibinfo{author}{Putirka, K.}, \bibinfo{year}{2016}.
\newblock \bibinfo{title}{{Rates and styles of planetary cooling on Earth,
  Moon, Mars, and Vesta, using new models for oxygen fugacity, ferric-ferrous
  ratios, olivine-liquid Fe-Mg exchange, and mantle potential temperature}}.
\newblock \bibinfo{journal}{American Mineralogist} \bibinfo{volume}{101},
  \bibinfo{pages}{819--840}.
\newblock \DOIprefix\doi{10.2138/am-2016-5402}.
%Type = Article
\bibitem[{Rai and van Westrenen(2013)}]{rai2013core}
\bibinfo{author}{Rai, N.}, \bibinfo{author}{van Westrenen, W.},
  \bibinfo{year}{2013}.
\newblock \bibinfo{title}{{Core-mantle differentiation in Mars}}.
\newblock \bibinfo{journal}{Journal of Geophysical Research: Planets}
  \bibinfo{volume}{118}, \bibinfo{pages}{1195--1203}.
\newblock \DOIprefix\doi{10.1002/jgre.20093}.
%Type = Article
\bibitem[{Raymond and Izidoro(2017)}]{raymond2017origin}
\bibinfo{author}{Raymond, S.N.}, \bibinfo{author}{Izidoro, A.},
  \bibinfo{year}{2017}.
\newblock \bibinfo{title}{{Origin of water in the inner Solar System:
  Planetesimals scattered inward during Jupiter and Saturn’s rapid gas
  accretion}}.
\newblock \bibinfo{journal}{Icarus} \bibinfo{volume}{297},
  \bibinfo{pages}{134--148}.
\newblock \DOIprefix\doi{10.1016/j.icarus.2017.06.030}.
%Type = Misc
\bibitem[{Righter(2017)}]{righter2017martian}
\bibinfo{author}{Righter, K.}, \bibinfo{year}{2017}.
\newblock \bibinfo{title}{{The Martian Meteorite Compendium}}.
\newblock \URLprefix \url{https://curator.jsc.nasa.gov/antmet/mmc/}.
%Type = Article
\bibitem[{Righter and Chabot(2011)}]{righter2011moderately}
\bibinfo{author}{Righter, K.}, \bibinfo{author}{Chabot, N.L.},
  \bibinfo{year}{2011}.
\newblock \bibinfo{title}{{Moderately and slightly siderophile element
  constraints on the depth and extent of melting in early Mars}}.
\newblock \bibinfo{journal}{Meteoritics \& Planetary Science}
  \bibinfo{volume}{46}, \bibinfo{pages}{157--176}.
\newblock \DOIprefix\doi{10.1111/j.1945-5100.2010.01140.x}.
%Type = Article
\bibitem[{Righter et~al.(2017)Righter, Nickodem, Pando, Danielson, Boujibar,
  Righter and Lapen}]{righter2017distribution}
\bibinfo{author}{Righter, K.}, \bibinfo{author}{Nickodem, K.},
  \bibinfo{author}{Pando, K.}, \bibinfo{author}{Danielson, L.},
  \bibinfo{author}{Boujibar, A.}, \bibinfo{author}{Righter, M.},
  \bibinfo{author}{Lapen, T.J.}, \bibinfo{year}{2017}.
\newblock \bibinfo{title}{{Distribution of Sb, As, Ge, and In between metal and
  silicate during accretion and core formation in the Earth}}.
\newblock \bibinfo{journal}{Geochimica et Cosmochimica Acta}
  \bibinfo{volume}{198}, \bibinfo{pages}{1--16}.
\newblock \DOIprefix\doi{10.1016/j.gca.2016.10.045}.
%Type = Article
\bibitem[{Ringwood(1966)}]{ringwood1966chemical}
\bibinfo{author}{Ringwood, A.E.}, \bibinfo{year}{1966}.
\newblock \bibinfo{title}{{Chemical evolution of the terrestrial planets}}.
\newblock \bibinfo{journal}{Geochimica et Cosmochimica Acta}
  \bibinfo{volume}{30}, \bibinfo{pages}{41--104}.
\newblock \DOIprefix\doi{10.1016/0016-7037(66)90090-1}.
%Type = Article
\bibitem[{Ringwood and Hibberson(1990)}]{ringwood1990system}
\bibinfo{author}{Ringwood, A.E.}, \bibinfo{author}{Hibberson, W.},
  \bibinfo{year}{1990}.
\newblock \bibinfo{title}{{The system Fe-FeO revisited}}.
\newblock \bibinfo{journal}{Physics and Chemistry of Minerals}
  \bibinfo{volume}{17}, \bibinfo{pages}{313--319}.
\newblock \DOIprefix\doi{10.1007/BF00200126}.
%Type = Article
\bibitem[{Rivoldini et~al.(2011)Rivoldini, Van~Hoolst, Verhoeven, Mocquet and
  Dehant}]{rivoldini2011geodesy}
\bibinfo{author}{Rivoldini, A.}, \bibinfo{author}{Van~Hoolst, T.},
  \bibinfo{author}{Verhoeven, O.}, \bibinfo{author}{Mocquet, A.},
  \bibinfo{author}{Dehant, V.}, \bibinfo{year}{2011}.
\newblock \bibinfo{title}{{Geodesy constraints on the interior structure and
  composition of Mars}}.
\newblock \bibinfo{journal}{Icarus} \bibinfo{volume}{213},
  \bibinfo{pages}{451--472}.
\newblock \DOIprefix\doi{10.1016/j.icarus.2011.03.024}.
%Type = Article
\bibitem[{Rose-Weston et~al.(2009)Rose-Weston, Brenan, Fei, Secco and
  Frost}]{rose2009effect}
\bibinfo{author}{Rose-Weston, L.}, \bibinfo{author}{Brenan, J.M.},
  \bibinfo{author}{Fei, Y.}, \bibinfo{author}{Secco, R.A.},
  \bibinfo{author}{Frost, D.J.}, \bibinfo{year}{2009}.
\newblock \bibinfo{title}{{Effect of pressure, temperature, and oxygen fugacity
  on the metal-silicate partitioning of Te, Se, and S: Implications for Earth
  differentiation}}.
\newblock \bibinfo{journal}{Geochimica et Cosmochimica Acta}
  \bibinfo{volume}{73}, \bibinfo{pages}{4598--4615}.
\newblock \DOIprefix\doi{10.1016/j.gca.2009.04.028}.
%Type = Article
\bibitem[{Rubie et~al.(2004)Rubie, Gessmann and Frost}]{rubie2004partitioning}
\bibinfo{author}{Rubie, D.C.}, \bibinfo{author}{Gessmann, C.K.},
  \bibinfo{author}{Frost, D.J.}, \bibinfo{year}{2004}.
\newblock \bibinfo{title}{{Partitioning of oxygen during core formation on the
  Earth and Mars}}.
\newblock \bibinfo{journal}{Nature} \bibinfo{volume}{429},
  \bibinfo{pages}{58--61}.
\newblock \DOIprefix\doi{10.1038/nature02473}.
%Type = Article
\bibitem[{Ruiz et~al.(2009)Ruiz, Williams, Dohm, Fern{\'a}ndez and
  L{\'o}pez}]{ruiz2009ancient}
\bibinfo{author}{Ruiz, J.}, \bibinfo{author}{Williams, J.P.},
  \bibinfo{author}{Dohm, J.M.}, \bibinfo{author}{Fern{\'a}ndez, C.},
  \bibinfo{author}{L{\'o}pez, V.}, \bibinfo{year}{2009}.
\newblock \bibinfo{title}{{Ancient heat flow and crustal thickness at Warrego
  rise, Thaumasia highlands, Mars: Implications for a stratified crust}}.
\newblock \bibinfo{journal}{Icarus} \bibinfo{volume}{203},
  \bibinfo{pages}{47--57}.
\newblock \DOIprefix\doi{10.1016/j.icarus.2009.05.008}.
%Type = Article
\bibitem[{Sanloup et~al.(1999)Sanloup, Jambon and Gillet}]{sanloup1999simple}
\bibinfo{author}{Sanloup, C.}, \bibinfo{author}{Jambon, A.},
  \bibinfo{author}{Gillet, P.}, \bibinfo{year}{1999}.
\newblock \bibinfo{title}{{A simple chondritic model of Mars}}.
\newblock \bibinfo{journal}{Physics of the Earth and Planetary Interiors}
  \bibinfo{volume}{112}, \bibinfo{pages}{43--54}.
\newblock \DOIprefix\doi{10.1016/S0031-9201(98)00175-7}.
%Type = Article
\bibitem[{Schmelz et~al.(2012)Schmelz, Reames, Von~Steiger and
  Basu}]{schmelz2012composition}
\bibinfo{author}{Schmelz, J.T.}, \bibinfo{author}{Reames, D.V.},
  \bibinfo{author}{Von~Steiger, R.}, \bibinfo{author}{Basu, S.},
  \bibinfo{year}{2012}.
\newblock \bibinfo{title}{{Composition of the solar corona, solar wind, and
  solar energetic particles}}.
\newblock \bibinfo{journal}{The Astrophysical Journal} \bibinfo{volume}{755},
  \bibinfo{pages}{33}.
\newblock \DOIprefix\doi{10.1088/0004-637X/755/1/33}.
%Type = Article
\bibitem[{Schmidt et~al.(2013)Schmidt, Schrader and McCoy}]{schmidt2013primary}
\bibinfo{author}{Schmidt, M.E.}, \bibinfo{author}{Schrader, C.M.},
  \bibinfo{author}{McCoy, T.J.}, \bibinfo{year}{2013}.
\newblock \bibinfo{title}{{The primary \textit{f}\ce{O2} of basalts examined by
  the Spirit rover in Gusev Crater, Mars: Evidence for multiple redox states in
  the Martian interior}}.
\newblock \bibinfo{journal}{Earth and Planetary Science Letters}
  \bibinfo{volume}{384}, \bibinfo{pages}{198--208}.
\newblock \DOIprefix\doi{10.1016/j.epsl.2013.10.005}.
%Type = Incollection
\bibitem[{Scott and Krot(2014)}]{scott2014chondrites}
\bibinfo{author}{Scott, E.R.D.}, \bibinfo{author}{Krot, A.N.},
  \bibinfo{year}{2014}.
\newblock \bibinfo{title}{{Chondrites and their Components}}, in:
  \bibinfo{editor}{Holland, H.D.}, \bibinfo{editor}{Turekian, K.K.} (Eds.),
  \bibinfo{booktitle}{Treatise on Geochemistry (Second Edition)}.
  \bibinfo{publisher}{Elsevier}, \bibinfo{address}{Oxford}.
  volume~\bibinfo{volume}{1}, pp. \bibinfo{pages}{65--137}.
\newblock \DOIprefix\doi{10.1016/B978-0-08-095975-7.00104-2}.
%Type = Article
\bibitem[{Seidelmann et~al.(2002)Seidelmann, Abalakin, Bursa, Davies, De~Bergh,
  Lieske, Oberst, Simon, Standish, Stooke and Thomas}]{seidelmann2002report}
\bibinfo{author}{Seidelmann, P.K.}, \bibinfo{author}{Abalakin, V.K.},
  \bibinfo{author}{Bursa, M.}, \bibinfo{author}{Davies, M.E.},
  \bibinfo{author}{De~Bergh, C.}, \bibinfo{author}{Lieske, J.H.},
  \bibinfo{author}{Oberst, J.}, \bibinfo{author}{Simon, J.L.},
  \bibinfo{author}{Standish, E.M.}, \bibinfo{author}{Stooke, P.},
  \bibinfo{author}{Thomas, P.C.}, \bibinfo{year}{2002}.
\newblock \bibinfo{title}{{Report of the IAU/IAG working group on cartographic
  coordinates and rotational elements of the planets and satellites: 2000}}.
\newblock \bibinfo{journal}{Celestial Mechanics and Dynamical Astronomy}
  \bibinfo{volume}{82}, \bibinfo{pages}{83--111}.
\newblock \DOIprefix\doi{10.1023/A:1013939327465}.
%Type = Article
\bibitem[{Shahar et~al.(2015)Shahar, Hillgren, Horan, Mesa-Garcia, Kaufman and
  Mock}]{shahar2015sulfur}
\bibinfo{author}{Shahar, A.}, \bibinfo{author}{Hillgren, V.J.},
  \bibinfo{author}{Horan, M.F.}, \bibinfo{author}{Mesa-Garcia, J.},
  \bibinfo{author}{Kaufman, L.A.}, \bibinfo{author}{Mock, T.D.},
  \bibinfo{year}{2015}.
\newblock \bibinfo{title}{{Sulfur-controlled iron isotope fractionation
  experiments of core formation in planetary bodies}}.
\newblock \bibinfo{journal}{Geochimica et Cosmochimica Acta}
  \bibinfo{volume}{150}, \bibinfo{pages}{253--264}.
\newblock \DOIprefix\doi{10.1016/j.gca.2014.08.011}.
%Type = Article
\bibitem[{Shibazaki et~al.(2009)Shibazaki, Ohtani, Terasaki, Suzuki and
  Funakoshi}]{shibazaki2009hydrogen}
\bibinfo{author}{Shibazaki, Y.}, \bibinfo{author}{Ohtani, E.},
  \bibinfo{author}{Terasaki, H.}, \bibinfo{author}{Suzuki, A.},
  \bibinfo{author}{Funakoshi, K.i.}, \bibinfo{year}{2009}.
\newblock \bibinfo{title}{{Hydrogen partitioning between iron and ringwoodite:
  Implications for water transport into the Martian core}}.
\newblock \bibinfo{journal}{Earth and Planetary Science Letters}
  \bibinfo{volume}{287}, \bibinfo{pages}{463--470}.
\newblock \DOIprefix\doi{10.1016/j.epsl.2009.08.034}.
%Type = Article
\bibitem[{Siebert et~al.(2013)Siebert, Badro, Antonangeli and
  Ryerson}]{siebert2013terrestrial}
\bibinfo{author}{Siebert, J.}, \bibinfo{author}{Badro, J.},
  \bibinfo{author}{Antonangeli, D.}, \bibinfo{author}{Ryerson, F.J.},
  \bibinfo{year}{2013}.
\newblock \bibinfo{title}{{Terrestrial accretion under oxidizing conditions}}.
\newblock \bibinfo{journal}{Science} \bibinfo{volume}{339},
  \bibinfo{pages}{1194--1197}.
\newblock \DOIprefix\doi{10.1126/science.122792}.
%Type = Article
\bibitem[{Smrekar et~al.(2019)Smrekar, Lognonn{\'e}, Spohn, Banerdt, Breuer,
  Christensen, Dehant, Drilleau, Folkner, Fuji, Garcia, Giardini, Golombek,
  Grott, Gudkova, Johnson, Khan, Langlais, Mittelholz, Mocquet, Myhill,
  Panning, Perrin, Pike, Plesa, Rivoldini, Samuel, St{\"a}hler, van Driel,
  Van~Hoolst, Verhoeven, Weber and Wieczorek}]{smrekar2019pre}
\bibinfo{author}{Smrekar, S.E.}, \bibinfo{author}{Lognonn{\'e}, P.},
  \bibinfo{author}{Spohn, T.}, \bibinfo{author}{Banerdt, W.B.},
  \bibinfo{author}{Breuer, D.}, \bibinfo{author}{Christensen, U.},
  \bibinfo{author}{Dehant, V.}, \bibinfo{author}{Drilleau, M.},
  \bibinfo{author}{Folkner, W.}, \bibinfo{author}{Fuji, N.},
  \bibinfo{author}{Garcia, R.F.}, \bibinfo{author}{Giardini, D.},
  \bibinfo{author}{Golombek, M.}, \bibinfo{author}{Grott, M.},
  \bibinfo{author}{Gudkova, T.}, \bibinfo{author}{Johnson, C.},
  \bibinfo{author}{Khan, A.}, \bibinfo{author}{Langlais, B.},
  \bibinfo{author}{Mittelholz, A.}, \bibinfo{author}{Mocquet, A.},
  \bibinfo{author}{Myhill, R.}, \bibinfo{author}{Panning, M.},
  \bibinfo{author}{Perrin, C.}, \bibinfo{author}{Pike, T.},
  \bibinfo{author}{Plesa, A.C.}, \bibinfo{author}{Rivoldini, A.},
  \bibinfo{author}{Samuel, H.}, \bibinfo{author}{St{\"a}hler, S.C.},
  \bibinfo{author}{van Driel, M.}, \bibinfo{author}{Van~Hoolst, T.},
  \bibinfo{author}{Verhoeven, O.}, \bibinfo{author}{Weber, R.},
  \bibinfo{author}{Wieczorek, M.}, \bibinfo{year}{2019}.
\newblock \bibinfo{title}{{Pre-mission InSights on the interior of Mars}}.
\newblock \bibinfo{journal}{Space Science Reviews} \bibinfo{volume}{215},
  \bibinfo{pages}{3}.
\newblock \DOIprefix\doi{10.1007/s11214-018-0563-9}.
%Type = Article
\bibitem[{Sohl and Spohn(1997)}]{sohl1997interior}
\bibinfo{author}{Sohl, F.}, \bibinfo{author}{Spohn, T.}, \bibinfo{year}{1997}.
\newblock \bibinfo{title}{{The interior structure of Mars: Implications from
  SNC meteorites}}.
\newblock \bibinfo{journal}{Journal of Geophysical Research: Planets}
  \bibinfo{volume}{102}, \bibinfo{pages}{1613--1635}.
\newblock \DOIprefix\doi{10.1029/96JE03419}.
%Type = Article
\bibitem[{Stixrude and Lithgow-Bertelloni(2005)}]{stixrude2005thermodynamics}
\bibinfo{author}{Stixrude, L.}, \bibinfo{author}{Lithgow-Bertelloni, C.},
  \bibinfo{year}{2005}.
\newblock \bibinfo{title}{{Thermodynamics of mantle minerals--I. Physical
  properties}}.
\newblock \bibinfo{journal}{Geophysical Journal International}
  \bibinfo{volume}{162}, \bibinfo{pages}{610--632}.
\newblock \DOIprefix\doi{10.1111/j.1365-246X.2005.02642.x}.
%Type = Article
\bibitem[{Stixrude and Lithgow-Bertelloni(2011)}]{stixrude2011thermodynamics}
\bibinfo{author}{Stixrude, L.}, \bibinfo{author}{Lithgow-Bertelloni, C.},
  \bibinfo{year}{2011}.
\newblock \bibinfo{title}{{Thermodynamics of mantle minerals-II. Phase
  equilibria}}.
\newblock \bibinfo{journal}{Geophysical Journal International}
  \bibinfo{volume}{184}, \bibinfo{pages}{1180--1213}.
\newblock \DOIprefix\doi{10.1111/j.1365-246X.2010.04890.x}.
%Type = Article
\bibitem[{Sun(1982)}]{sun1982chemical}
\bibinfo{author}{Sun, S.s.}, \bibinfo{year}{1982}.
\newblock \bibinfo{title}{{Chemical composition and origin of the Earth's
  primitive mantle}}.
\newblock \bibinfo{journal}{Geochimica et Cosmochimica Acta}
  \bibinfo{volume}{46}, \bibinfo{pages}{179--192}.
\newblock \DOIprefix\doi{10.1016/0016-7037(82)90245-9}.
%Type = Book
\bibitem[{Surkov(1977)}]{surkov1977gamma}
\bibinfo{author}{Surkov, Y.A.}, \bibinfo{year}{1977}.
\newblock \bibinfo{title}{{Gamma-Spectrometry in Cosmic Investigations}}.
\newblock \bibinfo{publisher}{Atomizdat}, \bibinfo{address}{Moscow}.
%Type = Article
\bibitem[{Surkov et~al.(1987)Surkov, Kirnozov, Glazov, Dunchenko, Tatsy and
  Sobornov}]{surkov1987uranium}
\bibinfo{author}{Surkov, Y.A.}, \bibinfo{author}{Kirnozov, F.F.},
  \bibinfo{author}{Glazov, V.N.}, \bibinfo{author}{Dunchenko, A.G.},
  \bibinfo{author}{Tatsy, L.P.}, \bibinfo{author}{Sobornov, O.P.},
  \bibinfo{year}{1987}.
\newblock \bibinfo{title}{{Uranium, thorium, and potassium in the Venusian
  rocks at the landing sites of Vega 1 and 2}}.
\newblock \bibinfo{journal}{Journal of Geophysical Research: Solid Earth}
  \bibinfo{volume}{92}, \bibinfo{pages}{E537--E540}.
\newblock \DOIprefix\doi{10.1029/JB092iB04p0E537}.
%Type = Article
\bibitem[{Surkov et~al.(1986)Surkov, Moskalyova, Kharyukova, Dudin, Smirnov and
  Zaitseva}]{surkov1986venus}
\bibinfo{author}{Surkov, Y.A.}, \bibinfo{author}{Moskalyova, L.P.},
  \bibinfo{author}{Kharyukova, V.P.}, \bibinfo{author}{Dudin, A.D.},
  \bibinfo{author}{Smirnov, G.G.}, \bibinfo{author}{Zaitseva, S.Y.},
  \bibinfo{year}{1986}.
\newblock \bibinfo{title}{{Venus rock composition at the Vega 2 landing site}}.
\newblock \bibinfo{journal}{Journal of Geophysical Research: Solid Earth}
  \bibinfo{volume}{91}, \bibinfo{pages}{E215--E218}.
\newblock \DOIprefix\doi{10.1029/JB091iB13p0E215}.
%Type = Article
\bibitem[{Symes et~al.(2008)Symes, Borg, Shearer and Irving}]{symes2008age}
\bibinfo{author}{Symes, S.J.K.}, \bibinfo{author}{Borg, L.E.},
  \bibinfo{author}{Shearer, C.K.}, \bibinfo{author}{Irving, A.J.},
  \bibinfo{year}{2008}.
\newblock \bibinfo{title}{{The age of the Martian meteorite Northwest Africa
  1195 and the differentiation history of the shergottites}}.
\newblock \bibinfo{journal}{Geochimica et Cosmochimica Acta}
  \bibinfo{volume}{72}, \bibinfo{pages}{1696--1710}.
\newblock \DOIprefix\doi{10.1016/j.gca.2007.12.022}.
%Type = Article
\bibitem[{Tait and Day(2018)}]{tait2018chondritic}
\bibinfo{author}{Tait, K.T.}, \bibinfo{author}{Day, J.M.D.},
  \bibinfo{year}{2018}.
\newblock \bibinfo{title}{{Chondritic late accretion to Mars and the nature of
  shergottite reservoirs}}.
\newblock \bibinfo{journal}{Earth and Planetary Science Letters}
  \bibinfo{volume}{494}, \bibinfo{pages}{99--108}.
\newblock \DOIprefix\doi{10.1016/j.epsl.2018.04.040}.
%Type = Article
\bibitem[{Takahashi and Kushiro(1983)}]{takahashi1983melting}
\bibinfo{author}{Takahashi, E.}, \bibinfo{author}{Kushiro, I.},
  \bibinfo{year}{1983}.
\newblock \bibinfo{title}{{Melting of a dry peridotite at high pressures and
  basalt magma genesis}}.
\newblock \bibinfo{journal}{American Mineralogist} \bibinfo{volume}{68},
  \bibinfo{pages}{859--879}.
%Type = Article
\bibitem[{Taylor(2013)}]{taylor2013bulk}
\bibinfo{author}{Taylor, G.J.}, \bibinfo{year}{2013}.
\newblock \bibinfo{title}{{The bulk composition of Mars}}.
\newblock \bibinfo{journal}{Chemie der Erde-Geochemistry} \bibinfo{volume}{73},
  \bibinfo{pages}{401--420}.
\newblock \DOIprefix\doi{10.1016/j.chemer.2013.09.006}.
%Type = Article
\bibitem[{Taylor et~al.(2006a)Taylor, Boynton, Brückner, Wänke, Dreibus,
  Kerry, Keller, Reedy, Evans, Starr, Squyres, Karunatillake, Gasnault,
  Maurice, d'Uston, Englert, Dohm, Baker, Hamara, Janes, Sprague, Kim and
  Drake}]{taylor2006bulk}
\bibinfo{author}{Taylor, G.J.}, \bibinfo{author}{Boynton, W.},
  \bibinfo{author}{Brückner, J.}, \bibinfo{author}{Wänke, H.},
  \bibinfo{author}{Dreibus, G.}, \bibinfo{author}{Kerry, K.},
  \bibinfo{author}{Keller, J.}, \bibinfo{author}{Reedy, R.},
  \bibinfo{author}{Evans, L.}, \bibinfo{author}{Starr, R.},
  \bibinfo{author}{Squyres, S.}, \bibinfo{author}{Karunatillake, S.},
  \bibinfo{author}{Gasnault, O.}, \bibinfo{author}{Maurice, S.},
  \bibinfo{author}{d'Uston, C.}, \bibinfo{author}{Englert, P.},
  \bibinfo{author}{Dohm, J.}, \bibinfo{author}{Baker, V.},
  \bibinfo{author}{Hamara, D.}, \bibinfo{author}{Janes, D.},
  \bibinfo{author}{Sprague, A.}, \bibinfo{author}{Kim, K.},
  \bibinfo{author}{Drake, D.}, \bibinfo{year}{2006}a.
\newblock \bibinfo{title}{{Bulk composition and early differentiation of
  Mars}}.
\newblock \bibinfo{journal}{Journal of Geophysical Research: Planets}
  \bibinfo{volume}{111}, \bibinfo{pages}{E03S10}.
\newblock \DOIprefix\doi{10.1029/2005JE002645}.
%Type = Article
\bibitem[{Taylor et~al.(2010)Taylor, Boynton, McLennan and
  Martel}]{taylor2010k}
\bibinfo{author}{Taylor, G.J.}, \bibinfo{author}{Boynton, W.V.},
  \bibinfo{author}{McLennan, S.M.}, \bibinfo{author}{Martel, L.M.V.},
  \bibinfo{year}{2010}.
\newblock \bibinfo{title}{{K and Cl concentrations on the Martian surface
  determined by the Mars Odyssey Gamma Ray Spectrometer: Implications for bulk
  halogen abundances in Mars}}.
\newblock \bibinfo{journal}{Geophysical Research Letters} \bibinfo{volume}{37},
  \bibinfo{pages}{L12204}.
\newblock \DOIprefix\doi{10.1029/2010GL043528}.
%Type = Article
\bibitem[{Taylor et~al.(2006b)Taylor, Stopar, Boynton, Karunatillake, Keller,
  Brückner, Wänke, Dreibus, Kerry, Reedy, Evans, Starr, Martel, Squyres,
  Gasnault, Maurice, d'Uston, Englert, Dohm, Baker, Hamara, Janes, Sprague,
  Kim, Drake, McLennan and Hahn}]{taylor2006variations}
\bibinfo{author}{Taylor, G.J.}, \bibinfo{author}{Stopar, J.D.},
  \bibinfo{author}{Boynton, W.V.}, \bibinfo{author}{Karunatillake, S.},
  \bibinfo{author}{Keller, J.M.}, \bibinfo{author}{Brückner, J.},
  \bibinfo{author}{Wänke, H.}, \bibinfo{author}{Dreibus, G.},
  \bibinfo{author}{Kerry, K.E.}, \bibinfo{author}{Reedy, R.C.},
  \bibinfo{author}{Evans, L.G.}, \bibinfo{author}{Starr, R.D.},
  \bibinfo{author}{Martel, L.M.V.}, \bibinfo{author}{Squyres, S.W.},
  \bibinfo{author}{Gasnault, O.}, \bibinfo{author}{Maurice, S.},
  \bibinfo{author}{d'Uston, C.}, \bibinfo{author}{Englert, P.},
  \bibinfo{author}{Dohm, J.M.}, \bibinfo{author}{Baker, V.R.},
  \bibinfo{author}{Hamara, D.}, \bibinfo{author}{Janes, D.},
  \bibinfo{author}{Sprague, A.L.}, \bibinfo{author}{Kim, K.J.},
  \bibinfo{author}{Drake, D.M.}, \bibinfo{author}{McLennan, S.M.},
  \bibinfo{author}{Hahn, B.C.}, \bibinfo{year}{2006}b.
\newblock \bibinfo{title}{{Variations in K/Th on Mars}}.
\newblock \bibinfo{journal}{Journal of Geophysical Research: Planets}
  \bibinfo{volume}{111}, \bibinfo{pages}{E03S06}.
\newblock \DOIprefix\doi{10.1029/2006JE002676}.
%Type = Book
\bibitem[{Taylor and McLennan(2009)}]{taylor2009planetary}
\bibinfo{author}{Taylor, S.R.}, \bibinfo{author}{McLennan, S.},
  \bibinfo{year}{2009}.
\newblock \bibinfo{title}{{Planetary Crusts: Their Composition, Origin and
  Evolution}}. volume~\bibinfo{volume}{10}.
\newblock \bibinfo{publisher}{Cambridge University Press}.
%Type = Article
\bibitem[{Treiman(2003)}]{treiman2003chemical}
\bibinfo{author}{Treiman, A.H.}, \bibinfo{year}{2003}.
\newblock \bibinfo{title}{{Chemical compositions of martian basalts
  (shergottites): Some inferences on basalt formation, mantle metasomatism, and
  differentiation in Mars}}.
\newblock \bibinfo{journal}{Meteoritics \& Planetary Science}
  \bibinfo{volume}{38}, \bibinfo{pages}{1849--1864}.
\newblock \DOIprefix\doi{10.1111/j.1945-5100.2003.tb00019.x}.
%Type = Article
\bibitem[{Treiman(2005)}]{treiman2005nakhlite}
\bibinfo{author}{Treiman, A.H.}, \bibinfo{year}{2005}.
\newblock \bibinfo{title}{{The nakhlite meteorites: Augite-rich igneous rocks
  from Mars}}.
\newblock \bibinfo{journal}{Chemie der Erde-Geochemistry} \bibinfo{volume}{65},
  \bibinfo{pages}{203--270}.
\newblock \DOIprefix\doi{10.1016/j.chemer.2005.01.004}.
%Type = Article
\bibitem[{Treiman and Lindstrom(1997)}]{treiman1997trace}
\bibinfo{author}{Treiman, A.H.}, \bibinfo{author}{Lindstrom, D.J.},
  \bibinfo{year}{1997}.
\newblock \bibinfo{title}{{Trace element geochemistry of Martian iddingsite in
  the Lafayette meteorite}}.
\newblock \bibinfo{journal}{Journal of Geophysical Research: Planets}
  \bibinfo{volume}{102}, \bibinfo{pages}{9153--9163}.
\newblock \DOIprefix\doi{10.1029/96JE03884}.
%Type = Article
\bibitem[{Tsuno et~al.(2011)Tsuno, Frost and Rubie}]{tsuno2011effects}
\bibinfo{author}{Tsuno, K.}, \bibinfo{author}{Frost, D.J.},
  \bibinfo{author}{Rubie, D.C.}, \bibinfo{year}{2011}.
\newblock \bibinfo{title}{{The effects of nickel and sulphur on the
  core--mantle partitioning of oxygen in Earth and Mars}}.
\newblock \bibinfo{journal}{Physics of the Earth and Planetary Interiors}
  \bibinfo{volume}{185}, \bibinfo{pages}{1--12}.
\newblock \DOIprefix\doi{10.1016/j.pepi.2010.11.009}.
%Type = Article
\bibitem[{Udry and Day(2018)}]{udry20181}
\bibinfo{author}{Udry, A.}, \bibinfo{author}{Day, J.M.D.},
  \bibinfo{year}{2018}.
\newblock \bibinfo{title}{{1.34 billion-year-old magmatism on Mars evaluated
  from the co-genetic nakhlite and chassignite meteorites}}.
\newblock \bibinfo{journal}{Geochimica et Cosmochimica Acta}
  \bibinfo{volume}{238}, \bibinfo{pages}{292--315}.
\newblock \DOIprefix\doi{10.1016/j.gca.2018.07.006}.
%Type = Article
\bibitem[{Umemoto and Hirose(2015)}]{umemoto2015liquid}
\bibinfo{author}{Umemoto, K.}, \bibinfo{author}{Hirose, K.},
  \bibinfo{year}{2015}.
\newblock \bibinfo{title}{{Liquid iron-hydrogen alloys at outer core conditions
  by first-principles calculations}}.
\newblock \bibinfo{journal}{Geophysical Research Letters} \bibinfo{volume}{42},
  \bibinfo{pages}{7513--7520}.
\newblock \DOIprefix\doi{10.1002/2015GL065899}.
%Type = Article
\bibitem[{Usui et~al.(2008)Usui, McSween and Floss}]{usui2008petrogenesis}
\bibinfo{author}{Usui, T.}, \bibinfo{author}{McSween, Jr., H.Y.},
  \bibinfo{author}{Floss, C.}, \bibinfo{year}{2008}.
\newblock \bibinfo{title}{{Petrogenesis of olivine-phyric shergottite Yamato
  980459, revisited}}.
\newblock \bibinfo{journal}{Geochimica et Cosmochimica Acta}
  \bibinfo{volume}{72}, \bibinfo{pages}{1711--1730}.
\newblock \DOIprefix\doi{10.1016/j.gca.2008.01.011}.
%Type = Article
\bibitem[{{Verhoeven} et~al.(2005){Verhoeven}, {Rivoldini}, {Vacher},
  {Mocquet}, {Choblet}, {Menvielle}, {Dehant}, {Van Hoolst}, {Sleewaegen},
  {Barriot} and {Lognonn{\'e}}}]{verhoeven2005interior}
\bibinfo{author}{{Verhoeven}, O.}, \bibinfo{author}{{Rivoldini}, A.},
  \bibinfo{author}{{Vacher}, P.}, \bibinfo{author}{{Mocquet}, A.},
  \bibinfo{author}{{Choblet}, G.}, \bibinfo{author}{{Menvielle}, M.},
  \bibinfo{author}{{Dehant}, V.}, \bibinfo{author}{{Van Hoolst}, T.},
  \bibinfo{author}{{Sleewaegen}, J.}, \bibinfo{author}{{Barriot}, J.P.},
  \bibinfo{author}{{Lognonn{\'e}}, P.}, \bibinfo{year}{2005}.
\newblock \bibinfo{title}{{Interior structure of terrestrial planets: Modeling
  Mars' mantle and its electromagnetic, geodetic, and seismic properties}}.
\newblock \bibinfo{journal}{Journal of Geophysical Research: Planets}
  \bibinfo{volume}{110}, \bibinfo{pages}{E04009}.
\newblock \DOIprefix\doi{10.1029/2004JE002271}.
%Type = Article
\bibitem[{Wade and Wood(2005)}]{wade2005core}
\bibinfo{author}{Wade, J.}, \bibinfo{author}{Wood, B.J.}, \bibinfo{year}{2005}.
\newblock \bibinfo{title}{{Core formation and the oxidation state of the
  Earth}}.
\newblock \bibinfo{journal}{Earth and Planetary Science Letters}
  \bibinfo{volume}{236}, \bibinfo{pages}{78--95}.
\newblock \DOIprefix\doi{10.1016/j.epsl.2005.05.017}.
%Type = Article
\bibitem[{Wadhwa(2001)}]{wadhwa2001redox}
\bibinfo{author}{Wadhwa, M.}, \bibinfo{year}{2001}.
\newblock \bibinfo{title}{{Redox state of Mars' upper mantle and crust from Eu
  anomalies in shergottite pyroxenes}}.
\newblock \bibinfo{journal}{Science} \bibinfo{volume}{291},
  \bibinfo{pages}{1527--1530}.
\newblock \DOIprefix\doi{10.1126/science.1057594}.
%Type = Incollection
\bibitem[{Wadhwa(2008)}]{wadhwa2008redox}
\bibinfo{author}{Wadhwa, M.}, \bibinfo{year}{2008}.
\newblock \bibinfo{title}{{Redox conditions on small bodies, the Moon and
  Mars}}, in: \bibinfo{editor}{MacPherson, G.J.} (Ed.),
  \bibinfo{booktitle}{Reviews in Mineralogy and Geochemistry}.
  \bibinfo{publisher}{Mineralogical Society of America}.
  volume~\bibinfo{volume}{68}, pp. \bibinfo{pages}{493--510}.
\newblock \DOIprefix\doi{10.2138/rmg.2008.68.17}.
%Type = Article
\bibitem[{Walsh et~al.(2011)Walsh, Morbidelli, Raymond, O'Brien and
  Mandell}]{walsh2011low}
\bibinfo{author}{Walsh, K.J.}, \bibinfo{author}{Morbidelli, A.},
  \bibinfo{author}{Raymond, S.N.}, \bibinfo{author}{O'Brien, D.P.},
  \bibinfo{author}{Mandell, A.M.}, \bibinfo{year}{2011}.
\newblock \bibinfo{title}{{A low mass for Mars from Jupiter's early gas-driven
  migration}}.
\newblock \bibinfo{journal}{Nature} \bibinfo{volume}{475},
  \bibinfo{pages}{206--209}.
\newblock \DOIprefix\doi{10.1038/nature10201}.
%Type = Article
\bibitem[{Walter(1998)}]{walter1998melting}
\bibinfo{author}{Walter, M.J.}, \bibinfo{year}{1998}.
\newblock \bibinfo{title}{{Melting of garnet peridotite and the origin of
  komatiite and depleted lithosphere}}.
\newblock \bibinfo{journal}{Journal of Petrology} \bibinfo{volume}{39},
  \bibinfo{pages}{29--60}.
\newblock \DOIprefix\doi{10.1093/petroj/39.1.29}.
%Type = Article
\bibitem[{Wang and Becker(2017)}]{wang2017chalcophile}
\bibinfo{author}{Wang, Z.}, \bibinfo{author}{Becker, H.}, \bibinfo{year}{2017}.
\newblock \bibinfo{title}{{Chalcophile elements in Martian meteorites indicate
  low sulfur content in the Martian interior and a volatile element-depleted
  late veneer}}.
\newblock \bibinfo{journal}{Earth and Planetary Science Letters}
  \bibinfo{volume}{463}, \bibinfo{pages}{56--68}.
\newblock \DOIprefix\doi{10.1016/j.epsl.2017.01.023}.
%Type = Article
\bibitem[{Wang et~al.(2015)Wang, Becker and Wombacher}]{wang2015mass}
\bibinfo{author}{Wang, Z.}, \bibinfo{author}{Becker, H.},
  \bibinfo{author}{Wombacher, F.}, \bibinfo{year}{2015}.
\newblock \bibinfo{title}{{Mass fractions of S, Cu, Se, Mo, Ag, Cd, In, Te, Ba,
  Sm, W, Tl and Bi in geological reference materials and selected carbonaceous
  chondrites determined by isotope dilution ICP-MS}}.
\newblock \bibinfo{journal}{Geostandards and Geoanalytical Research}
  \bibinfo{volume}{39}, \bibinfo{pages}{185--208}.
\newblock \DOIprefix\doi{10.1111/j.1751-908X.2014.00312.x}.
%Type = Article
\bibitem[{W{\"a}nke(1981)}]{wanke1981constitution}
\bibinfo{author}{W{\"a}nke, H.}, \bibinfo{year}{1981}.
\newblock \bibinfo{title}{{Constitution of terrestrial planets}}.
\newblock \bibinfo{journal}{Philosophical Transactions of the Royal Society of
  London. Series A: Mathematical and Physical Sciences} \bibinfo{volume}{303},
  \bibinfo{pages}{287--302}.
\newblock \DOIprefix\doi{10.1098/rsta.1981.0203}.
%Type = Article
\bibitem[{W{\"a}nke(1987)}]{wanke1987chemistry}
\bibinfo{author}{W{\"a}nke, H.}, \bibinfo{year}{1987}.
\newblock \bibinfo{title}{{Chemistry and accretion of Earth and Mars}}.
\newblock \bibinfo{journal}{Bulletin de la Soci{\'e}t{\'e} G{\'e}ologique de
  France} \bibinfo{volume}{3}, \bibinfo{pages}{13--19}.
%Type = Article
\bibitem[{W{\"a}nke and Dreibus(1988)}]{wanke1988chemical}
\bibinfo{author}{W{\"a}nke, H.}, \bibinfo{author}{Dreibus, G.},
  \bibinfo{year}{1988}.
\newblock \bibinfo{title}{{Chemical composition and accretion history of
  terrestrial planets}}.
\newblock \bibinfo{journal}{Philosophical Transactions of the Royal Society of
  London. Series A, Mathematical and Physical Sciences} \bibinfo{volume}{325},
  \bibinfo{pages}{545--557}.
\newblock \DOIprefix\doi{10.1098/rsta.1988.0067}.
%Type = Article
\bibitem[{W{\"a}nke and Dreibus(1994)}]{wanke1994chemistry}
\bibinfo{author}{W{\"a}nke, H.}, \bibinfo{author}{Dreibus, G.},
  \bibinfo{year}{1994}.
\newblock \bibinfo{title}{{Chemistry and accretion history of Mars}}.
\newblock \bibinfo{journal}{Philosophical Transactions of the Royal Society of
  London. Series A, Mathematical and Physical Sciences} \bibinfo{volume}{349},
  \bibinfo{pages}{285--293}.
\newblock \DOIprefix\doi{10.1098/rsta.1994.0132}.
%Type = Article
\bibitem[{Warren(2011)}]{warren2011stable}
\bibinfo{author}{Warren, P.H.}, \bibinfo{year}{2011}.
\newblock \bibinfo{title}{{Stable-isotopic anomalies and the accretionary
  assemblage of the Earth and Mars: A subordinate role for carbonaceous
  chondrites}}.
\newblock \bibinfo{journal}{Earth and Planetary Science Letters}
  \bibinfo{volume}{311}, \bibinfo{pages}{93--100}.
\newblock \DOIprefix\doi{10.1016/j.epsl.2011.08.047}.
%Type = Article
\bibitem[{Wasson and Kallemeyn(1988)}]{wasson1988compositions}
\bibinfo{author}{Wasson, J.T.}, \bibinfo{author}{Kallemeyn, G.W.},
  \bibinfo{year}{1988}.
\newblock \bibinfo{title}{{Compositions of chondrites}}.
\newblock \bibinfo{journal}{Philosophical Transactions of the Royal Society of
  London A: Mathematical, Physical and Engineering Sciences}
  \bibinfo{volume}{325}, \bibinfo{pages}{535--544}.
\newblock \DOIprefix\doi{10.1098/rsta.1988.0066}.
%Type = Article
\bibitem[{Wasylenki et~al.(2003)Wasylenki, Baker, Kent and
  Stolper}]{wasylenki2003near}
\bibinfo{author}{Wasylenki, L.E.}, \bibinfo{author}{Baker, M.B.},
  \bibinfo{author}{Kent, A.J.}, \bibinfo{author}{Stolper, E.M.},
  \bibinfo{year}{2003}.
\newblock \bibinfo{title}{{Near-solidus melting of the shallow upper mantle:
  partial melting experiments on depleted peridotite}}.
\newblock \bibinfo{journal}{Journal of Petrology} \bibinfo{volume}{44},
  \bibinfo{pages}{1163--1191}.
\newblock \DOIprefix\doi{10.1093/petrology/44.7.1163}.
%Type = Article
\bibitem[{Weinke(1978)}]{weinke1978chemical}
\bibinfo{author}{Weinke, H.H.}, \bibinfo{year}{1978}.
\newblock \bibinfo{title}{{Chemical and mineralogical examination of the Nakhla
  achondrite}}.
\newblock \bibinfo{journal}{Meteoritics} \bibinfo{volume}{13},
  \bibinfo{pages}{660--664}.
%Type = Article
\bibitem[{Wieczorek and Zuber(2004)}]{wieczorek2004thickness}
\bibinfo{author}{Wieczorek, M.A.}, \bibinfo{author}{Zuber, M.T.},
  \bibinfo{year}{2004}.
\newblock \bibinfo{title}{{Thickness of the Martian crust: Improved constraints
  from geoid-to-topography ratios}}.
\newblock \bibinfo{journal}{Journal of Geophysical Research: Planets}
  \bibinfo{volume}{109}.
\newblock \DOIprefix\doi{10.1029/2003JE002153}.
%Type = Article
\bibitem[{Wipperfurth et~al.(2018)Wipperfurth, Guo, {\v{S}}r{\'a}mek and
  McDonough}]{wipperfurth2018earth}
\bibinfo{author}{Wipperfurth, S.A.}, \bibinfo{author}{Guo, M.},
  \bibinfo{author}{{\v{S}}r{\'a}mek, O.}, \bibinfo{author}{McDonough, W.F.},
  \bibinfo{year}{2018}.
\newblock \bibinfo{title}{{Earth's chondritic Th/U: Negligible fractionation
  during accretion, core formation, and crust-mantle differentiation}}.
\newblock \bibinfo{journal}{Earth and Planetary Science Letters}
  \bibinfo{volume}{498}, \bibinfo{pages}{196--202}.
\newblock \DOIprefix\doi{10.1016/j.epsl.2018.06.029}.
%Type = Article
\bibitem[{Wood(1993)}]{wood1993carbon}
\bibinfo{author}{Wood, B.J.}, \bibinfo{year}{1993}.
\newblock \bibinfo{title}{{Carbon in the core}}.
\newblock \bibinfo{journal}{Earth and Planetary Science Letters}
  \bibinfo{volume}{117}, \bibinfo{pages}{593--607}.
\newblock \DOIprefix\doi{10.1016/0012-821X(93)90105-I}.
%Type = Article
\bibitem[{Wood et~al.(2013)Wood, Li and Shahar}]{wood2013carbon}
\bibinfo{author}{Wood, B.J.}, \bibinfo{author}{Li, J.},
  \bibinfo{author}{Shahar, A.}, \bibinfo{year}{2013}.
\newblock \bibinfo{title}{{Carbon in the core: its influence on the properties
  of core and mantle}}.
\newblock \bibinfo{journal}{Reviews in Mineralogy and Geochemistry}
  \bibinfo{volume}{75}, \bibinfo{pages}{231--250}.
\newblock \DOIprefix\doi{10.2138/rmg.2013.75.8}.
%Type = Article
\bibitem[{Wood et~al.(2019)Wood, Smythe and Harrison}]{wood2019condensation}
\bibinfo{author}{Wood, B.J.}, \bibinfo{author}{Smythe, D.J.},
  \bibinfo{author}{Harrison, T.}, \bibinfo{year}{2019}.
\newblock \bibinfo{title}{{The condensation temperatures of the elements: A
  reappraisal}}.
\newblock \bibinfo{journal}{American Mineralogist} \bibinfo{volume}{104},
  \bibinfo{pages}{844--856}.
\newblock \DOIprefix\doi{10.2138/am-2019-6852CCBY}.
%Type = Article
\bibitem[{Wood et~al.(2008)Wood, Wade and Kilburn}]{wood2008core}
\bibinfo{author}{Wood, B.J.}, \bibinfo{author}{Wade, J.},
  \bibinfo{author}{Kilburn, M.R.}, \bibinfo{year}{2008}.
\newblock \bibinfo{title}{{Core formation and the oxidation state of the Earth:
  Additional constraints from Nb, V and Cr partitioning}}.
\newblock \bibinfo{journal}{Geochimica et Cosmochimica Acta}
  \bibinfo{volume}{72}, \bibinfo{pages}{1415--1426}.
\newblock \DOIprefix\doi{10.1016/j.gca.2007.11.036}.
%Type = Article
\bibitem[{Wood et~al.(2006)Wood, Walter and Wade}]{wood2006accretion}
\bibinfo{author}{Wood, B.J.}, \bibinfo{author}{Walter, M.J.},
  \bibinfo{author}{Wade, J.}, \bibinfo{year}{2006}.
\newblock \bibinfo{title}{{Accretion of the Earth and segregation of its
  core}}.
\newblock \bibinfo{journal}{Nature} \bibinfo{volume}{441},
  \bibinfo{pages}{825--833}.
\newblock \DOIprefix\doi{10.1038/nature04763}.
%Type = Article
\bibitem[{Yang et~al.(2015)Yang, Humayun, Righter, Jefferson, Fields and
  Irving}]{yang2015siderophile}
\bibinfo{author}{Yang, S.}, \bibinfo{author}{Humayun, M.},
  \bibinfo{author}{Righter, K.}, \bibinfo{author}{Jefferson, G.},
  \bibinfo{author}{Fields, D.}, \bibinfo{author}{Irving, A.J.},
  \bibinfo{year}{2015}.
\newblock \bibinfo{title}{{Siderophile and chalcophile element abundances in
  shergottites: Implications for Martian core formation}}.
\newblock \bibinfo{journal}{Meteoritics \& Planetary Science}
  \bibinfo{volume}{50}, \bibinfo{pages}{691--714}.
\newblock \DOIprefix\doi{10.1111/maps.12384}.
%Type = Article
\bibitem[{Zambardi et~al.(2013)Zambardi, Poitrasson, Corgne, M{\'e}heut,
  Quitt{\'e} and Anand}]{zambardi2013silicon}
\bibinfo{author}{Zambardi, T.}, \bibinfo{author}{Poitrasson, F.},
  \bibinfo{author}{Corgne, A.}, \bibinfo{author}{M{\'e}heut, M.},
  \bibinfo{author}{Quitt{\'e}, G.}, \bibinfo{author}{Anand, M.},
  \bibinfo{year}{2013}.
\newblock \bibinfo{title}{{Silicon isotope variations in the inner solar
  system: Implications for planetary formation, differentiation and
  composition}}.
\newblock \bibinfo{journal}{Geochimica et Cosmochimica Acta}
  \bibinfo{volume}{121}, \bibinfo{pages}{67--83}.
\newblock \DOIprefix\doi{10.1016/j.gca.2013.06.040}.
%Type = Article
\bibitem[{Zharkov(1996)}]{zharkov1996internal}
\bibinfo{author}{Zharkov, V.N.}, \bibinfo{year}{1996}.
\newblock \bibinfo{title}{{The internal structure of Mars: a key to
  understanding the origin of terrestrial planets}}.
\newblock \bibinfo{journal}{Solar System Research} \bibinfo{volume}{30},
  \bibinfo{pages}{456--466}.
%Type = Article
\bibitem[{Zharkov and Gudkova(2005)}]{zharkov2005construction}
\bibinfo{author}{Zharkov, V.N.}, \bibinfo{author}{Gudkova, T.V.},
  \bibinfo{year}{2005}.
\newblock \bibinfo{title}{{Construction of Martian interior model}}.
\newblock \bibinfo{journal}{Solar System Research} \bibinfo{volume}{39},
  \bibinfo{pages}{343--373}.
\newblock \DOIprefix\doi{10.1007/s11208-005-0049-7}.
%Type = Article
\bibitem[{Zuber et~al.(2000)Zuber, Solomon, Phillips, Smith, Tyler, Aharonson,
  Balmino, Banerdt, Head, Johnson, Lemoine, McGovern, Neumann, Rowlands and
  Zhong}]{zuber2000internal}
\bibinfo{author}{Zuber, M.T.}, \bibinfo{author}{Solomon, S.C.},
  \bibinfo{author}{Phillips, R.J.}, \bibinfo{author}{Smith, D.E.},
  \bibinfo{author}{Tyler, G.L.}, \bibinfo{author}{Aharonson, O.},
  \bibinfo{author}{Balmino, G.}, \bibinfo{author}{Banerdt, W.B.},
  \bibinfo{author}{Head, J.W.}, \bibinfo{author}{Johnson, C.L.},
  \bibinfo{author}{Lemoine, F.G.}, \bibinfo{author}{McGovern, P.J.},
  \bibinfo{author}{Neumann, G.A.}, \bibinfo{author}{Rowlands, D.D.},
  \bibinfo{author}{Zhong, S.}, \bibinfo{year}{2000}.
\newblock \bibinfo{title}{{Internal structure and early thermal evolution of
  Mars from Mars Global Surveyor topography and gravity}}.
\newblock \bibinfo{journal}{Science} \bibinfo{volume}{287},
  \bibinfo{pages}{1788--1793}.
\newblock \DOIprefix\doi{10.1126/science.287.5459.1788}.

\end{thebibliography}

\begin{appendices}
	
	\setcounter{figure}{0}
	\setcounter{table}{0}
	\renewcommand{\thefigure}{A\arabic{figure}}
	\renewcommand{\thetable}{A\arabic{table}}
	\renewcommand{\thesection}{A\arabic{section}}
	
	\clearpage
	
	\subsection*{Appendix A: Supplementary data}
	
	\begin{figure}[h]
		\centering
		\includegraphics[width=1\linewidth]{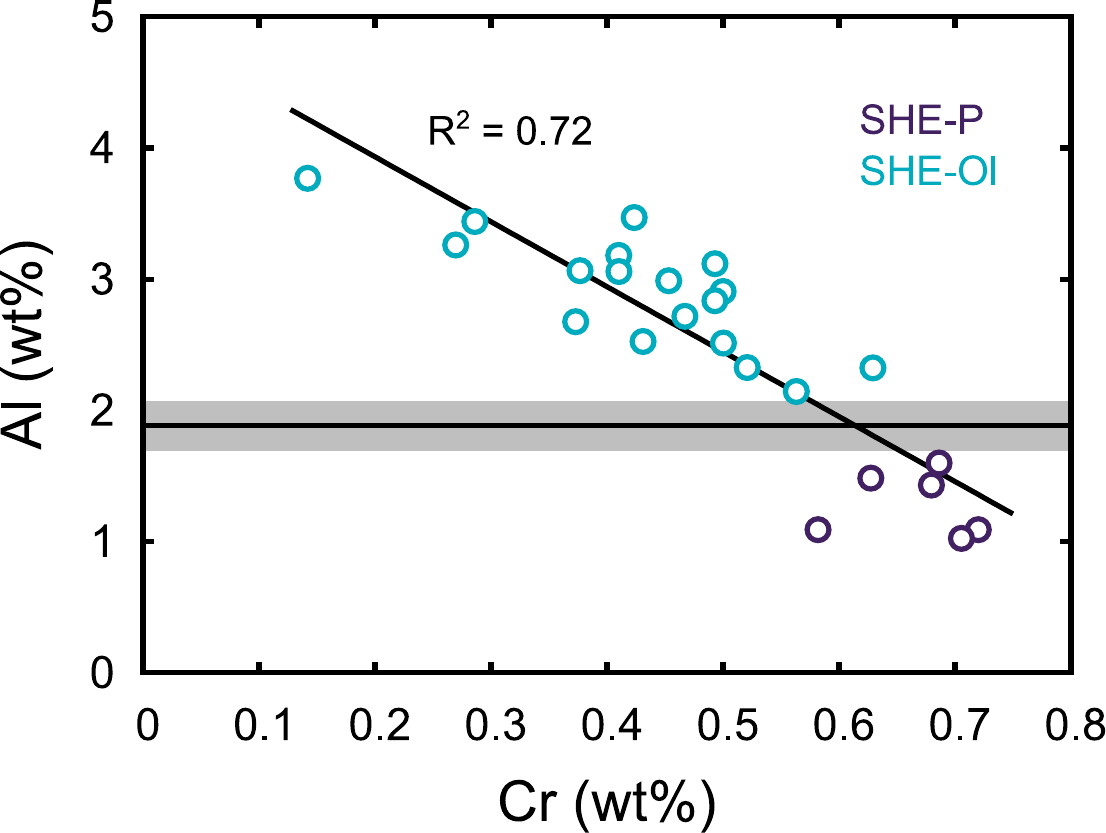}
		\caption{Chromium versus Al contents in poikilitic (SHE-P) and olivine-phyric (SHE-Ol) shergottites. Horizontal line and gray band show the BSM abundance of Al and its uncertainty, respectively (\cref{tab:BSM_all}).}
		\label{fig:Cr}
	\end{figure}
	\clearpage	
	
	\begin{figure}[h]
		\centering
		\includegraphics[width=1\linewidth]{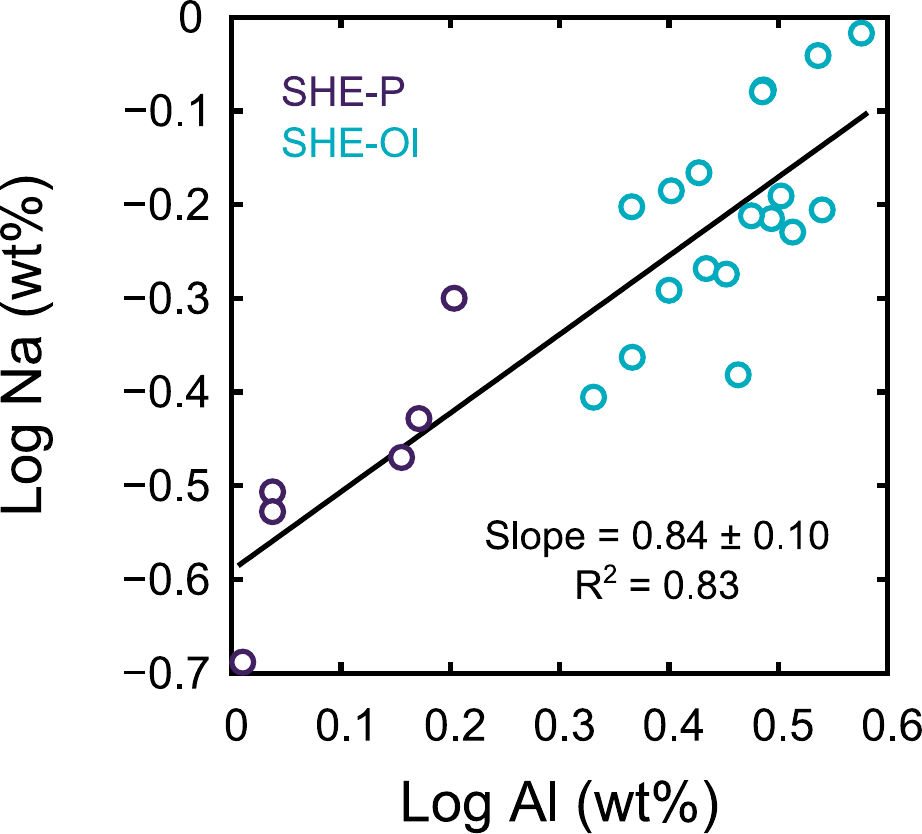}
		\caption{Log-log plot of Na and Al in poikilitic and olivine-phyric shergottites.}
		\label{fig:Na}
	\end{figure}
	\clearpage
	
	\begin{figure}[h]
		\centering
		\includegraphics[width=1\linewidth]{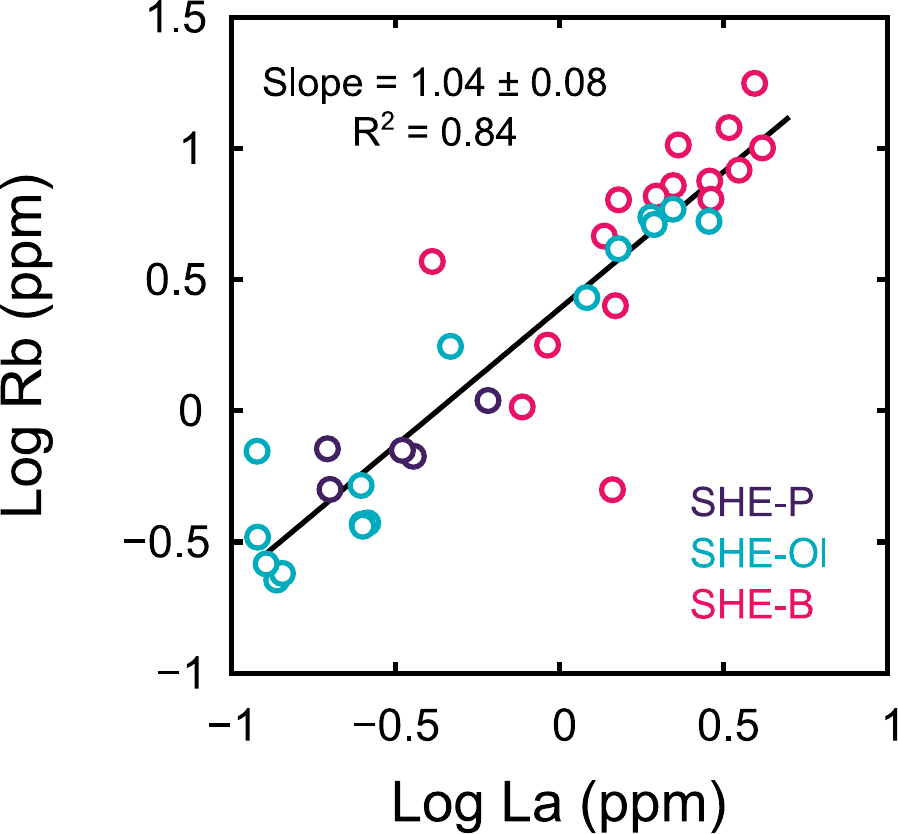}
		\caption{Log-log plot of Rb and La in poikilitic, olivine-phyric and basaltic (SHE-B) shergottites.}
		\label{fig:Rb}
	\end{figure}
	\clearpage
	
	\begin{figure}[h]
		\centering
		\includegraphics[width=1\linewidth]{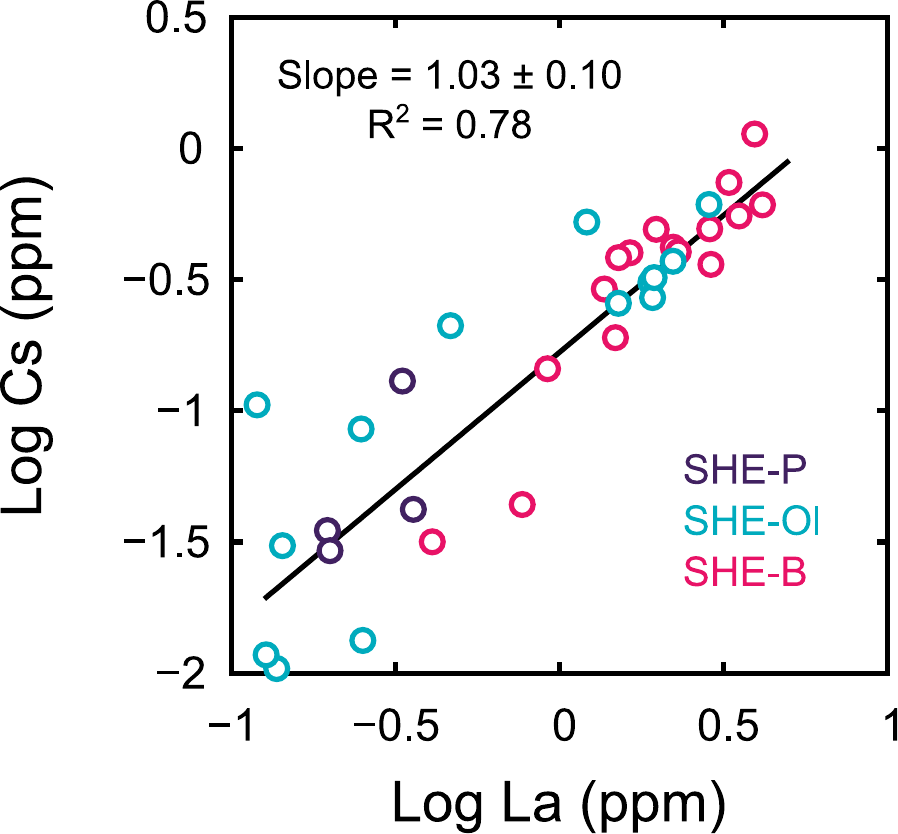}
		\caption{Log-log plot of Cs and La in poikilitic, olivine-phyric and basaltic shergottites.}
		\label{fig:Cs}
	\end{figure}
	\clearpage
	
	\begin{figure}[h]
		\centering
		\includegraphics[width=1\linewidth]{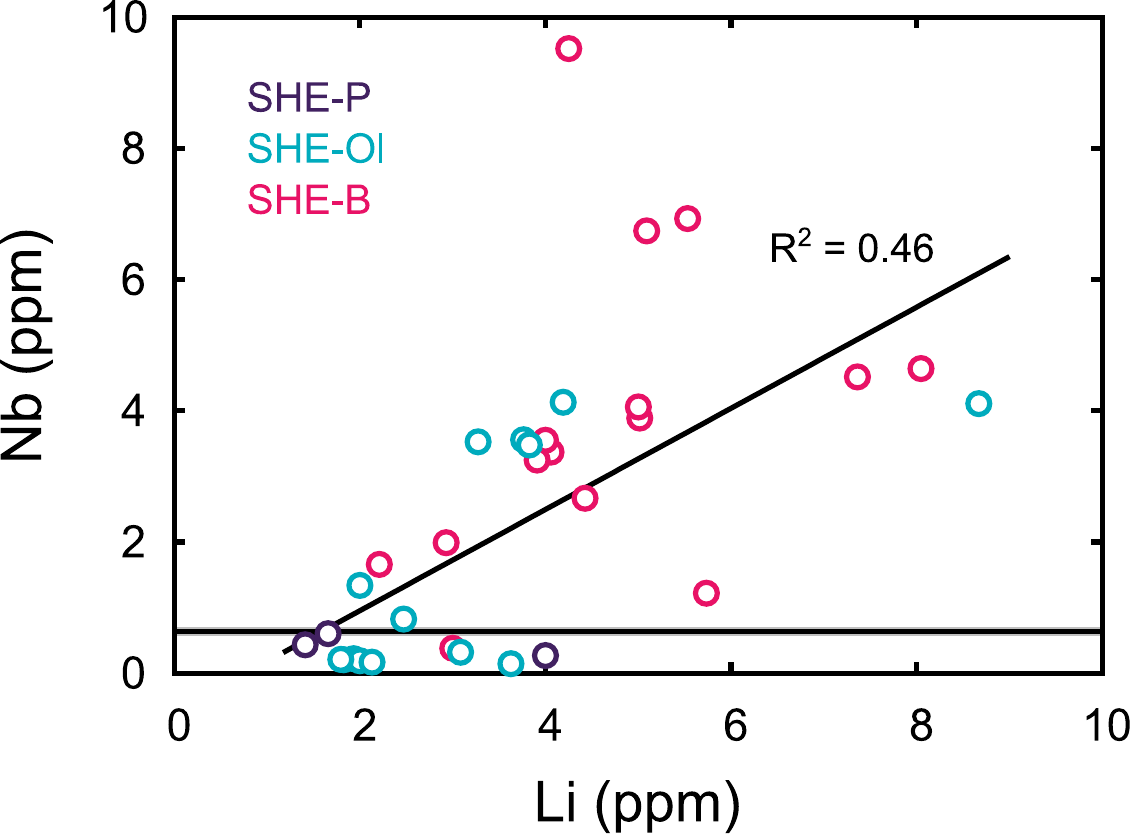}
		\caption{Lithium versus Nb contents in poikilitic, olivine-phyric and basaltic shergottites. Horizontal line and gray band show the BSM abundance of Nb and its uncertainty, respectively (\cref{tab:BSM_all}).}
		\label{fig:Li}
	\end{figure}
	\clearpage	
	
	\begin{figure}[h]
		\centering
		\includegraphics[width=1\linewidth]{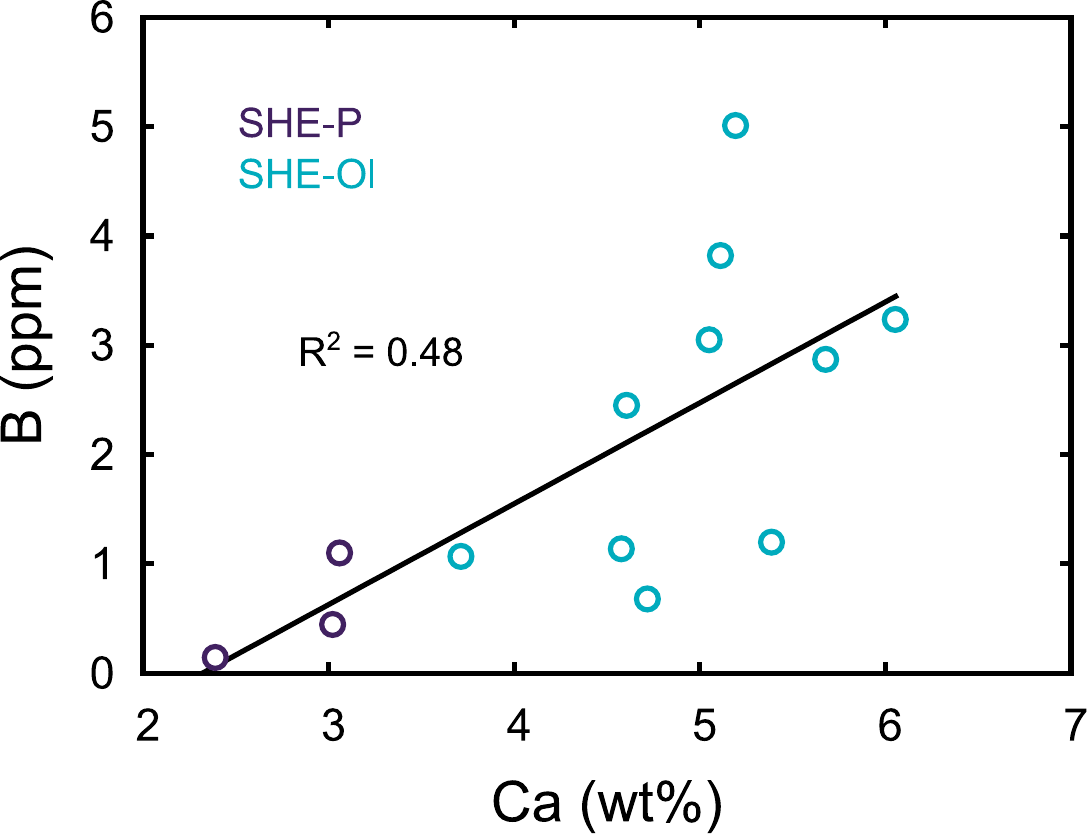}
		\caption{Boron versus Ca contents in poikilitic and olivine-phyric shergottites.}
		\label{fig:B}
	\end{figure}
	\clearpage	
	
	\begin{figure}[h]
		\centering
		\includegraphics[width=1\linewidth]{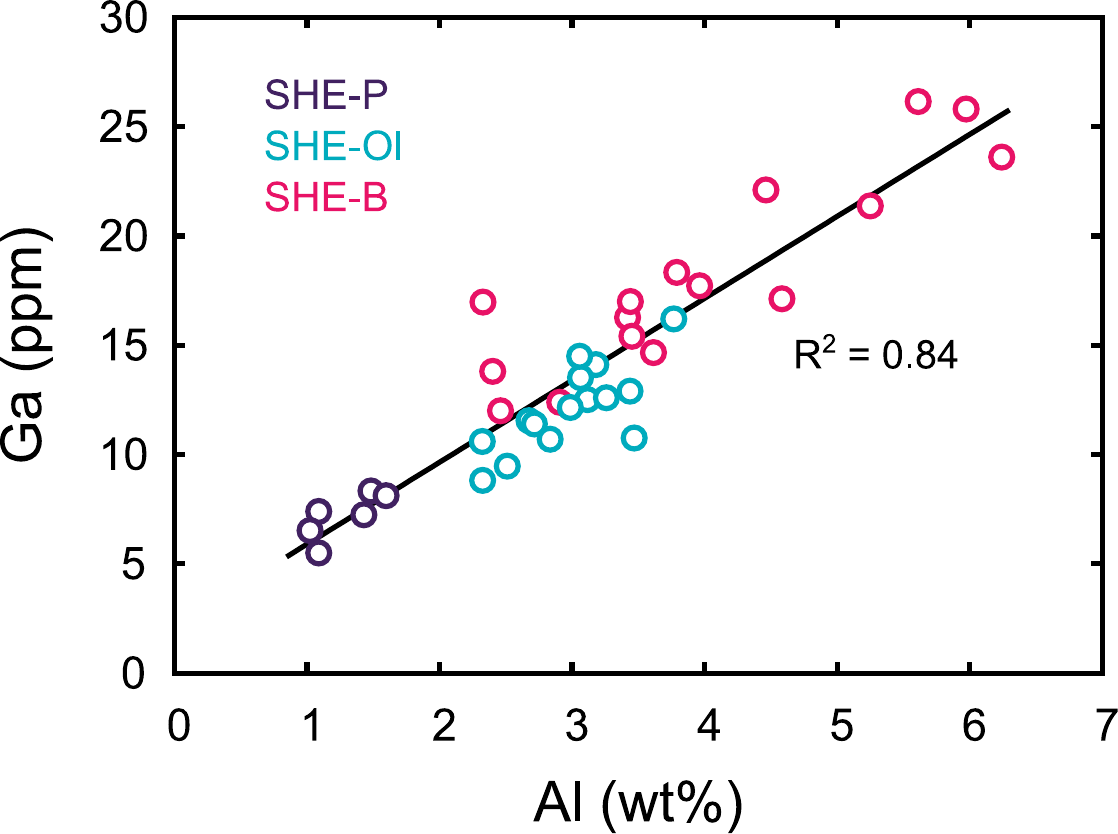}
		\caption{Gallium versus Al contents in poikilitic, olivine-phyric and basaltic shergottites.}
		\label{fig:Ga}
	\end{figure}
	\clearpage	
	
	\begin{figure}[h]
		\centering
		\includegraphics[width=1\linewidth]{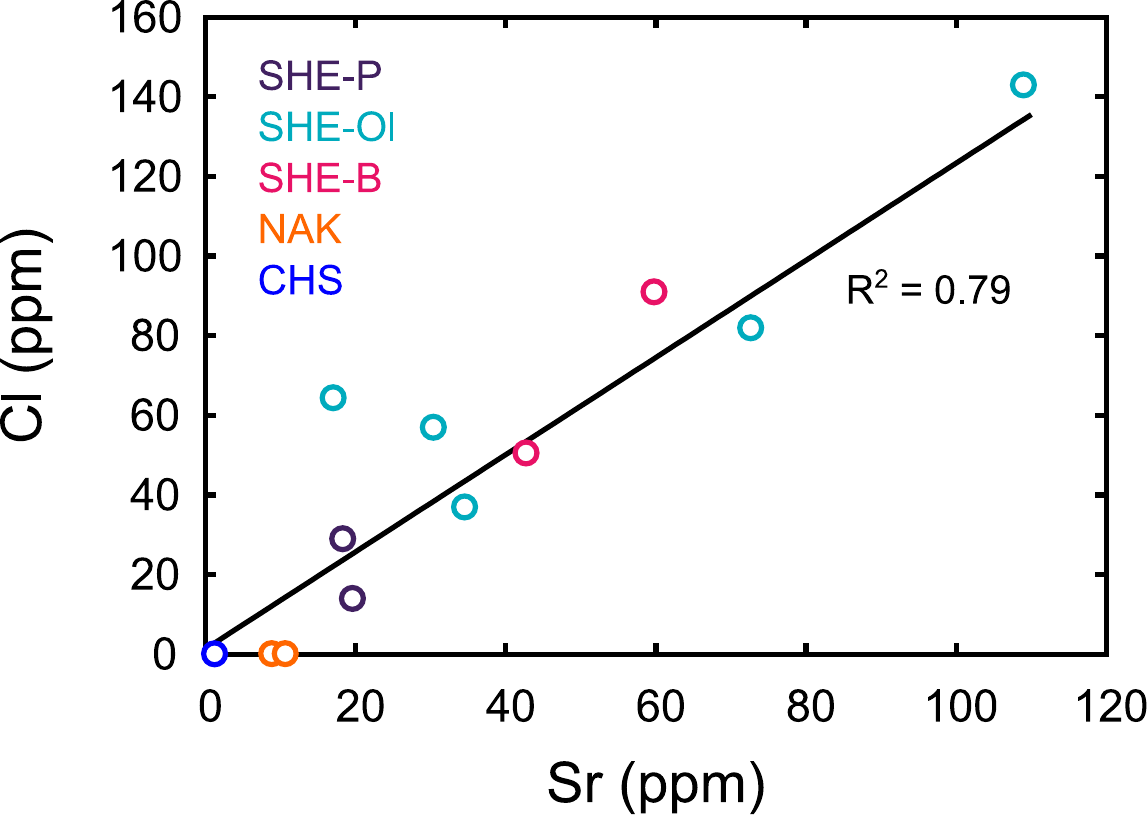}
		\caption{Chlorine versus Sr contents in Martian meteorites.}
		\label{fig:Cl}
	\end{figure}
	\clearpage	
	
	\begin{figure}[h]
		\centering
		\includegraphics[width=1\linewidth]{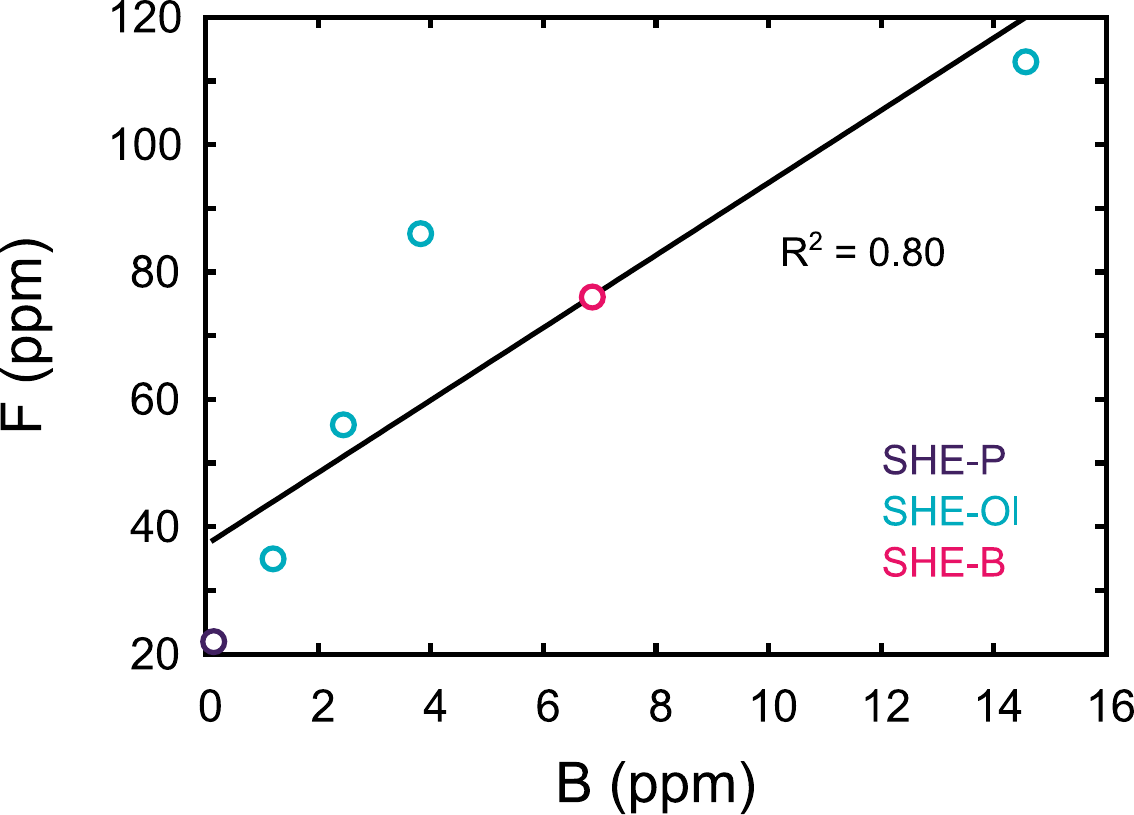}
		\caption{Fluorine versus B contents in poikilitic, olivine-phyric and basaltic shergottites.}
		\label{fig:F}
	\end{figure}
	\clearpage
	
	\begin{figure}[h]
		\centering
		\includegraphics[width=1\linewidth]{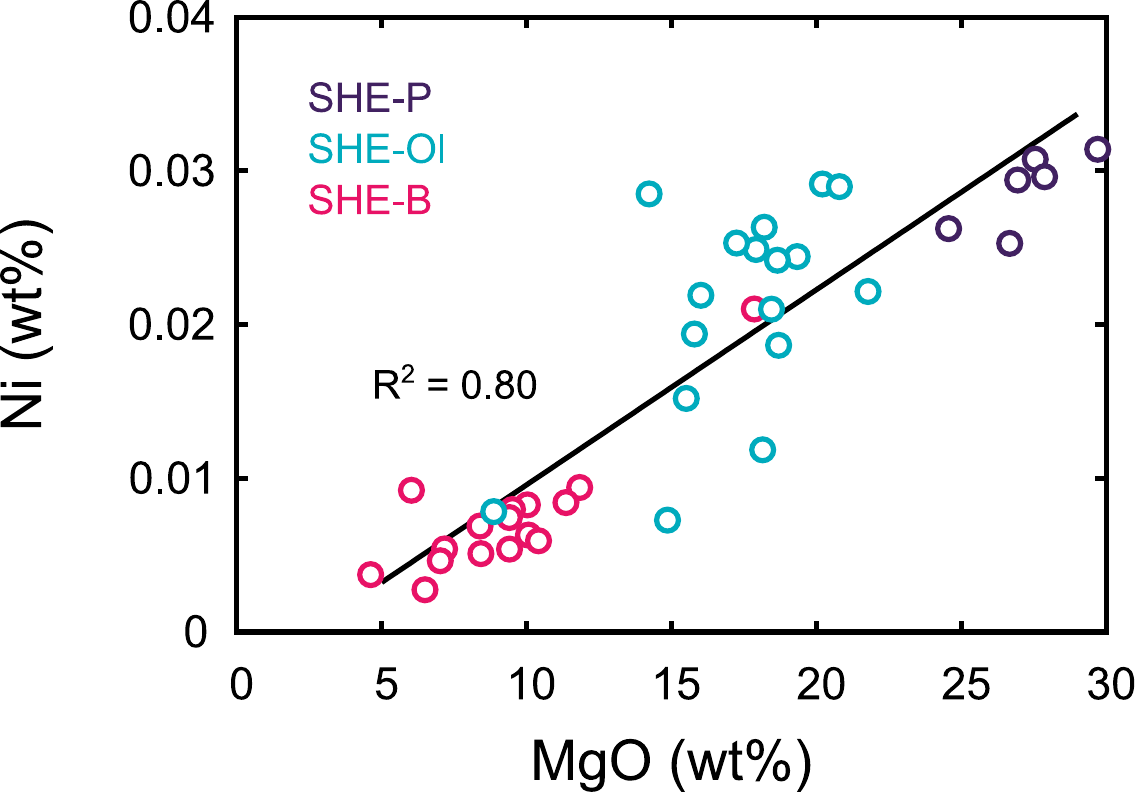}
		\caption{Nickel versus MgO contents in poikilitic, olivine-phyric and basaltic shergottites.}
		\label{fig:Ni}
	\end{figure}
	\clearpage	
	
	\begin{figure}[h]
		\centering
		\includegraphics[width=1\linewidth]{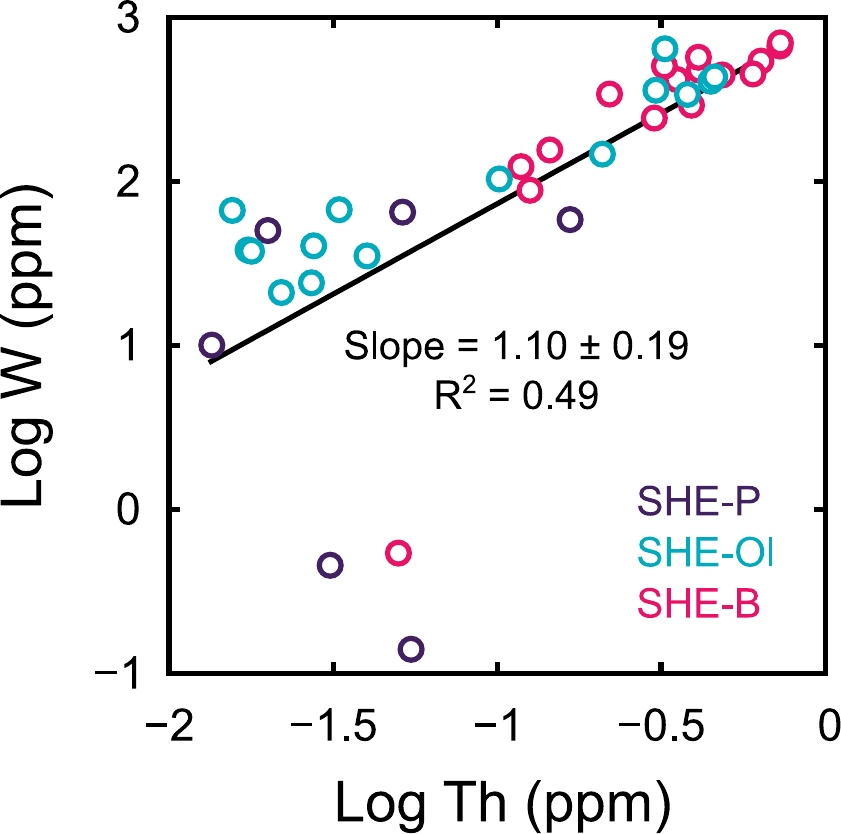}
		\caption{Log-log plot of W and Th in poikilitic, olivine-phyric and basaltic shergottites.}
		\label{fig:W}
	\end{figure}
	\clearpage
	
	\begin{figure}[h]
		\centering
		\includegraphics[width=1\linewidth]{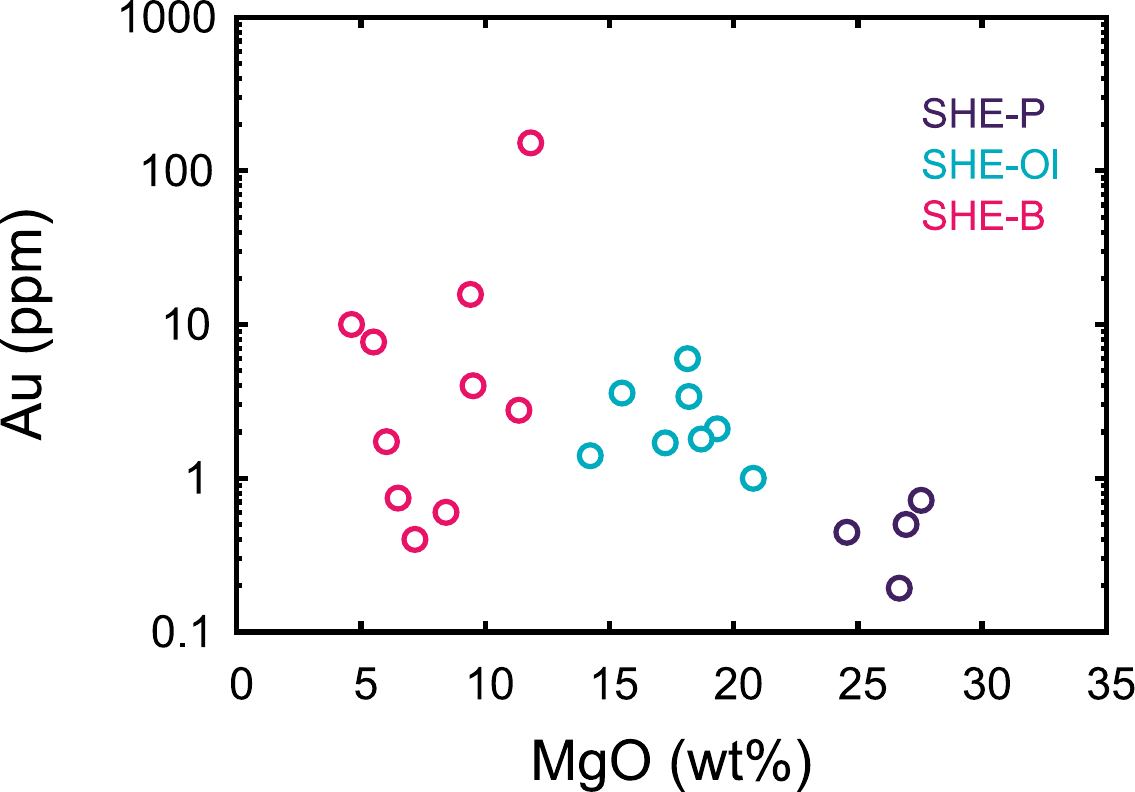}
		\caption{Gold versus MgO contents in poikilitic, olivine-phyric and basaltic shergottites.}
		\label{fig:Au}
	\end{figure}
	\clearpage	
	
	\begin{figure}[h]
		\centering
		\includegraphics[width=1\linewidth]{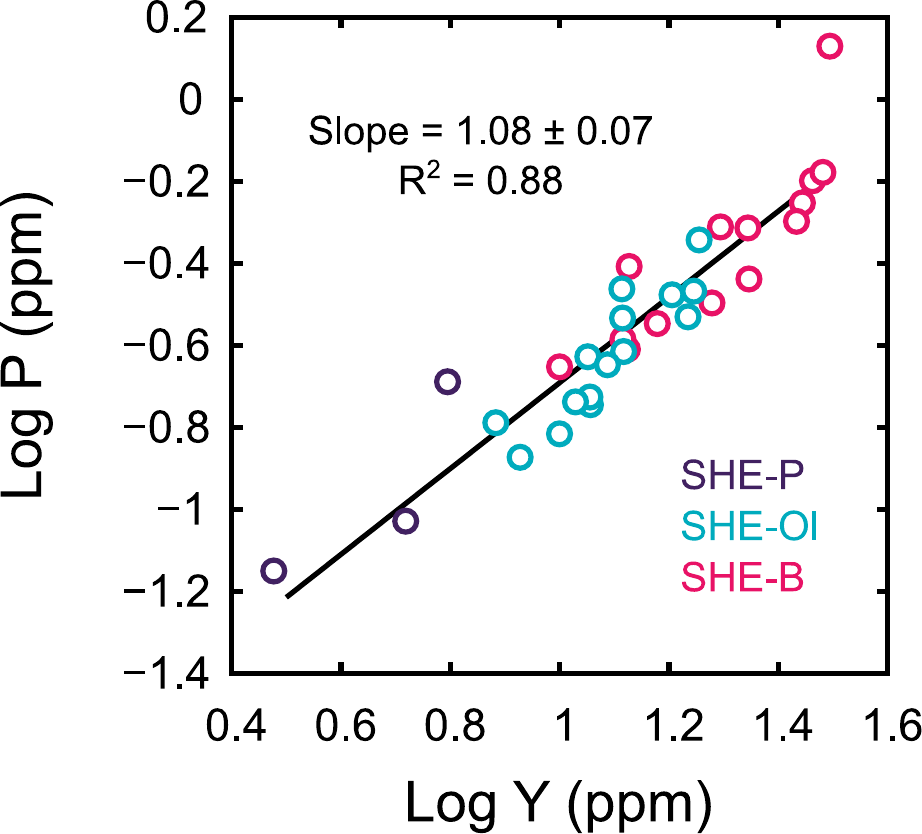}
		\caption{Log-log plot of P and Y in poikilitic, olivine-phyric and basaltic shergottites.}
		\label{fig:P}
	\end{figure}
	\clearpage
	
	\begin{figure}[h]
		\centering
		\includegraphics[width=1\linewidth]{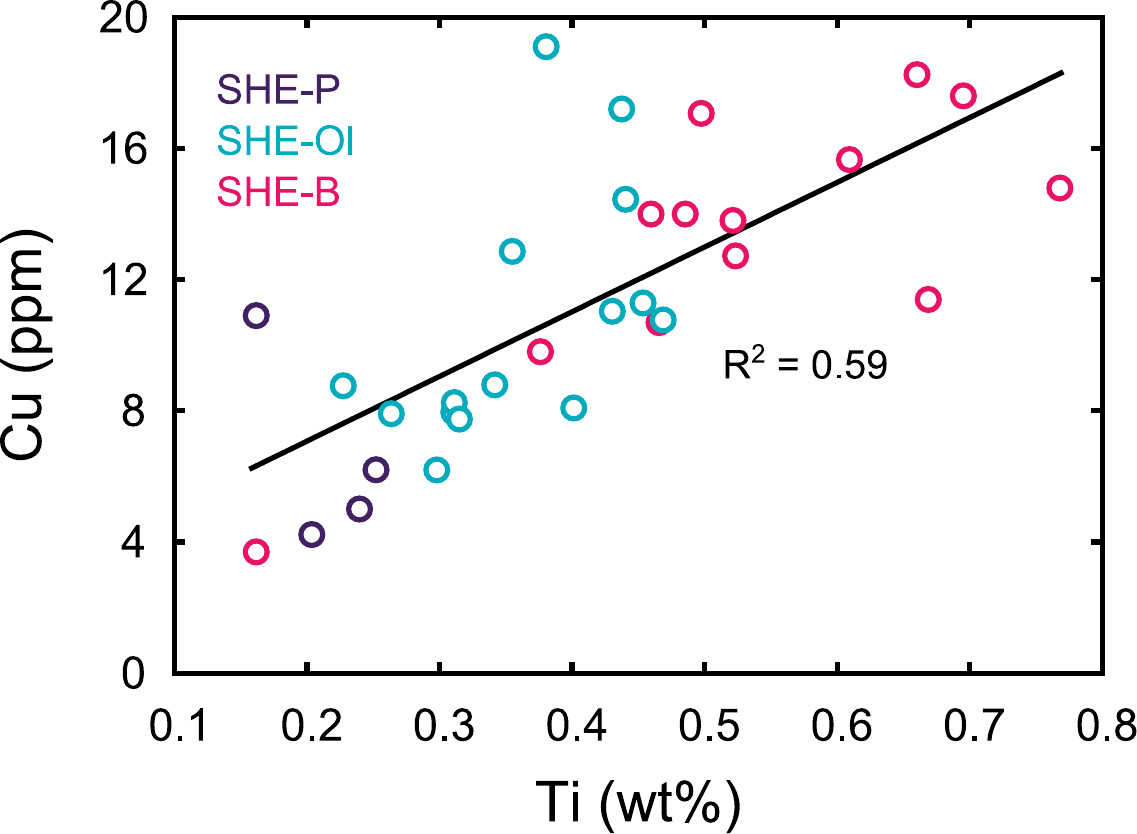}
		\caption{Copper versus Ti contents in poikilitic, olivine-phyric and basaltic shergottites.}
		\label{fig:Cu}
	\end{figure}
	\clearpage	
	
	\begin{figure}[h]
		\centering
		\includegraphics[width=1\linewidth]{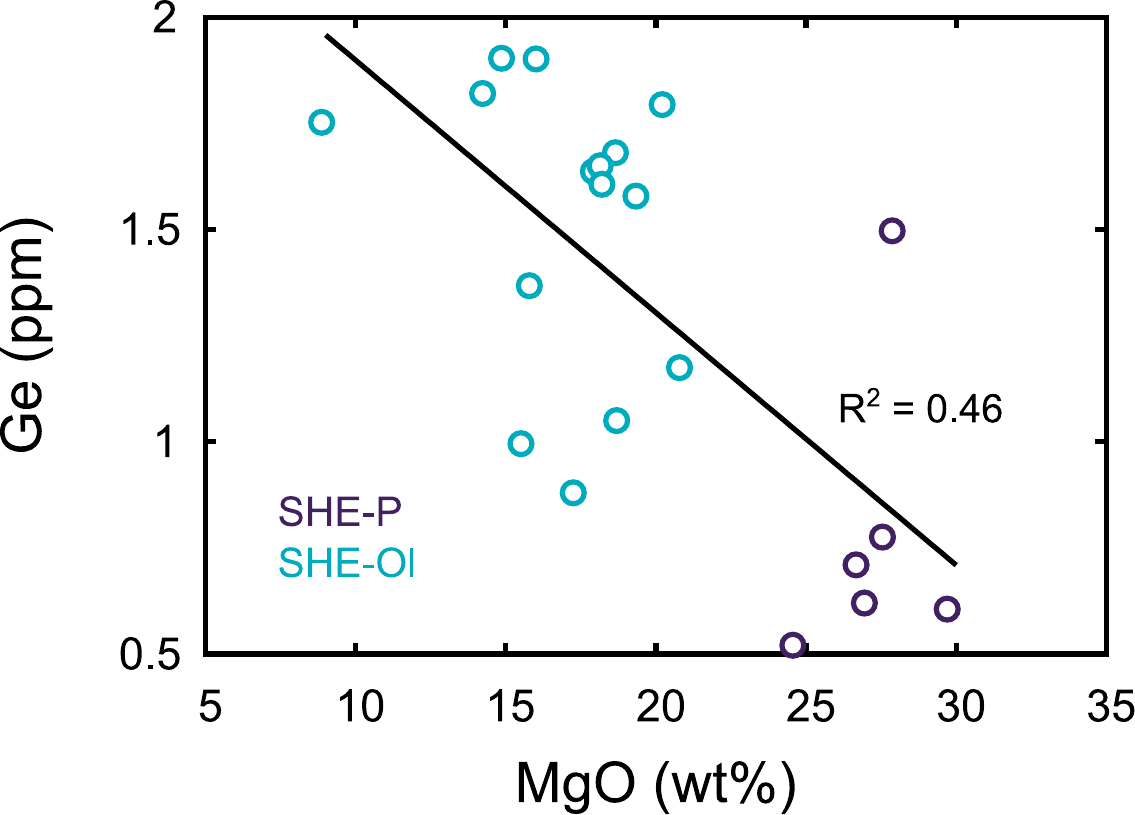}
		\caption{Germanium versus MgO contents in poikilitic and olivine-phyric shergottites.}
		\label{fig:Ge}
	\end{figure}
	\clearpage	
	
	\begin{figure}[h]
		\centering
		\includegraphics[width=1\linewidth]{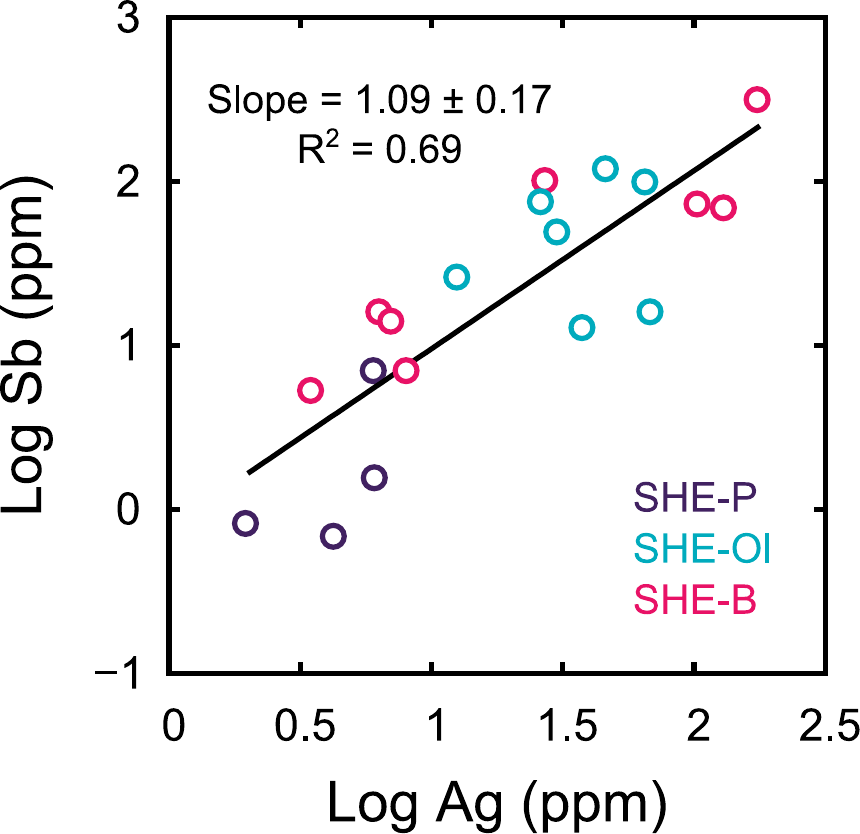}
		\caption{Log-log plot of Sb and Ag in poikilitic, olivine-phyric and basaltic shergottites.}
		\label{fig:Sb}
	\end{figure}
	\clearpage
	
	\begin{figure}[h]
		\centering
		\includegraphics[width=1\linewidth]{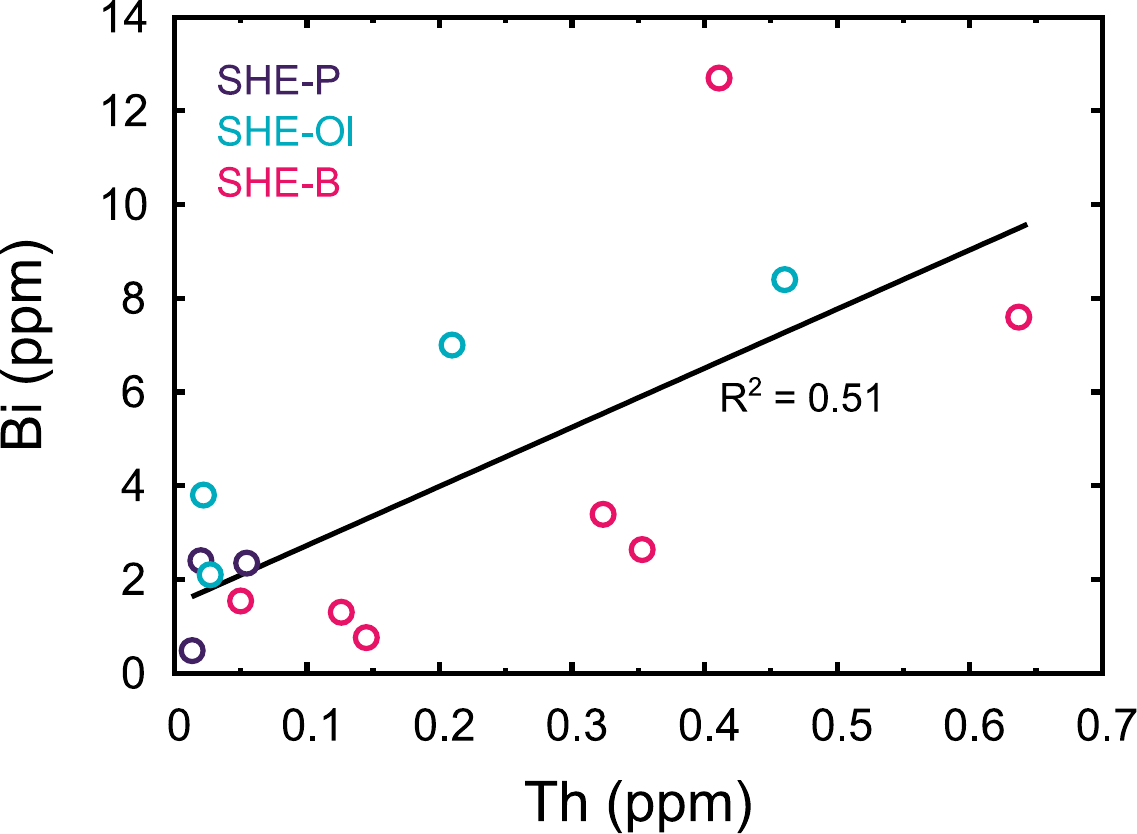}
		\caption{Bismuth versus Th contents in poikilitic, olivine-phyric and basaltic shergottites.}
		\label{fig:Bi}
	\end{figure}
	\clearpage	
	
	\begin{figure}[h]
		\centering
		\includegraphics[width=0.8\linewidth]{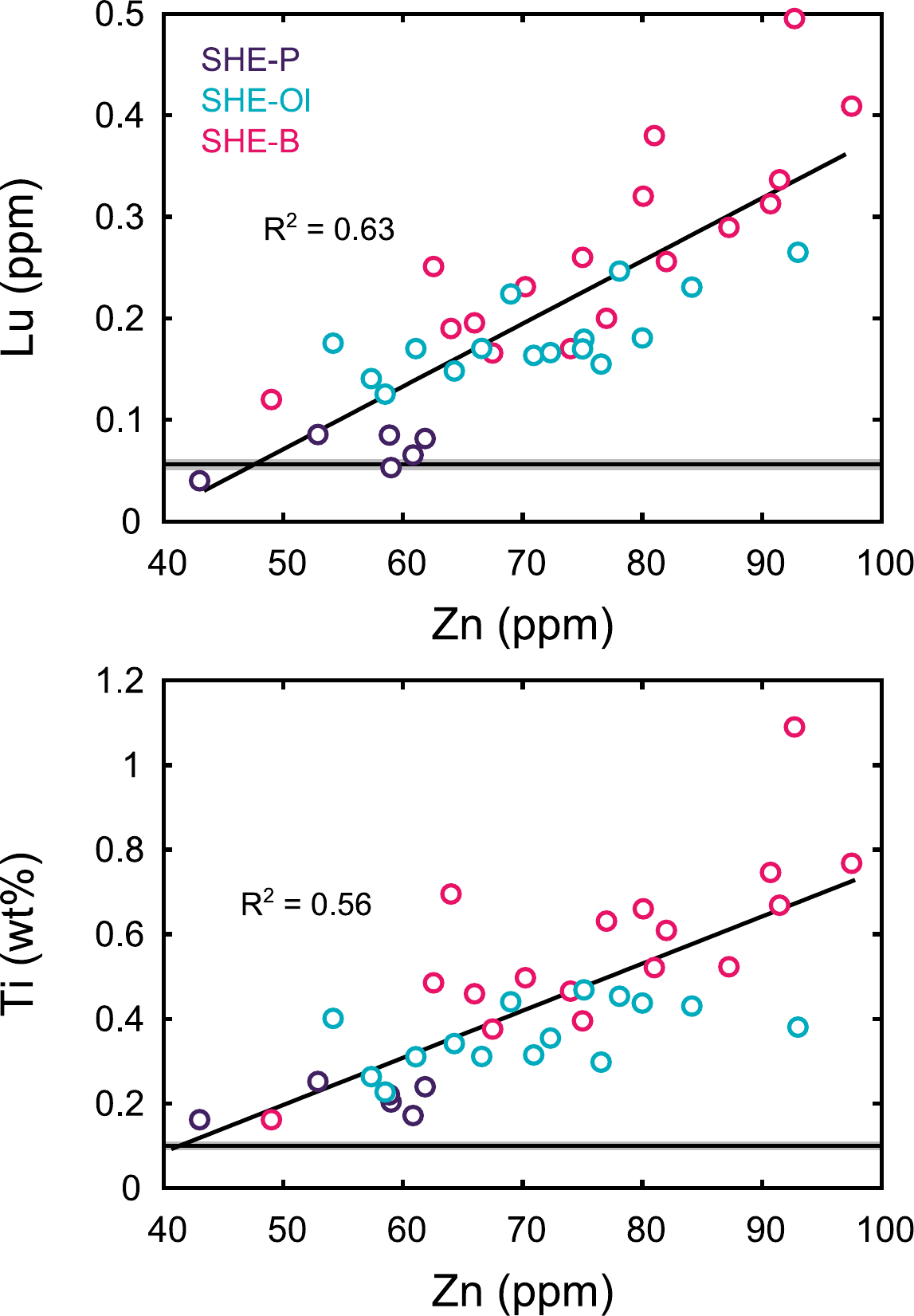}
		\caption{Zinc versus Lu (top) and Zn versus Ti (bottom) contents in poikilitic, olivine-phyric and basaltic shergottites. Horizontal lines and gray bands present the BSM abundances of Lu and Ti, and their uncertainties, respectively (\cref{tab:BSM_all}).}
		\label{fig:Zn}
	\end{figure}
	\clearpage	
	
	\begin{figure}[h]
		\centering
		\includegraphics[width=0.8\linewidth]{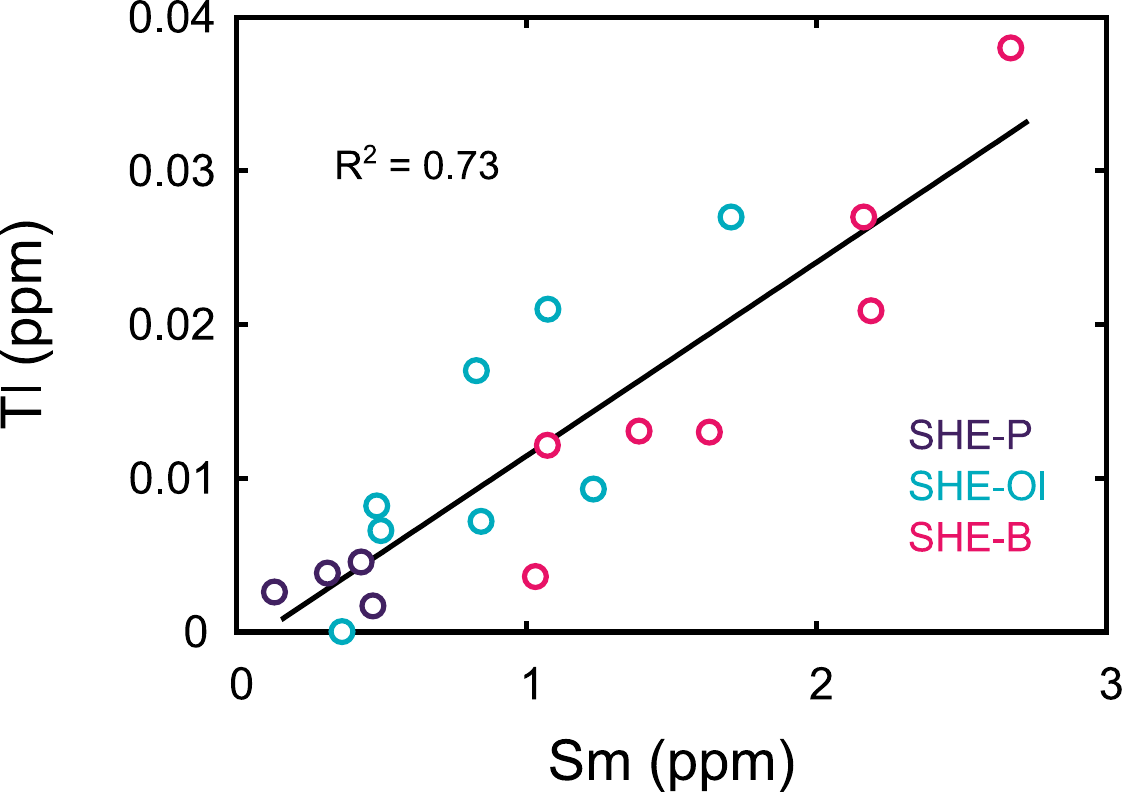}
		\caption{Thallium versus Sm contents in poikilitic, olivine-phyric and basaltic shergottites.}
		\label{fig:Tl}
	\end{figure}
	\clearpage	
	
	\begin{figure}[h]
		\centering
		\includegraphics[width=0.8\linewidth]{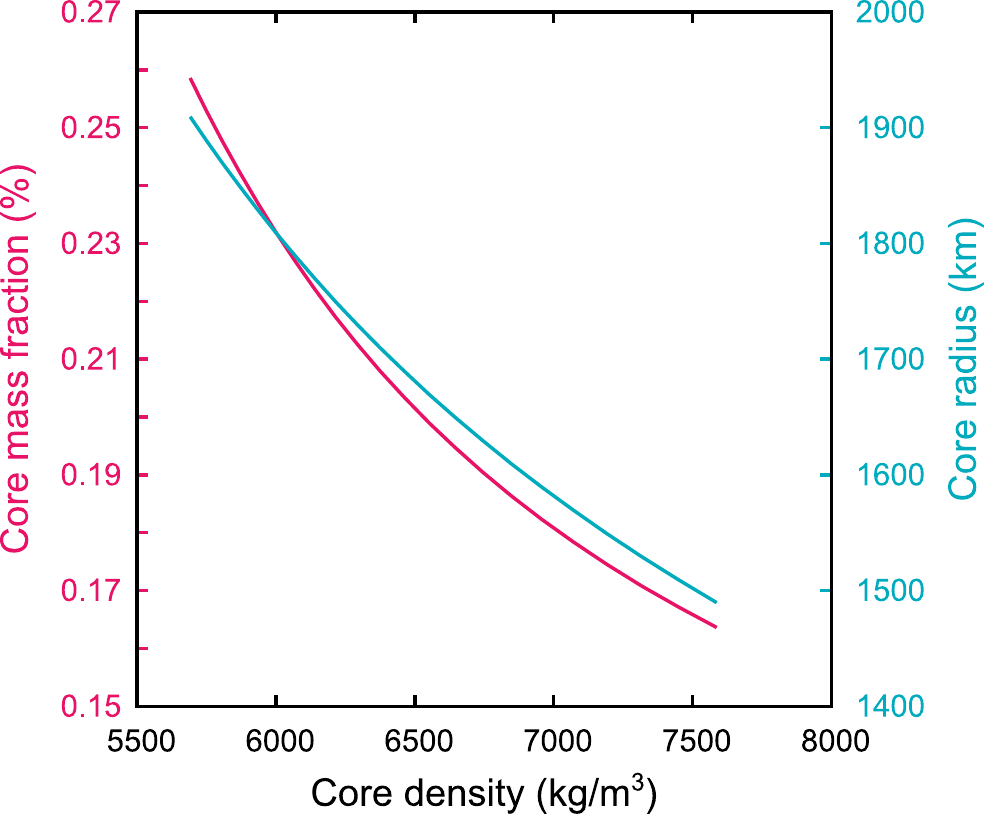}
		\caption{Mass fraction and radius of the Martian core as a function of its density that are consistent with the BSM model presented in this study.}
		\label{fig:tradeoffs}
	\end{figure}
	\clearpage

\end{appendices}

\end{document}